\documentclass[%
 reprint,prb,
 amsmath,amssymb,
 aps,
]{revtex4-2}
\setcitestyle{super}

\usepackage{comment}
\usepackage{graphicx}
\usepackage{dcolumn}
\usepackage{bm}
\usepackage{xcolor}
\usepackage{hyperref}

\begin{document}

\title{Common errors in BoltzTraP-based calculations}

\author{Øven A. Grimenes}
 \email{oven.andreas.grimenes@nmbu.no}
\affiliation{
 Department of Mechanical Engineering and Technology Management, \\ Norwegian University of Life Sciences, NO-1432 Ås, Norway}

\author{Kristian Berland}
\email{kristian.berland@nmbu.no}
\affiliation{
 Department of Mechanical Engineering and Technology Management, \\ Norwegian University of Life Sciences, NO-1432 Ås, Norway}

\date{\today}

\begin{abstract}
Boltzmann transport calculations based on band structures computed from first principles play an important role in modern thermoelectric materials research. Among available codes, the \textsc{BoltzTraP} code is the most widely adopted, but many recent studies contain systematic mistakes. We identify three error modes: (1) inserting the electronic thermal conductivity at zero electric field, $\kappa_0$, in place of the electronic thermal conductivity at zero electric current, $\kappa_e$, (2) computing the figure of merit $zT$ by combining a constant relaxation time of unity while keeping the lattice thermal conductivity $\kappa_\ell$ in standard units, and (3) doing both errors at once. We have found many examples of the third error, but since the first two are simpler, we suspect they are also present in the literature. For the single parabolic band model, we derive exact analytical limits in the non-degenerate and near-degenerate regimes, and we show how mistakes appear for the realistic case study of ZrNiSn. Our results illustrate how faulty calculations can appear reasonable at certain temperatures and Fermi levels, and we provide practical guidance for identifying faulty results and avoiding such pitfalls in thermoelectric transport studies.
\end{abstract}

\maketitle

\section{Introduction}

Thermoelectric materials have long attracted sustained interest for energy conversion and solid-state cooling applications.\cite{snyderComplexThermoelectricMaterials2008,heAdvancesThermoelectricMaterials2017} 
Increasingly, \textit{ab initio} calculations combined with Boltzmann transport theory have become central to the search and analysis of thermoelectric materials. In turn, these methods have been adopted beyond computational solid-state physics to broader engineering and materials design contexts. While this broader adoption can accelerate discovery, it also carries the risk that core physical assumptions become obscured or misapplied, especially when computational tools are treated as black boxes. Combined with time-constrained or non-expert peer review, methodological errors can spread. 

The problem is particularly evident for the case of \textsc{BoltzTraP} (version~1) \cite{madsenBoltzTraPCodeCalculating2006b}, a cornerstone method in modern thermoelectric research. It can straightforwardly compute electronic transport coefficients from electronic band structures calculated with density functional theory (DFT). Many recent results, however, are incorrect. 

In this work, we first review Boltzmann transport theory in the relaxation-time approximation, derive generic results for the parabolic-band system, and analyze  results both correct and incorrect expressions. Using ZrNiSn as a case study, we reproduce representative errors from the literature and contrast them with correct calculations. Our aims are twofold: (i) to provide a reference for new users to avoid pitfalls in computing transport properties, and (ii) to equip journal reviewers and editors without direct experience in \textsc{BoltzTraP} with practical diagnostic tools to identify faulty results. We conclude with recommendations for best practice.

\section{Theory}
\subsection{Boltzmann transport theory}

The charge current density $\mathbf{J}$ and electronic heat current density $\mathbf{J}^Q$ in a material under an electric field $\mathbf{E}$ and a temperature gradient $\nabla T$ can be written in the coupled linear-response form \cite{mahanFigureMeritThermoelectrics1989}
\begin{align}
    \mathbf{J} &= \sigma \mathbf{E} + \sigma S\,(-\nabla T), \\
    \mathbf{J}_Q &= T\, (\sigma S) \mathbf{E} + \kappa_{0}\,(-\nabla T) .
\end{align}
Here $\sigma$ is the electronic conductivity, $S$ is the Seebeck coefficient, and $\kappa_0$ is the electronic thermal conductivity at zero electric field. 

In the following, we consider transport along a single Cartesian direction ($x$) and isotropic systems. For a full tensors-based description, see Ref.~[\onlinecite{biesThermoelectricPropertiesAnisotropic2002}].
In the linearized Boltzmann transport theory within the relaxation time approximation, the generalized transport coefficients can be written as
\begin{equation}
    \mathcal{L}_m(\mu_\mathrm{F}, T)
    = q^2 \int_{-\infty}^{\infty}  d\varepsilon\,\Sigma(\varepsilon)\,
      W_m(\varepsilon; \mu_\mathrm{F}, T)
\end{equation}
where $\mu_\mathrm{F}$ is the Fermi level and $q>0$ is the elementary charge. The transport distribution $\Sigma(\varepsilon)$ depends only on the band structure and the scattering model\cite{mahanBestThermoelectric1996, madsenBoltzTraPCodeCalculating2006b}:
\begin{equation}
    \Sigma(\varepsilon) = g_s
    \int_{\mathrm{BZ}} \frac{d\mathbf{k}}{8\pi^3} \sum_n
    \left(v^{x}_{n\mathbf{k}}\right)^2\,
    \tau_{n\mathbf{k}}\,
    \delta\!\left(\varepsilon - \varepsilon_{n\mathbf{k}}\right),
\end{equation}
where $v^{x}_{n\mathbf{k}}$ is the group velocity, $\tau_{n\mathbf{k}}$, the relaxation time, and $\delta$ is the Dirac delta function, and $g_s$ is the spin degeneracy. For isotropic systems, averaging over the three directions gives
\begin{equation}
    \Sigma(\varepsilon) =
   \frac{g_s}{3} \int_{\mathrm{BZ}} \frac{d\mathbf{k}}{8\pi^3} \sum_n
    \| v_{n\mathbf{k}}\|^2\,
    \tau_{n\mathbf{k}}\,
    \delta\!\left(\varepsilon - \varepsilon_{n\mathbf{k}}\right), \label{eq:sigma_iso}
\end{equation}
The temperature and Fermi-level dependence enter through the selection functions
\begin{equation}
    W_m(\varepsilon; \mu_\mathrm{F}, T) =
    (\varepsilon - \mu_\mathrm{F})^m
    \left(-\frac{\partial f(\varepsilon; \mu_\mathrm{F}, T)}{\partial \varepsilon}\right)\,,
\end{equation}
where $f$ is the Fermi–Dirac function.

From $\mathcal{L}_m$, one can express
\begin{equation}
    \sigma = \mathcal{L}_0, \qquad
    S= \frac{\mathcal{L}_1}{qT \mathcal{L}_0}, \qquad
    \kappa_{0} = \frac{\mathcal{L}_2}{q^2 T}\,.
\end{equation}

For materials with a non-zero Seebeck coefficient, a temperature gradient induces an internal electronic current which reduces the net heat conduction, so that the physically relevant coefficient is the electronic thermal conductivity at zero electric current, $\kappa_e$ \cite{madsenBoltzTraP2ProgramInterpolating2018b, mahanFigureMeritThermoelectrics1989}, given by 
\begin{equation}
    \kappa_{e} = \kappa_{0} - S^2 \sigma T
    = \frac{1}{q^2 T} \left[
      \mathcal{L}_2
      - \frac{\mathcal{L}_1^2}{\mathcal{L}_0}
    \right]\,,
\end{equation}
and this is the appropriate one to use for the thermoelectric figure of merit,\cite{mahanFigureMeritThermoelectrics1989} which is given by
\begin{equation}
    zT = \frac{S^2 \sigma T}{\kappa_{e} + \kappa_\ell},
\end{equation}
where $\kappa_\ell$ is the lattice thermal conductivity.

\subsection{Error modes}

In our literature review we have encountered three error modes:

\subsubsection{Confusing $\kappa_0$ and $\kappa_e$}
In \textsc{BoltzTraP}, the default output for electronic thermal conductivity is $\kappa_0$, whereas $\kappa_e$, the physically relevant quantity, must be computed explicitly by the user. Many newcomers to both thermoelectricity and the code might have this error, although it can sometimes be caught when attempting to reproduce literature data, underlining the importance of correct literature values. Since $\kappa_0$ is consistently larger than $\kappa_e$, the error reduces the computed value of $zT$. We label the resulting figure of merit as
\begin{equation}
    (zT)_< = \frac{S^2 \sigma T}{\kappa_0 + \kappa_\ell}. \label{eq:ZT<}
\end{equation}
If $\kappa_\ell$ is large, this error may go unnoticed, especially at low temperature. In fact, the resulting $zT$ values are conservative and even appear in closer agreement with experiment than a correct computation,
for instance, when using the standard constant electron relaxation lifetime of $\tau = 10 \mathrm{fs}$.

\subsubsection{Failing to provide a relaxation time}
Because electron relaxation lifetimes are difficult to compute, many authors report relaxation-time independent transport properties such as $\sigma/\tau$, $\kappa_e/\tau$, and $S^2\sigma/\tau$,
and this is how the default output from \textsc{BoltzTraP}. 
Done deliberately, this choice in itself is perfectly legitimate and circumvents the need to provide a value of $\tau$. 
 
However, if plugged into an expression for $zT$ without scaling $\kappa_\ell$ accordingly, the result for $zT$ corresponds to the limit $\kappa_e >> \kappa_\ell$ (by about 14 orders of magnitude). This leads to an expression for the so-called electronic figure of merit.
 \begin{equation}
    \label{eq: zTe}
    \left(zT\right)_\mathrm{e} = \frac{S^2 \frac{\sigma}{\tau}T}{\frac{\kappa_e}{\tau} + \kappa_\ell}
    \rightarrow \frac{S^2 \sigma T}{\kappa_e}  =  \frac{\mathcal{L}_1 ^2}{\mathcal{L}_0 \mathcal{L}_2 - \mathcal{L}_1^2}\,,
\end{equation}
which is generally much larger than the conventional $zT$. In the non-degenerate case, the expression scales as $\sim S^2/L$, where $L$ is the Lorenz number, and hence diverges as $T \rightarrow 0$. Some authors consciously report the $(zT)_\mathrm{e}$, instead of estimating $zT$. We caution that this limit is only an appropriate stand-in or reflection of the potential of the material, for very heavily doped semiconductors and metals at (sufficiently) high temperature. 

In the near-degenerate limit, with the Fermi level set to the band edge, however, $(zT)_\mathrm{e}$ does not diverge. Specifically, in parabolic band model, detailed in appendix \ref{sec:parabolic}, gives the following value
\begin{align}
    \left(zT\right)_\mathrm{e} \approx 3.11\,.
\end{align}
Note that $zT$ values of $3$ or more can result from perfectly valid \textsc{BoltzTraP} calculations (within the assumptions made) and such values can arise for instance in high-throughput screening of hypothetical materials. 
Thus, there is a risk that that values of this magnitude computed incorrectly are reported. 


\subsubsection{Using $\kappa_0$ in place of $\kappa_\mathrm{e}$ and failing to provide $\tau$}

A particularly pervasive error mode combines the two errors discussed above, effectively calculating
 \begin{equation}
    \label{eq: zT0}
    \left(zT\right)_\mathrm{0} = \frac{S^2 \sigma T}{ \kappa_0} = \frac{\mathcal{L}_1 ^2}{\mathcal{L}_0 \mathcal{L}_2}\,.
\end{equation}
In the non-degenerate low-temperature limit, in the parabolic band approximation (detailed in App.~\ref{sec:parabolic}), this error gives the exact limit   
\begin{equation}
    \left(zT\right)_0 \rightarrow 1\,.
\end{equation}
In the near-degenerate limit, i.e., $\mu_\mathrm{F} = 0$, the parabolic band model gives (for all temperatures) 
\begin{equation}
    \left(zT\right)_0 \approx 0.757\,.
\end{equation}

\section{Case study: $\textrm{ZrNiSn}$}
To illustrate what types of results and corresponding figures 
emerge in the different error modes, we here consider the case of ZrNiSn, a widely studied thermoelectric material. 

\subsection{Computational details}
The DFT calculations were based on VASP \cite{kresseUltrasoftPseudopotentialsProjector1999,vasp4} code (version 6.4.1) using the Zr${\_}$sv, Ni, and Sn${\_}$d pseudopotentials. Exchange-correlation functional choices are not always stated in papers, but is an import detail to ensure reproducibility. We used the  vdW-DF-cx functional \cite{behy14} to relax the unit cell dimensions to forces within 10$^{-4}$\;eV/Å and the PBE functional \cite{pbe1996} for electronic structure. In all DFT calculations, a plane-wave energy cutoff of 520\;eV and an electronic self-consistency criterion of 10$^{-7}$\;eV. A $\mathbf{k}$-point grid of  12$\times$12$\times$12 was used for relaxations and a 40$\times$40$\times$40 for the electronic band structure calculations. The electronic transport properties were calculated in \textsc{BoltzTraP2},\cite{madsenBoltzTraP2ProgramInterpolating2018b} with a band interpolation factor of 5. In computing $zT$, we set $\kappa_\ell = 4$ W/mK.

\subsection{Results}

\begin{figure}[h!]
    \centering
    \includegraphics[width=\linewidth]{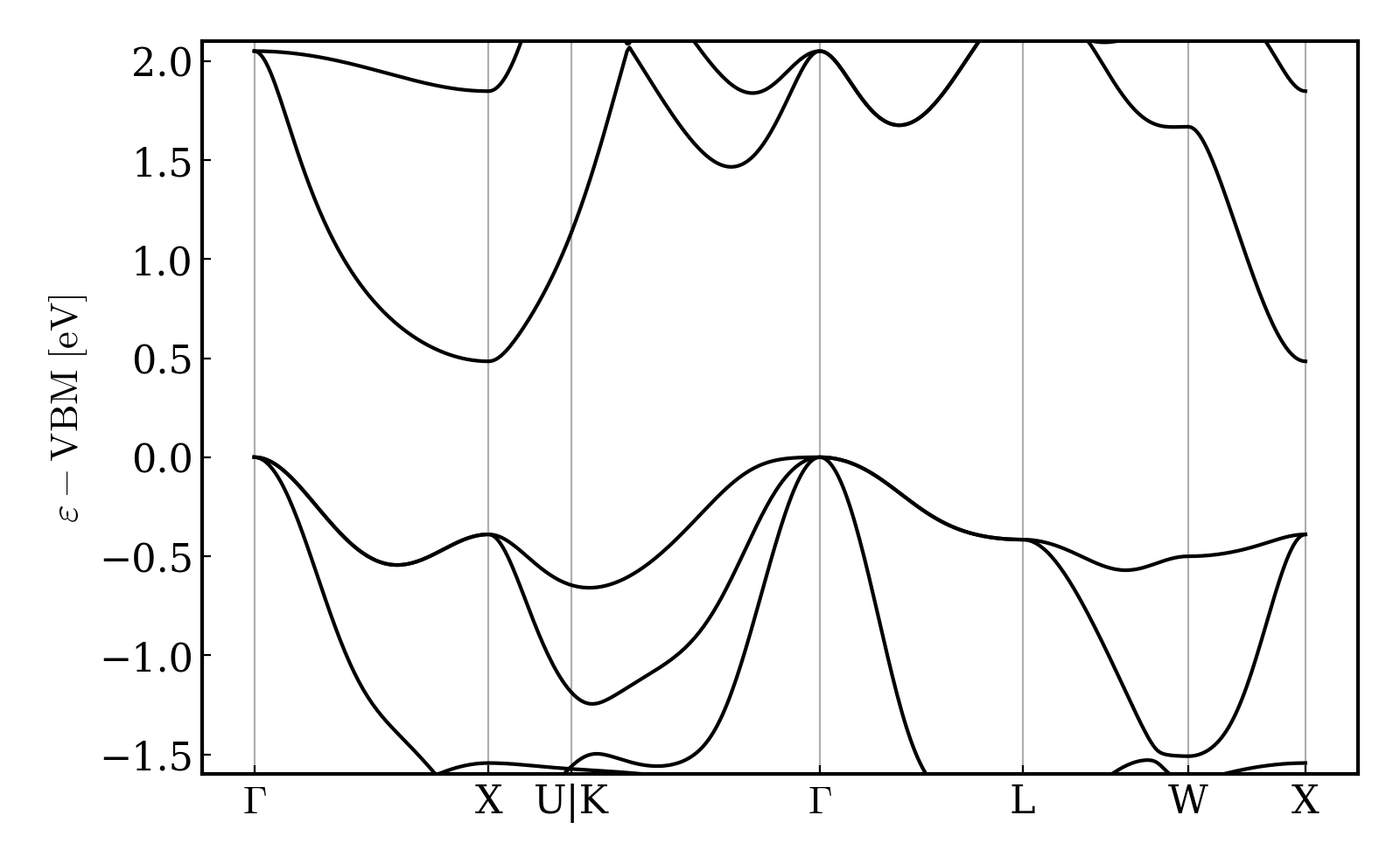}
    \caption{
    Electronic band structure of ZrNiSn calculated with DFT. Band structures are frequently shown in transport studies to help readers connect with the underlying electronic structure and to lend legitimacy to the results. However, this legitimacy applies only to the DFT band structure itself—not necessarily to derived transport properties.
    \label{fig: bands}}
\end{figure}

\begin{figure}[h!]
    \centering
    \includegraphics[width=\linewidth]{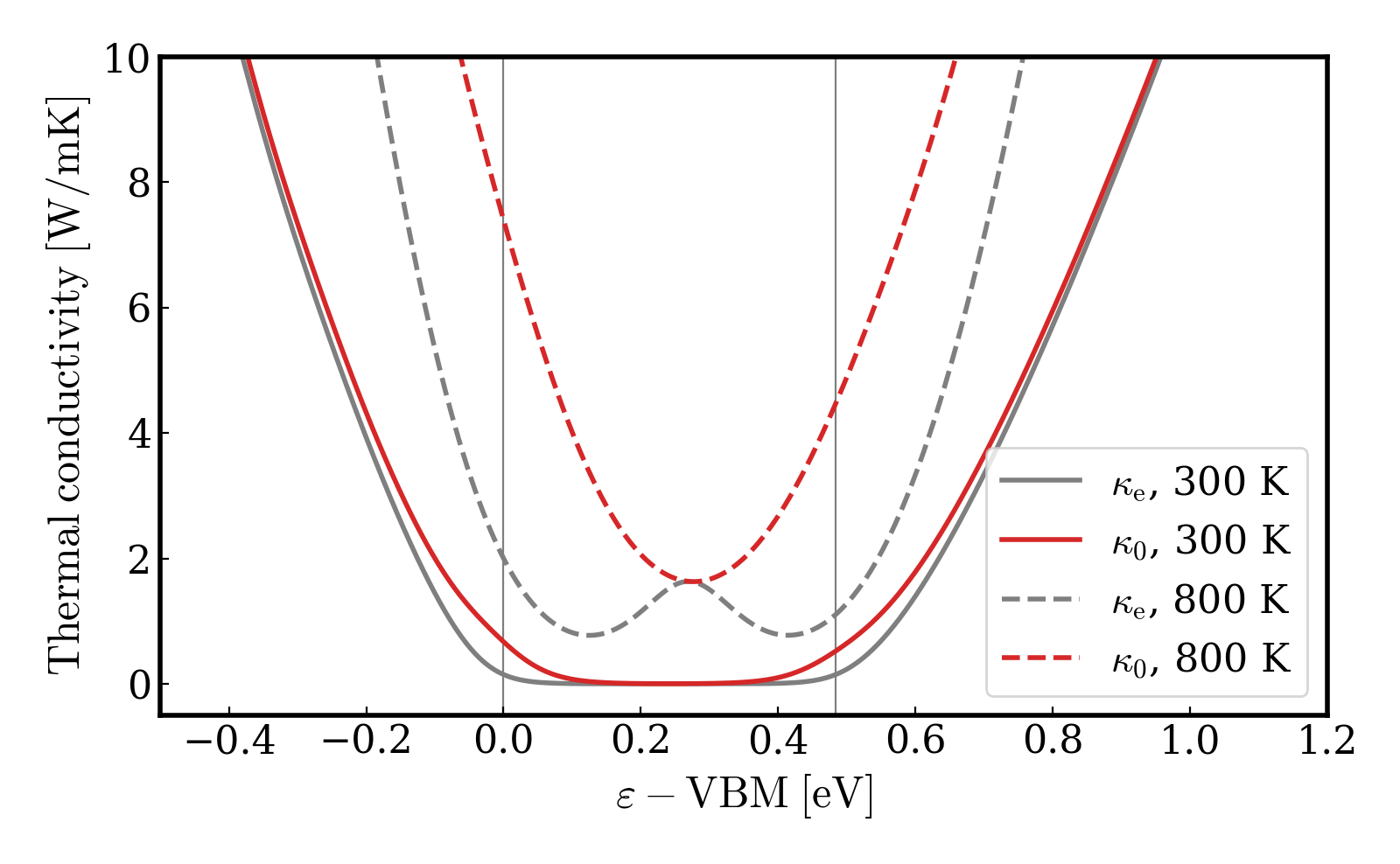}
    \caption{
    Electronic thermal conductivity of ZrNiSn at 300 K and 800 K. For both near- and non-degenerate cases $\kappa_\mathrm{e}$ is significantly smaller than $\kappa_0$. At high temperatures, using $\kappa_0$ may even lead to a fortuitous compensation that counteracts the neglect of $\kappa_\ell$.
    \label{fig: kappa}}
\end{figure}
 
Fig.~\ref{fig: bands} shows The electronic band structure of ZrNiSn and Fig.~\ref{fig: kappa}, the corresponding electronic thermal conductivities. The value of $\kappa_0 - \kappa_e$ at high temperature, with $\mu_\mathrm{F}$ set close to close to the Fermi level, can be quite comparable to the typical range of  $\kappa_\ell$, e.g., $0.5-10 \,{\rm W/mK}$ in  many thermoelectric materials. 

\begin{figure}[h!]
    \centering
    \includegraphics[width=\linewidth]{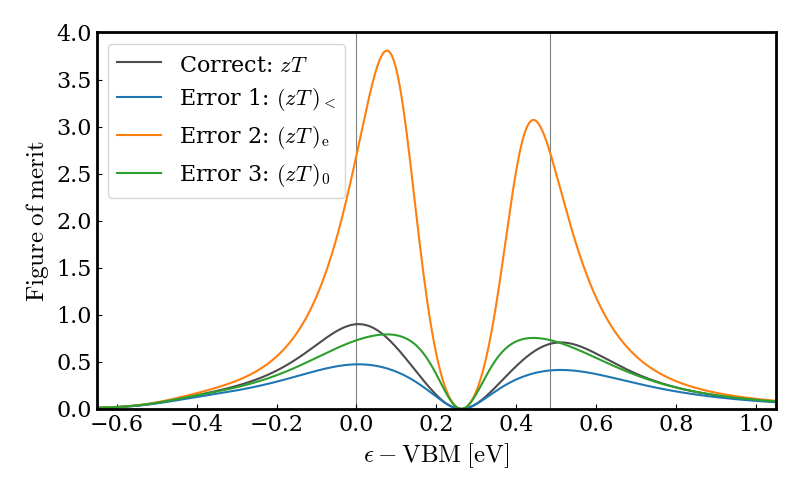}
    \caption{
    Different correct and incorrect expressions for $zT$ at 800 K, corresponding to the three error modes discussed.
    \label{fig: three_zT}}
\end{figure}

Figure~\ref{fig: three_zT} shows the figure of merit of ZrNiSn at 800 K calculated according to the three error modes. At this temperature, each result may fall within ranges seen for thermoelectric materials. The $(zT)_e$ value is large and users making the error are likely to correct it, at least if they consider results at lower temperature as well, but there is a danger that for less viable thermoelectric materials, where the value of $(zT)_\mathrm{e}$ might be more modest, that the error goes unspotted. As shown, error mode 3 ($(zT)_0$) is particularly insidious because its values can be very similar those of correctly evaluated $zT$.

\begin{figure*}[t!]
    \centering
    \includegraphics[width=0.8\linewidth]{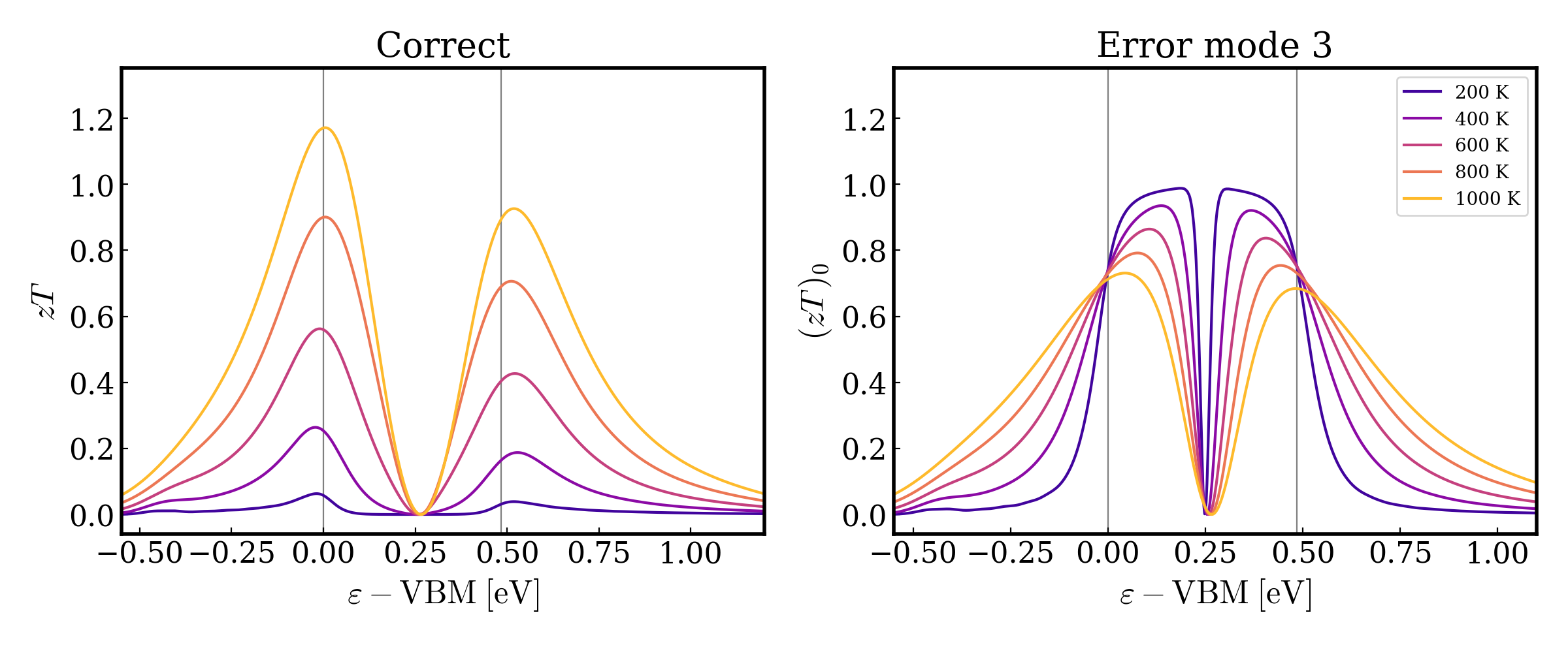}
    \caption{(a) $zT$ and (b) $zT_0$ of ZrNiSn at temperatures 200--1000\;K. 
    \label{fig: zTek}}
\end{figure*}

Figure~\ref{fig: zTek} plot $(zT)_0$ (error mode 3) against the chemical potential at temperatures from 200--1000 K. At low temperatures the $zT$ curve approaches 1 and starts to flatten in the band gap, with a sharp dip where the $S$ changes sign. Curves reminiscent of this shape have been observed in many recent works, including Refs. 
\cite{Aaouita_2025,Abraham_2025,Ahsan_2025,Al-Daraghmeh_2025,Al-Qaisi_2025,Ali_2025a,Amghar_2025,Anuradha_2025,Arbaoui_2025,Benahmedi_2025a,Boutramine_2025,Caid_2025,Choudhary_2025,Dixit_2025,ElGoutni_2025,Fatima_2025,Gurunani_2025,Hongjie_2025,Hongjie_2025a,Iram_2025,Iram_2025a,Jolayemi_2025,Karthikraja_2025,Ketfi_2025,Khan_2025,Khatar_2025,Kumari_2025,Kumari_2025a,Luo_2025,Manzoor_2025,Manzoor_2025a,Manzoor_2025b,Nazir_2025,Nazir_2025a,Nazir_2025b,Nazir_2025c,Nazir_2025d,Owais_2025,Rached_2025,Ramzan_2025a,Righi_2025,Sofi_2025,Srivastava_2025,Tufail_2025,Ullah_2025,Yasin_2025a,Yutomo_2025,Zayed_2025} 
and our case study suggests they arise from error mode 3. Some of them are reported as $zT$ and others as $(zT)_\mathrm{e}$. It is worth noting that in some cases $(zT)_0$ can obtain values higher than 1 at certain energies, but this is unusual. Particularly revealing is that they all have $zT$-curves that center around 0.76 for Fermi levels set to VBM or the conduction band minimum (CBM). This numerical value is in close agreement with the parabolic band results in the near-degenerate limit.

\begin{figure}[h!]
    \centering
    \includegraphics[width=\linewidth]{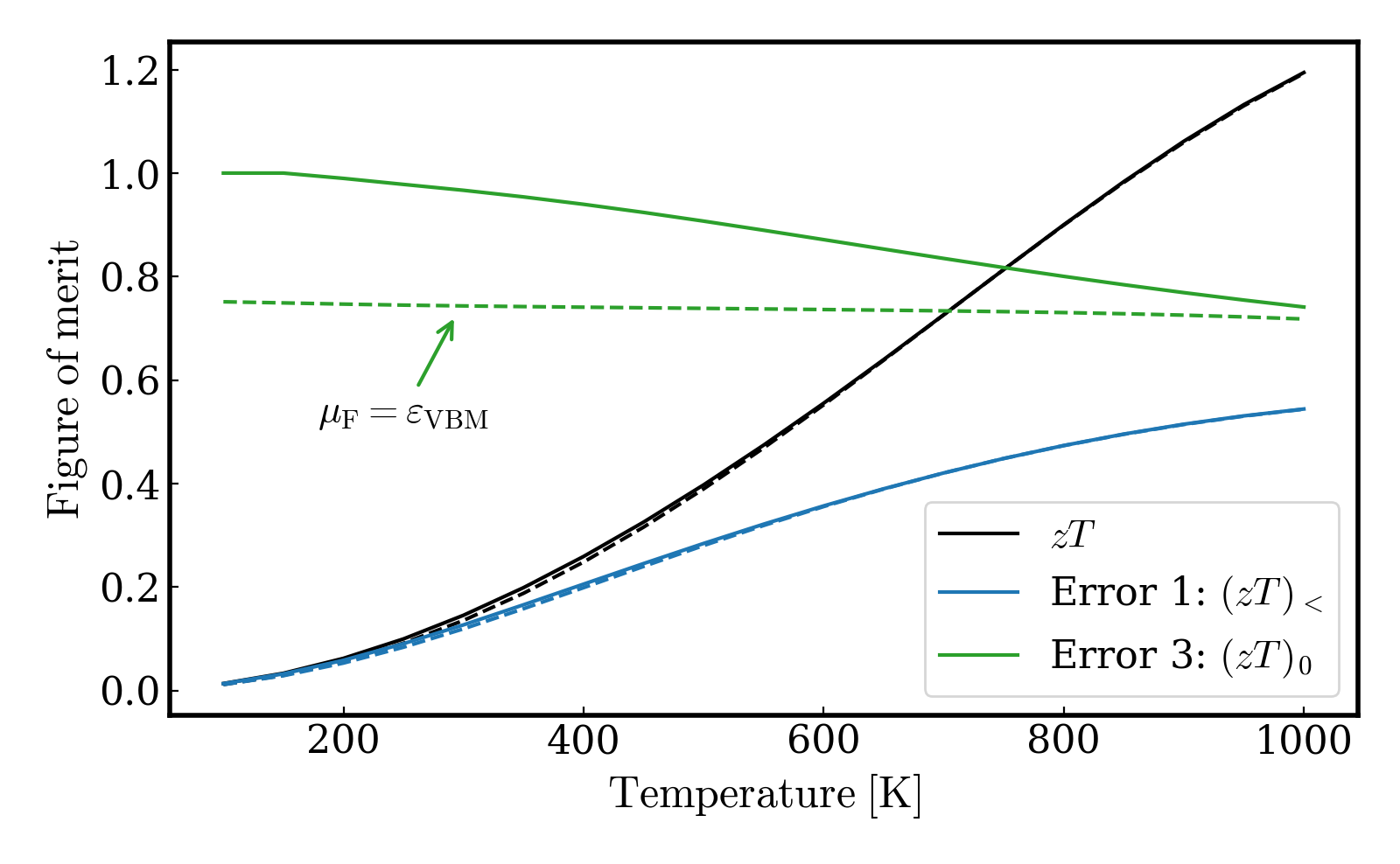}
    \caption{Maximum (p-type) figure of merit of ZrNiSn at different temperatures 100--1000\;K given by full curves, dashed curves gives results with $\mu_\mathrm{F}$ set to the VBM.
    \label{fig: zT_T}}
\end{figure}

Figure \ref{fig: zT_T} plots the maximum $zT$ with $\mu_\mathrm{F}$ re-optimized for each temperature and compared with $\mu_\mathrm{F}$ set to the VBM (dashed curve), where a correct calculation is compared with error modes 1 and 3. As before, replacing $\kappa_e$ with $\kappa_0$ provides results that appear correct but conservative. The pervasive error mode 3, however, clearly gives an incorrect temperature dependence. 

Curves that go to unity at low temperature, consistent with this error mode, have been observed in the following papers 
\cite{Abbas_2025,Al-Anazy_2025,Al-Anazy_2025a,Al-Hmoud_2025,Albalawi_2025,Ali_2025,Almeshal_2025,Ayyaz_2025,Bellahcene_2025,Benahmedi_2025,Dahane_2025,Das_2025,Farah_2025,Lemnawar_2025,Mechehoud_2025,Mechehoud_2025a,Mohamed_2025,Ramzan_2025,SaadH.-E._2025,SaadH.-E._2025a,Shah_2025,Yasin_2025}. 
The dashed curves in this case become almost a flat, consistent with the near-degenerate parabolic-band limit of 0.76. Curves of similar magnitudes and temperature dependence have been observed in the following papers
\cite{Ahmed_2025,Albalawi_2025a,Algethami_2025,Alsaiari_2025,AlShaikhMohammad_2025,Ashik_2025,Ayyaz_2025a,Ayyaz_2025b,Ayyaz_2025c,Baaalla_2025,Benahmedi_2025b,Bouferrache_2025,Bourahla_2025,Chelh_2025,ElAkkel_2025,Farooq_2025,Hamza_2025,Israr_2025,Khatar_2025a,Mbilo_2025,Mbilo_2025a,Musanyi_2025,Mustafa_2025,Nasarullah_2025,Nasarullah_2025a,Nazir_2025e,Noor_2025,Noor_2025a,Saeed_2025,Safdar_2025,Yaseen_2025,Yasir_2025,Yasir_2025a}.

Our literature survey was limited to 2025, however, this error has also been observed in earlier work. We emphasize that although the types of figure of merit curves illustrated here are incorrect, other results in these works may still be valid and physically meaningful.

\section{Discussion}
We speculate that the pervasive errors seen in published \textsc{BoltzTraP}-based calculations stem from a mindset where numerical results are treated as arbitrary computer output, rather than as the product of deterministic algorithms, numerically exact within the assumptions of the method and the reliability of the underlying code.

Other common issues in the field include insufficient convergence studies, blurred boundaries between computational and experimental results, and a lack of computational details sufficient to reproduce published work. The trend of relegating methodological descriptions to appendices or supplementary materials may further compound the problem by downplaying the importance of clear and transparent method descriptions. 

How should such issues be addressed? Broader efforts to improve scientific practices, such as data-management schemes or unifying notation, while valuable, do not address the key issue. What is needed is a working mentality that emphasizes physical intuition or even adversarial curiosity, where one systematically attempts to uncover flaws in one's own work, for example by plotting curves beyond the limits of immediate experimental interest. The starting point of any simulation project with unfamiliar code should be to reproduce benchmark results from the early literature.

We have identified and analyzed three recurring error modes in \textsc{BoltzTraP}-based thermoelectric calculations. Although only one of these appears frequently in recent literature, the others are simpler and therefore likely present as well. Using both analytic limits of the parabolic-band model and a case study of ZrNiSn, we have shown how such mistakes can yield results that look numerically reasonable but are physically inconsistent. The main lesson is that reproducibility, sanity checks, and awareness of the correct definitions of $\kappa_e$, $\kappa_0$, and the role of $\tau$ are essential for reliable transport modeling. We hope that the diagnostic guidance presented here will help both authors and reviewers to identify faulty calculations and thereby strengthen the reliability of thermoelectric research.

\begin{acknowledgments}
This work is funded by the Research Council of Norway through the Allotherm project (Project No. 314778). Computational resources were provided by the Norwegian e-infrastructure for research and education, Sigma2, through grants No. nn9711k.
\end{acknowledgments}

\appendix
\subsection{Parabolic band results \label{sec:parabolic}} 

Here we provide details of the isotropic parabolic band model in the constant relaxation time approximation.  
The model provides exact limits which can be used to analyze the different error modes. 
Given, 
\begin{equation}
    \epsilon_k = \hbar^2 k^2/2m\,,
\end{equation}
we get
\begin{equation}
    v_g = \frac{1}{\hbar} \left|\nabla \varepsilon_k\right|= \hbar k/m\,\,,
\end{equation}
and
\begin{equation}
    \left(v_g\right)^2 = 2 m \epsilon_k\,.
\end{equation} 

Evaluating Eq.~\ref{eq:sigma_iso}
 setting $g_s=2$, we get in the constant relaxation time approximation,
\begin{equation}
    \Sigma(\varepsilon) =  C\varepsilon^{3/2}.
\end{equation}
with $C \equiv \frac{2}{3\pi^2}\frac{\sqrt{2 m}}{\hbar^3}\,\tau$,

In terms of the (unnormalized) complete Fermi integrals \cite{limeletteThermopowerFigureMerit2021} 
\begin{equation}
    F_j(\eta)\equiv \int_{0}^{\infty}\frac{t^j}{1+e^{t-\eta}}\,dt.
\end{equation}
the generalized transport coefficient can be expressed in terms of the reduced parameter
$\eta \equiv \mu_\mathrm{F}/(k_B T)$,
as follows:
\begin{align}
    \mathcal{L}_0 & =  q^2 C \,(k_B T)^{3/2} \left[ \frac{3}{2} F_{1/2}(\eta) \right],\\
    \mathcal{L}_1&= q^2 C \,(k_B T)^{5/2} 
    \left[\frac{5}{2}F_{3/2}(\eta) - \frac{3}{2}\eta F_{1/2}(\eta)\right],\\
    \mathcal{L}_2 &= q^2 C \,(k_B T)^{7/2}
    \left[\frac{7}{2}F_{5/2}(\eta) - 5\eta F_{3/2}(\eta) + \frac{3}{2}\eta^{2} F_{1/2}(\eta)\right].
\end{align}
 It is practical to express these with $\mathcal{L}_0$ as denominator: 
\begin{equation}
    \frac{\mathcal{L}_1}{ \mathcal{L}_0} =\frac{5}{3} \frac{F_{3/2}(\eta)}{ F_{1/2}(\eta)} k_\mathrm{B}T - \mu_\mathrm{F}
\end{equation}
\begin{equation}
    \frac{\mathcal{L}_2}{\mathcal{L}_0} = 
    (k_\mathrm{B} T)^2 \frac{7}{3}\frac{ F_{5/2}(\eta)}{F_{1/2}(\eta)} -\frac{10}{3}(k_\mathrm{B} T) \mu_\mathrm{F}  \frac{F_{3/2}(\eta)} {F_{1/2}(\eta)}
    -\mu_\mathrm{F}^2\,.
\end{equation}
as these expressions are related to $S$ and the Lorenz number, and figure of merit expressions with $\kappa_\ell =0$ in Eqs.~\ref{eq: zTe} and \ref{eq: zT0}.

\subsubsection{Non-degenerate limit}

In the non-degenerate limit, $\eta \rightarrow -\infty$, the Fermi-Dirac integral is given by a polylog expansion
\begin{equation}
    F_j(\eta) = \Gamma(j+1) \left( e^{\eta}
    \;-\; \frac{e^{2\eta}}{2^{\,j+1}}  + \ldots \right)
\end{equation}
and to lowest order 
\begin{equation}
    \frac{F_{3/2}(\eta)}{F_{1/2}(\eta)} = \frac{3}{2}   \,,\qquad\frac{F_{5/2}(\eta)}{F_{1/2}(\eta)} =  \left(\frac{5}{2} \right)\left(\frac{3}{2}\right)  = \frac{15}{4} 
 \end{equation}

We can thus express
\begin{equation}
    \frac{\mathcal{L}_1 }{ \mathcal{L}_0 }=  \left(\frac{5}{2} k_\mathrm{B}T -\mu_{\mathrm{F}}  \right) 
\end{equation}
and
\begin{equation}
    \frac{\mathcal{L}_2} { \mathcal{L}_0 }= \frac{35}{4} (k_\mathrm{B}T)^2 - 5(k_\mathrm{B}T) \mu_\mathrm{F} +      \mu_{\mathrm{F}} ^2 \,.
\end{equation}
From these result, we get to lowest order in $T$, the limit $(zT)_0=1$. 

Other expressions of interest in this limit are
\begin{equation}
    \sigma =  \frac{3}{2}Cq^2  \,(k_B T)^{3/2}  \exp\left({\mu/k_\mathrm{B}T}\right)
\end{equation}
\begin{equation}
    S =  -\mu_\mathrm{F}/qT
\end{equation}
\begin{equation}
    \kappa_0 = \frac{3C}{2}\mu_\mathrm{F}^2 k_\mathrm{B}^{3/2} T^{1/2} \exp\left({\mu/k_\mathrm{B}T}\right)\,
\end{equation}
while 
\begin{equation}
    \kappa_e = 5 \frac{k_\mathrm{B}|\mu_\mathrm{F}|}{q^2} \sigma \,.
\end{equation}

\subsubsection{Near-degenerate limit}

It was early realized that a Fermi level close to the band edge could maximize $zT$, i.e., $\eta\approx0 $.\cite{mahanFigureMeritThermoelectrics1989} This limit corresponds to Taylor expanding the Fermi-integrals in $\eta$, 
which take the form: 
\begin{equation}
    F_j(0)=\Gamma(j+1)\,\bigl(1-2^{-j}\bigr)\,\zeta(j+1).
\end{equation}
We only the $\eta=0$ limit, for which
\begin{align}
    F_{1/2}(0) &\approx 0.6780939\\
    F_{3/2}(0)  &\approx 1.1528038 \\
    F_{5/2}(0) &\approx 3.0825861
\end{align}

We can express
\begin{equation}
    \frac{\mathcal{L}_1}{ \mathcal{L}_0} =\frac{5}{3} \frac{F_{3/2}(\eta)}{ F_{1/2}(\eta)} (k_\mathrm{B}T) =   A\, (k_\mathrm{B}T) 
\end{equation}
where $A\approx 2.833442$, and
\begin{equation}
    \frac{\mathcal{L}_2}{\mathcal{L}_0} = 
    \frac{7}{3}\frac{ F_{5/2}(\eta)}{F_{1/2}(\eta)} (k_\mathrm{B} T)^2= \,B (k_\mathrm{B}T)^2
\end{equation}
where  $B \approx 10.607234$.

From these limits, the temperature-independent limits of 
$(zT)_\mathrm{e}= 3.11$ and $(zT)_0= 0.757$ emerge. 

For completeness, we note that the electronic thermal conductivity, in this limit is given by 
\begin{equation}
    \frac{\kappa_e}{\sigma T} =  L \, \frac{k_\mathrm{B}^2}{q^2}
\end{equation}
with $L = B - A^2 \approx  2.578840 $.

\bibliography{ref,kb,zT_0.76,zT_1_curve,zT_1_T}

\begin{thebibliography}{115}%
\makeatletter
\providecommand \@ifxundefined [1]{%
 \@ifx{#1\undefined}
}%
\providecommand \@ifnum [1]{%
 \ifnum #1\expandafter \@firstoftwo
 \else \expandafter \@secondoftwo
 \fi
}%
\providecommand \@ifx [1]{%
 \ifx #1\expandafter \@firstoftwo
 \else \expandafter \@secondoftwo
 \fi
}%
\providecommand \natexlab [1]{#1}%
\providecommand \enquote  [1]{``#1''}%
\providecommand \bibnamefont  [1]{#1}%
\providecommand \bibfnamefont [1]{#1}%
\providecommand \citenamefont [1]{#1}%
\providecommand \href@noop [0]{\@secondoftwo}%
\providecommand \href [0]{\begingroup \@sanitize@url \@href}%
\providecommand \@href[1]{\@@startlink{#1}\@@href}%
\providecommand \@@href[1]{\endgroup#1\@@endlink}%
\providecommand \@sanitize@url [0]{\catcode `\\12\catcode `\$12\catcode `\&12\catcode `\#12\catcode `\^12\catcode `\_12\catcode `\%12\relax}%
\providecommand \@@startlink[1]{}%
\providecommand \@@endlink[0]{}%
\providecommand \url  [0]{\begingroup\@sanitize@url \@url }%
\providecommand \@url [1]{\endgroup\@href {#1}{\urlprefix }}%
\providecommand \urlprefix  [0]{URL }%
\providecommand \Eprint [0]{\href }%
\providecommand \doibase [0]{https://doi.org/}%
\providecommand \selectlanguage [0]{\@gobble}%
\providecommand \bibinfo  [0]{\@secondoftwo}%
\providecommand \bibfield  [0]{\@secondoftwo}%
\providecommand \translation [1]{[#1]}%
\providecommand \BibitemOpen [0]{}%
\providecommand \bibitemStop [0]{}%
\providecommand \bibitemNoStop [0]{.\EOS\space}%
\providecommand \EOS [0]{\spacefactor3000\relax}%
\providecommand \BibitemShut  [1]{\csname bibitem#1\endcsname}%
\let\auto@bib@innerbib\@empty
\bibitem [{\citenamefont {Snyder}\ and\ \citenamefont {Toberer}(2008)}]{snyderComplexThermoelectricMaterials2008}%
  \BibitemOpen
  \bibfield  {author} {\bibinfo {author} {\bibfnamefont {G.~J.}\ \bibnamefont {Snyder}}\ and\ \bibinfo {author} {\bibfnamefont {E.~S.}\ \bibnamefont {Toberer}},\ }\bibfield  {title} {\bibinfo {title} {Complex thermoelectric materials},\ }\href {https://doi.org/10.1038/nmat2090} {\bibfield  {journal} {\bibinfo  {journal} {Nature Materials}\ }\textbf {\bibinfo {volume} {7}},\ \bibinfo {pages} {105} (\bibinfo {year} {2008})}\BibitemShut {NoStop}%
\bibitem [{\citenamefont {He}\ and\ \citenamefont {Tritt}(2017)}]{heAdvancesThermoelectricMaterials2017}%
  \BibitemOpen
  \bibfield  {author} {\bibinfo {author} {\bibfnamefont {J.}~\bibnamefont {He}}\ and\ \bibinfo {author} {\bibfnamefont {T.~M.}\ \bibnamefont {Tritt}},\ }\bibfield  {title} {\bibinfo {title} {Advances in thermoelectric materials research: {{Looking}} back and moving forward},\ }\href {https://doi.org/10.1126/science.aak9997} {\bibfield  {journal} {\bibinfo  {journal} {Science}\ }\textbf {\bibinfo {volume} {357}},\ \bibinfo {pages} {eaak9997} (\bibinfo {year} {2017})}\BibitemShut {NoStop}%
\bibitem [{\citenamefont {Madsen}\ and\ \citenamefont {Singh}(2006)}]{madsenBoltzTraPCodeCalculating2006b}%
  \BibitemOpen
  \bibfield  {author} {\bibinfo {author} {\bibfnamefont {G.~K.}\ \bibnamefont {Madsen}}\ and\ \bibinfo {author} {\bibfnamefont {D.~J.}\ \bibnamefont {Singh}},\ }\bibfield  {title} {\bibinfo {title} {{{BoltzTraP}}. {{A}} code for calculating band-structure dependent quantities},\ }\href {https://doi.org/10.1016/j.cpc.2006.03.007} {\bibfield  {journal} {\bibinfo  {journal} {Computer Physics Communications}\ }\textbf {\bibinfo {volume} {175}},\ \bibinfo {pages} {67} (\bibinfo {year} {2006})}\BibitemShut {NoStop}%
\bibitem [{\citenamefont {Mahan}(1989)}]{mahanFigureMeritThermoelectrics1989}%
  \BibitemOpen
  \bibfield  {author} {\bibinfo {author} {\bibfnamefont {G.~D.}\ \bibnamefont {Mahan}},\ }\bibfield  {title} {\bibinfo {title} {Figure of merit for thermoelectrics},\ }\href {https://doi.org/10.1063/1.342976} {\bibfield  {journal} {\bibinfo  {journal} {Journal of Applied Physics}\ }\textbf {\bibinfo {volume} {65}},\ \bibinfo {pages} {1578} (\bibinfo {year} {1989})}\BibitemShut {NoStop}%
\bibitem [{\citenamefont {Bies}\ \emph {et~al.}(2002)\citenamefont {Bies}, \citenamefont {Radtke}, \citenamefont {Ehrenreich},\ and\ \citenamefont {Runge}}]{biesThermoelectricPropertiesAnisotropic2002}%
  \BibitemOpen
  \bibfield  {author} {\bibinfo {author} {\bibfnamefont {W.~E.}\ \bibnamefont {Bies}}, \bibinfo {author} {\bibfnamefont {R.~J.}\ \bibnamefont {Radtke}}, \bibinfo {author} {\bibfnamefont {H.}~\bibnamefont {Ehrenreich}},\ and\ \bibinfo {author} {\bibfnamefont {E.}~\bibnamefont {Runge}},\ }\bibfield  {title} {\bibinfo {title} {Thermoelectric properties of anisotropic semiconductors},\ }\href {https://doi.org/10.1103/PhysRevB.65.085208} {\bibfield  {journal} {\bibinfo  {journal} {Phys. Rev. B}\ }\textbf {\bibinfo {volume} {65}},\ \bibinfo {pages} {085208} (\bibinfo {year} {2002})}\BibitemShut {NoStop}%
\bibitem [{\citenamefont {Mahan}\ and\ \citenamefont {Sofo}(1996)}]{mahanBestThermoelectric1996}%
  \BibitemOpen
  \bibfield  {author} {\bibinfo {author} {\bibfnamefont {G.~D.}\ \bibnamefont {Mahan}}\ and\ \bibinfo {author} {\bibfnamefont {J.~O.}\ \bibnamefont {Sofo}},\ }\bibfield  {title} {\bibinfo {title} {The best thermoelectric.},\ }\href {https://doi.org/10.1073/pnas.93.15.7436} {\bibfield  {journal} {\bibinfo  {journal} {Proc. Natl. Acad. Sci. U.S.A.}\ }\textbf {\bibinfo {volume} {93}},\ \bibinfo {pages} {7436} (\bibinfo {year} {1996})}\BibitemShut {NoStop}%
\bibitem [{\citenamefont {Madsen}\ \emph {et~al.}(2018)\citenamefont {Madsen}, \citenamefont {Carrete},\ and\ \citenamefont {Verstraete}}]{madsenBoltzTraP2ProgramInterpolating2018b}%
  \BibitemOpen
  \bibfield  {author} {\bibinfo {author} {\bibfnamefont {G.~K.}\ \bibnamefont {Madsen}}, \bibinfo {author} {\bibfnamefont {J.}~\bibnamefont {Carrete}},\ and\ \bibinfo {author} {\bibfnamefont {M.~J.}\ \bibnamefont {Verstraete}},\ }\bibfield  {title} {\bibinfo {title} {{{BoltzTraP2}}, a program for interpolating band structures and calculating semi-classical transport coefficients},\ }\href {https://doi.org/10.1016/j.cpc.2018.05.010} {\bibfield  {journal} {\bibinfo  {journal} {Computer Physics Communications}\ }\textbf {\bibinfo {volume} {231}},\ \bibinfo {pages} {140} (\bibinfo {year} {2018})}\BibitemShut {NoStop}%
\bibitem [{\citenamefont {Kresse}\ and\ \citenamefont {Joubert}(1999)}]{kresseUltrasoftPseudopotentialsProjector1999}%
  \BibitemOpen
  \bibfield  {author} {\bibinfo {author} {\bibfnamefont {G.}~\bibnamefont {Kresse}}\ and\ \bibinfo {author} {\bibfnamefont {D.}~\bibnamefont {Joubert}},\ }\bibfield  {title} {\bibinfo {title} {From ultrasoft pseudopotentials to the projector augmented-wave method},\ }\href {https://doi.org/10.1103/PhysRevB.59.1758} {\bibfield  {journal} {\bibinfo  {journal} {Phys. Rev. B}\ }\textbf {\bibinfo {volume} {59}},\ \bibinfo {pages} {1758} (\bibinfo {year} {1999})}\BibitemShut {NoStop}%
\bibitem [{\citenamefont {Kresse}\ and\ \citenamefont {Furthm{\"u}ller}(1996)}]{vasp4}%
  \BibitemOpen
  \bibfield  {author} {\bibinfo {author} {\bibfnamefont {G.}~\bibnamefont {Kresse}}\ and\ \bibinfo {author} {\bibfnamefont {J.}~\bibnamefont {Furthm{\"u}ller}},\ }\bibfield  {title} {\bibinfo {title} {Efficient iterative schemes for {\emph{ab initio}} total-energy calculations using a plane-wave basis set},\ }\href@noop {} {\bibfield  {journal} {\bibinfo  {journal} {Phys. Rev. B}\ }\textbf {\bibinfo {volume} {54}},\ \bibinfo {pages} {11169} (\bibinfo {year} {1996})}\BibitemShut {NoStop}%
\bibitem [{\citenamefont {Berland}\ and\ \citenamefont {Hyldgaard}(2014)}]{behy14}%
  \BibitemOpen
  \bibfield  {author} {\bibinfo {author} {\bibfnamefont {K.}~\bibnamefont {Berland}}\ and\ \bibinfo {author} {\bibfnamefont {P.}~\bibnamefont {Hyldgaard}},\ }\bibfield  {title} {\bibinfo {title} {Exchange functional that tests the robustness of the plasmon description of the van der {{Waals}} density functional},\ }\href@noop {} {\bibfield  {journal} {\bibinfo  {journal} {Phys. Rev. B}\ }\textbf {\bibinfo {volume} {89}},\ \bibinfo {pages} {035412} (\bibinfo {year} {2014})}\BibitemShut {NoStop}%
\bibitem [{\citenamefont {Perdew}\ \emph {et~al.}(1996)\citenamefont {Perdew}, \citenamefont {Burke},\ and\ \citenamefont {Ernzerhof}}]{pbe1996}%
  \BibitemOpen
  \bibfield  {author} {\bibinfo {author} {\bibfnamefont {J.~P.}\ \bibnamefont {Perdew}}, \bibinfo {author} {\bibfnamefont {K.}~\bibnamefont {Burke}},\ and\ \bibinfo {author} {\bibfnamefont {M.}~\bibnamefont {Ernzerhof}},\ }\bibfield  {title} {\bibinfo {title} {Generalized {{Gradient Approximation Made Simple}}},\ }\href {https://doi.org/10.1103/PhysRevLett.77.3865} {\bibfield  {journal} {\bibinfo  {journal} {Phys. Rev. Lett.}\ }\textbf {\bibinfo {volume} {77}},\ \bibinfo {pages} {3865} (\bibinfo {year} {1996})}\BibitemShut {NoStop}%
\bibitem [{\citenamefont {Aaouita}\ \emph {et~al.}(2025)\citenamefont {Aaouita}, \citenamefont {Mediane}, \citenamefont {Goumrhar}, \citenamefont {Drissi},\ and\ \citenamefont {Laamara}}]{Aaouita_2025}%
  \BibitemOpen
  \bibfield  {author} {\bibinfo {author} {\bibfnamefont {K.~E.}\ \bibnamefont {Aaouita}}, \bibinfo {author} {\bibfnamefont {N.}~\bibnamefont {Mediane}}, \bibinfo {author} {\bibfnamefont {F.}~\bibnamefont {Goumrhar}}, \bibinfo {author} {\bibfnamefont {L.}~\bibnamefont {Drissi}},\ and\ \bibinfo {author} {\bibfnamefont {R.~A.}\ \bibnamefont {Laamara}},\ }\bibfield  {title} {\bibinfo {title} {Enhanced spintronic and thermoelectric performance in vanadium-doped {{CaTiO3}}: {{An}} ab-initio study},\ }\href {https://doi.org/10.1016/j.cocom.2024.e00999} {\bibfield  {journal} {\bibinfo  {journal} {Computational Condensed Matter}\ }\textbf {\bibinfo {volume} {42}},\ \bibinfo {pages} {e00999} (\bibinfo {year} {2025})}\BibitemShut {NoStop}%
\bibitem [{\citenamefont {Abraham}\ \emph {et~al.}(2025)\citenamefont {Abraham}, \citenamefont {Saxena}, \citenamefont {Kumbhakar}, \citenamefont {Srivastava}, \citenamefont {Ibrahim}, \citenamefont {{El-Meligy}}, \citenamefont {Manzoor},\ and\ \citenamefont {Sharma}}]{Abraham_2025}%
  \BibitemOpen
  \bibfield  {author} {\bibinfo {author} {\bibfnamefont {J.~A.}\ \bibnamefont {Abraham}}, \bibinfo {author} {\bibfnamefont {A.}~\bibnamefont {Saxena}}, \bibinfo {author} {\bibfnamefont {J.}~\bibnamefont {Kumbhakar}}, \bibinfo {author} {\bibfnamefont {A.}~\bibnamefont {Srivastava}}, \bibinfo {author} {\bibfnamefont {A.~A.}\ \bibnamefont {Ibrahim}}, \bibinfo {author} {\bibfnamefont {M.}~\bibnamefont {{El-Meligy}}}, \bibinfo {author} {\bibfnamefont {M.}~\bibnamefont {Manzoor}},\ and\ \bibinfo {author} {\bibfnamefont {R.}~\bibnamefont {Sharma}},\ }\bibfield  {title} {\bibinfo {title} {First {{Principles Calculation}} on the {{Physical Attributes}} of {{Cubic Perovskites SbXO3}} ({{X}} = {{Al}} and {{Ga}}) for {{Renewable Energy Devices Application}}},\ }\href {https://doi.org/10.1007/s10904-025-03628-4} {\bibfield  {journal} {\bibinfo  {journal} {Journal of Inorganic and Organometallic Polymers and Materials}\ }\textbf {\bibinfo {volume} {35}},\ \bibinfo {pages} {5876} (\bibinfo {year} {2025})}\BibitemShut
  {NoStop}%
\bibitem [{\citenamefont {Ahsan}\ \emph {et~al.}(2025)\citenamefont {Ahsan}, \citenamefont {Rather}, \citenamefont {Sultan}, \citenamefont {Zriouel},\ and\ \citenamefont {Hlil}}]{Ahsan_2025}%
  \BibitemOpen
  \bibfield  {author} {\bibinfo {author} {\bibfnamefont {J.~U.}\ \bibnamefont {Ahsan}}, \bibinfo {author} {\bibfnamefont {M.~R.}\ \bibnamefont {Rather}}, \bibinfo {author} {\bibfnamefont {K.}~\bibnamefont {Sultan}}, \bibinfo {author} {\bibfnamefont {S.}~\bibnamefont {Zriouel}},\ and\ \bibinfo {author} {\bibfnamefont {E.-k.}\ \bibnamefont {Hlil}},\ }\bibfield  {title} {\bibinfo {title} {In-depth study of double perovskite {{Sr}} 2 {{NiTaO}} 6 : {{Structural}}, electronic, thermoelectric, and spintronic properties for sustainable and high-performance applications},\ }\href {https://doi.org/10.1016/j.cocom.2025.e01026} {\bibfield  {journal} {\bibinfo  {journal} {Computational Condensed Matter}\ }\textbf {\bibinfo {volume} {43}},\ \bibinfo {pages} {e01026} (\bibinfo {year} {2025})}\BibitemShut {NoStop}%
\bibitem [{\citenamefont {{Al-Daraghmeh}}\ \emph {et~al.}(2025)\citenamefont {{Al-Daraghmeh}}, \citenamefont {Mustafa}, \citenamefont {Younas}, \citenamefont {Zayed}, \citenamefont {Bouzgarrou}, \citenamefont {Boukhris}, \citenamefont {{Al-Anazy}}, \citenamefont {{Al-Buriahi}},\ and\ \citenamefont {Mahmood}}]{Al-Daraghmeh_2025}%
  \BibitemOpen
  \bibfield  {author} {\bibinfo {author} {\bibfnamefont {T.~M.}\ \bibnamefont {{Al-Daraghmeh}}}, \bibinfo {author} {\bibfnamefont {G.~M.}\ \bibnamefont {Mustafa}}, \bibinfo {author} {\bibfnamefont {B.}~\bibnamefont {Younas}}, \bibinfo {author} {\bibfnamefont {O.}~\bibnamefont {Zayed}}, \bibinfo {author} {\bibfnamefont {S.}~\bibnamefont {Bouzgarrou}}, \bibinfo {author} {\bibfnamefont {I.}~\bibnamefont {Boukhris}}, \bibinfo {author} {\bibfnamefont {M.~M.}\ \bibnamefont {{Al-Anazy}}}, \bibinfo {author} {\bibfnamefont {M.~S.}\ \bibnamefont {{Al-Buriahi}}},\ and\ \bibinfo {author} {\bibfnamefont {Q.}~\bibnamefont {Mahmood}},\ }\bibfield  {title} {\bibinfo {title} {Study of optoelectronic and transport properties of {{SrCd2X2}} ({{X}} = {{P}}, {{As}}, {{Sb}}) zintl-phase for renewable energy applications},\ }\href {https://doi.org/10.1007/s11082-025-08253-6} {\bibfield  {journal} {\bibinfo  {journal} {Optical and Quantum Electronics}\ }\textbf {\bibinfo {volume} {57}},\ \bibinfo {pages} {320} (\bibinfo {year}
  {2025})}\BibitemShut {NoStop}%
\bibitem [{\citenamefont {{Al-Qaisi}}\ \emph {et~al.}(2025)\citenamefont {{Al-Qaisi}}, \citenamefont {Iram}, \citenamefont {Sfina}, \citenamefont {Boutramine}, \citenamefont {Jappor}, \citenamefont {Alfaifi}, \citenamefont {Alzahrani}, \citenamefont {Rached}, \citenamefont {Ali},\ and\ \citenamefont {Murtaza}}]{Al-Qaisi_2025}%
  \BibitemOpen
  \bibfield  {author} {\bibinfo {author} {\bibfnamefont {S.}~\bibnamefont {{Al-Qaisi}}}, \bibinfo {author} {\bibfnamefont {N.}~\bibnamefont {Iram}}, \bibinfo {author} {\bibfnamefont {N.}~\bibnamefont {Sfina}}, \bibinfo {author} {\bibfnamefont {A.}~\bibnamefont {Boutramine}}, \bibinfo {author} {\bibfnamefont {H.~R.}\ \bibnamefont {Jappor}}, \bibinfo {author} {\bibfnamefont {A.~H.}\ \bibnamefont {Alfaifi}}, \bibinfo {author} {\bibfnamefont {H.~S.}\ \bibnamefont {Alzahrani}}, \bibinfo {author} {\bibfnamefont {H.}~\bibnamefont {Rached}}, \bibinfo {author} {\bibfnamefont {M.~A.}\ \bibnamefont {Ali}},\ and\ \bibinfo {author} {\bibfnamefont {G.}~\bibnamefont {Murtaza}},\ }\bibfield  {title} {\bibinfo {title} {Comprehensive {{DFT}} study of {{K2TlZI6}} ({{Z}} = {{Al}}, {{In}}) double perovskites: {{Structural}} stability and potential for optoelectronic and thermoelectric energy harvesting},\ }\href {https://doi.org/10.1016/j.physb.2025.417239} {\bibfield  {journal} {\bibinfo  {journal} {Physica B: Condensed Matter}\
  }\textbf {\bibinfo {volume} {710}},\ \bibinfo {pages} {417239} (\bibinfo {year} {2025})}\BibitemShut {NoStop}%
\bibitem [{\citenamefont {Ali}\ \emph {et~al.}(2025{\natexlab{a}})\citenamefont {Ali}, \citenamefont {Hussain}, \citenamefont {Asghar}, \citenamefont {Ullah},\ and\ \citenamefont {Khan}}]{Ali_2025a}%
  \BibitemOpen
  \bibfield  {author} {\bibinfo {author} {\bibfnamefont {F.}~\bibnamefont {Ali}}, \bibinfo {author} {\bibfnamefont {M.}~\bibnamefont {Hussain}}, \bibinfo {author} {\bibfnamefont {M.}~\bibnamefont {Asghar}}, \bibinfo {author} {\bibfnamefont {H.}~\bibnamefont {Ullah}},\ and\ \bibinfo {author} {\bibfnamefont {S.~A.}\ \bibnamefont {Khan}},\ }\href {https://doi.org/10.2139/ssrn.5372503} {\bibinfo {title} {Tuning the {{Physical Properties}} of {{Cs2ptcl6 Via External Strain}} for {{Energy Storage Applications}}}} (\bibinfo {year} {2025}{\natexlab{a}})\BibitemShut {NoStop}%
\bibitem [{\citenamefont {Amghar}\ \emph {et~al.}(2025)\citenamefont {Amghar}, \citenamefont {Lamrani}, \citenamefont {Maskar}, \citenamefont {El~Hat}, \citenamefont {Laamara}, \citenamefont {Yvaz},\ and\ \citenamefont {Rai}}]{Amghar_2025}%
  \BibitemOpen
  \bibfield  {author} {\bibinfo {author} {\bibfnamefont {W.}~\bibnamefont {Amghar}}, \bibinfo {author} {\bibfnamefont {A.~F.}\ \bibnamefont {Lamrani}}, \bibinfo {author} {\bibfnamefont {E.}~\bibnamefont {Maskar}}, \bibinfo {author} {\bibfnamefont {A.}~\bibnamefont {El~Hat}}, \bibinfo {author} {\bibfnamefont {R.~A.}\ \bibnamefont {Laamara}}, \bibinfo {author} {\bibfnamefont {A.}~\bibnamefont {Yvaz}},\ and\ \bibinfo {author} {\bibfnamefont {D.~P.}\ \bibnamefont {Rai}},\ }\href {https://doi.org/10.2139/ssrn.5200610} {\bibinfo {title} {Study of {{Optoelectronic}}, {{Thermoelectric}} and {{Induced Half-Metallicity}} in {{Wurtzite Zns Via Codoped}} ({{EU}}, {{Cr}}): {{A Dft Method}}}} (\bibinfo {year} {2025})\BibitemShut {NoStop}%
\bibitem [{\citenamefont {{Anuradha}}\ \emph {et~al.}(2025)\citenamefont {{Anuradha}}, \citenamefont {Kumawat}, \citenamefont {Chauhan}, \citenamefont {Joshi}, \citenamefont {Verma},\ and\ \citenamefont {Arora}}]{Anuradha_2025}%
  \BibitemOpen
  \bibfield  {author} {\bibinfo {author} {\bibnamefont {{Anuradha}}}, \bibinfo {author} {\bibfnamefont {S.}~\bibnamefont {Kumawat}}, \bibinfo {author} {\bibfnamefont {M.~S.}\ \bibnamefont {Chauhan}}, \bibinfo {author} {\bibfnamefont {T.~K.}\ \bibnamefont {Joshi}}, \bibinfo {author} {\bibfnamefont {A.~S.}\ \bibnamefont {Verma}},\ and\ \bibinfo {author} {\bibfnamefont {G.}~\bibnamefont {Arora}},\ }\bibfield  {title} {\bibinfo {title} {Investigation of the physical characteristics of {{Ba2AlXO6}} ({{X}} = {{V}}, {{Nb}}, {{Ta}}) double perovskite oxides for energy conversion applications},\ }\href {https://doi.org/10.1016/j.mtcomm.2025.112010} {\bibfield  {journal} {\bibinfo  {journal} {Materials Today Communications}\ }\textbf {\bibinfo {volume} {44}},\ \bibinfo {pages} {112010} (\bibinfo {year} {2025})}\BibitemShut {NoStop}%
\bibitem [{\citenamefont {Arbaoui}\ \emph {et~al.}(2025)\citenamefont {Arbaoui}, \citenamefont {Boutahir}, \citenamefont {Fakrach}, \citenamefont {Chadli}, \citenamefont {Rahmani},\ and\ \citenamefont {Rahmani}}]{Arbaoui_2025}%
  \BibitemOpen
  \bibfield  {author} {\bibinfo {author} {\bibfnamefont {Z.}~\bibnamefont {Arbaoui}}, \bibinfo {author} {\bibfnamefont {M.}~\bibnamefont {Boutahir}}, \bibinfo {author} {\bibfnamefont {B.}~\bibnamefont {Fakrach}}, \bibinfo {author} {\bibfnamefont {H.}~\bibnamefont {Chadli}}, \bibinfo {author} {\bibfnamefont {A.}~\bibnamefont {Rahmani}},\ and\ \bibinfo {author} {\bibfnamefont {A.}~\bibnamefont {Rahmani}},\ }\bibfield  {title} {\bibinfo {title} {First-principles analysis of the optoelectronic and thermoelectric properties of black phosphorene for energy harvesting applications},\ }\href {https://doi.org/10.1080/01411594.2025.2526527} {\bibfield  {journal} {\bibinfo  {journal} {Phase Transitions}\ ,\ \bibinfo {pages} {1}} (\bibinfo {year} {2025})}\BibitemShut {NoStop}%
\bibitem [{\citenamefont {Benahmedi}\ \emph {et~al.}(2025{\natexlab{a}})\citenamefont {Benahmedi}, \citenamefont {Besbes}, \citenamefont {Djelti}, \citenamefont {Bendehiba},\ and\ \citenamefont {Aissani}}]{Benahmedi_2025a}%
  \BibitemOpen
  \bibfield  {author} {\bibinfo {author} {\bibfnamefont {L.}~\bibnamefont {Benahmedi}}, \bibinfo {author} {\bibfnamefont {A.}~\bibnamefont {Besbes}}, \bibinfo {author} {\bibfnamefont {R.}~\bibnamefont {Djelti}}, \bibinfo {author} {\bibfnamefont {S.}~\bibnamefont {Bendehiba}},\ and\ \bibinfo {author} {\bibfnamefont {A.}~\bibnamefont {Aissani}},\ }\bibfield  {title} {\bibinfo {title} {First-principles investigation of inorganic antiperovskite {{A}}{\textsubscript{3}} {{SbAs}} ({{A}} = {{Ba}}, {{Sr}}, and {{Ca}}): Insights into thermoelectric and optoelectronic potential},\ }\href {https://doi.org/10.1039/D5NJ00017C} {\bibfield  {journal} {\bibinfo  {journal} {New Journal of Chemistry}\ }\textbf {\bibinfo {volume} {49}},\ \bibinfo {pages} {6741} (\bibinfo {year} {2025}{\natexlab{a}})}\BibitemShut {NoStop}%
\bibitem [{\citenamefont {Boutramine}\ \emph {et~al.}(2025)\citenamefont {Boutramine}, \citenamefont {{Al-Qaisi}}, \citenamefont {Sfina}, \citenamefont {Lamine}, \citenamefont {Chaib}, \citenamefont {Archi}, \citenamefont {Alsalmi},\ and\ \citenamefont {Rabhi}}]{Boutramine_2025}%
  \BibitemOpen
  \bibfield  {author} {\bibinfo {author} {\bibfnamefont {A.}~\bibnamefont {Boutramine}}, \bibinfo {author} {\bibfnamefont {S.}~\bibnamefont {{Al-Qaisi}}}, \bibinfo {author} {\bibfnamefont {N.}~\bibnamefont {Sfina}}, \bibinfo {author} {\bibfnamefont {L.~A.}\ \bibnamefont {Lamine}}, \bibinfo {author} {\bibfnamefont {H.}~\bibnamefont {Chaib}}, \bibinfo {author} {\bibfnamefont {M.}~\bibnamefont {Archi}}, \bibinfo {author} {\bibfnamefont {O.}~\bibnamefont {Alsalmi}},\ and\ \bibinfo {author} {\bibfnamefont {S.}~\bibnamefont {Rabhi}},\ }\bibfield  {title} {\bibinfo {title} {Insights into the thermodynamic, optoelectronic, and thermoelectric properties of ternary transition metal chalcogenides {{BiIrQ}} ({{Q}} = {{S}}, {{Se}}, {{Te}}) for next-generation optoelectronic and energy harvesting technologies: {{A DFT}} and {{AIMD}} study},\ }\href {https://doi.org/10.1016/j.surfin.2025.107269} {\bibfield  {journal} {\bibinfo  {journal} {Surfaces and Interfaces}\ }\textbf {\bibinfo {volume} {72}},\ \bibinfo {pages}
  {107269} (\bibinfo {year} {2025})}\BibitemShut {NoStop}%
\bibitem [{\citenamefont {Caid}\ \emph {et~al.}(2025)\citenamefont {Caid}, \citenamefont {Delig{\"o}z}, \citenamefont {Rached}, \citenamefont {Rached}, \citenamefont {Mansour}, \citenamefont {Ozisik},\ and\ \citenamefont {Rached}}]{Caid_2025}%
  \BibitemOpen
  \bibfield  {author} {\bibinfo {author} {\bibfnamefont {M.}~\bibnamefont {Caid}}, \bibinfo {author} {\bibfnamefont {E.}~\bibnamefont {Delig{\"o}z}}, \bibinfo {author} {\bibfnamefont {D.}~\bibnamefont {Rached}}, \bibinfo {author} {\bibfnamefont {H.}~\bibnamefont {Rached}}, \bibinfo {author} {\bibfnamefont {H.}~\bibnamefont {Mansour}}, \bibinfo {author} {\bibfnamefont {H.}~\bibnamefont {Ozisik}},\ and\ \bibinfo {author} {\bibfnamefont {Y.}~\bibnamefont {Rached}},\ }\bibfield  {title} {\bibinfo {title} {{{Pb2CoMoO6}} as a {{Promising Energy Material}}: {{A First-Principles Perspective}}},\ }\bibfield  {journal} {\bibinfo  {journal} {Journal of Inorganic and Organometallic Polymers and Materials}\ }\href {https://doi.org/10.1007/s10904-025-03962-7} {10.1007/s10904-025-03962-7} (\bibinfo {year} {2025})\BibitemShut {NoStop}%
\bibitem [{\citenamefont {Choudhary}\ \emph {et~al.}(2025)\citenamefont {Choudhary}, \citenamefont {Kumar}, \citenamefont {Sahariya},\ and\ \citenamefont {Soni}}]{Choudhary_2025}%
  \BibitemOpen
  \bibfield  {author} {\bibinfo {author} {\bibfnamefont {N.}~\bibnamefont {Choudhary}}, \bibinfo {author} {\bibfnamefont {K.}~\bibnamefont {Kumar}}, \bibinfo {author} {\bibfnamefont {J.}~\bibnamefont {Sahariya}},\ and\ \bibinfo {author} {\bibfnamefont {A.}~\bibnamefont {Soni}},\ }\bibfield  {title} {\bibinfo {title} {Investigation of {{Double Halide Perovskites Materials Rb2AgMCl6}} ({{M}} = {{As}}, {{Co}}, {{Rh}}) for {{Their Electronic}}, {{Optoelectronic}}, and {{Thermoelectric Properties}}},\ }\bibfield  {journal} {\bibinfo  {journal} {Journal of Inorganic and Organometallic Polymers and Materials}\ }\href {https://doi.org/10.1007/s10904-025-03743-2} {10.1007/s10904-025-03743-2} (\bibinfo {year} {2025})\BibitemShut {NoStop}%
\bibitem [{\citenamefont {Dixit}\ \emph {et~al.}(2025)\citenamefont {Dixit}, \citenamefont {Saxena}, \citenamefont {Manzoor}, \citenamefont {Sahid}, \citenamefont {Sharma}, \citenamefont {Ibrahim}, \citenamefont {{El-Meligy}}, \citenamefont {Abdullaeva},\ and\ \citenamefont {Srivastava}}]{Dixit_2025}%
  \BibitemOpen
  \bibfield  {author} {\bibinfo {author} {\bibfnamefont {A.}~\bibnamefont {Dixit}}, \bibinfo {author} {\bibfnamefont {A.}~\bibnamefont {Saxena}}, \bibinfo {author} {\bibfnamefont {M.}~\bibnamefont {Manzoor}}, \bibinfo {author} {\bibfnamefont {S.}~\bibnamefont {Sahid}}, \bibinfo {author} {\bibfnamefont {R.}~\bibnamefont {Sharma}}, \bibinfo {author} {\bibfnamefont {A.~A.}\ \bibnamefont {Ibrahim}}, \bibinfo {author} {\bibfnamefont {M.~A.}\ \bibnamefont {{El-Meligy}}}, \bibinfo {author} {\bibfnamefont {B.}~\bibnamefont {Abdullaeva}},\ and\ \bibinfo {author} {\bibfnamefont {V.}~\bibnamefont {Srivastava}},\ }\bibfield  {title} {\bibinfo {title} {{{DFT-Based Investigation}} of {{Opto-Electronic}}, {{Mechanical}}, and {{Thermoelectric Properties}} of {{Sr2YBiO6}} for {{Green Energy Applications}}},\ }\bibfield  {journal} {\bibinfo  {journal} {Journal of Inorganic and Organometallic Polymers and Materials}\ }\href {https://doi.org/10.1007/s10904-025-03710-x} {10.1007/s10904-025-03710-x} (\bibinfo {year}
  {2025})\BibitemShut {NoStop}%
\bibitem [{\citenamefont {El~Goutni}\ \emph {et~al.}(2025)\citenamefont {El~Goutni}, \citenamefont {Remil}, \citenamefont {Saidi}, \citenamefont {Batouche}, \citenamefont {Seddik},\ and\ \citenamefont {Khenata}}]{ElGoutni_2025}%
  \BibitemOpen
  \bibfield  {author} {\bibinfo {author} {\bibfnamefont {M.~E.~A.}\ \bibnamefont {El~Goutni}}, \bibinfo {author} {\bibfnamefont {A.}~\bibnamefont {Remil}}, \bibinfo {author} {\bibfnamefont {M.}~\bibnamefont {Saidi}}, \bibinfo {author} {\bibfnamefont {M.}~\bibnamefont {Batouche}}, \bibinfo {author} {\bibfnamefont {T.}~\bibnamefont {Seddik}},\ and\ \bibinfo {author} {\bibfnamefont {R.}~\bibnamefont {Khenata}},\ }\bibfield  {title} {\bibinfo {title} {Computational investigation of {{Rb2ReX6}} ({{X}}= {{Cl}}, {{Br}}, {{I}}) vacancy-ordered double perovskites: {{From}} structural stability to thermoelectric performance},\ }\href {https://doi.org/10.1016/j.cocom.2025.e01038} {\bibfield  {journal} {\bibinfo  {journal} {Computational Condensed Matter}\ }\textbf {\bibinfo {volume} {43}},\ \bibinfo {pages} {e01038} (\bibinfo {year} {2025})}\BibitemShut {NoStop}%
\bibitem [{\citenamefont {Fatima}\ \emph {et~al.}(2025)\citenamefont {Fatima}, \citenamefont {Maabed}, \citenamefont {Mebarki}, \citenamefont {Halit}, \citenamefont {Matta},\ and\ \citenamefont {Ben Kamri~Ahmed}}]{Fatima_2025}%
  \BibitemOpen
  \bibfield  {author} {\bibinfo {author} {\bibfnamefont {G.}~\bibnamefont {Fatima}}, \bibinfo {author} {\bibfnamefont {S.}~\bibnamefont {Maabed}}, \bibinfo {author} {\bibfnamefont {H.}~\bibnamefont {Mebarki}}, \bibinfo {author} {\bibfnamefont {M.}~\bibnamefont {Halit}}, \bibinfo {author} {\bibfnamefont {C.~F.}\ \bibnamefont {Matta}},\ and\ \bibinfo {author} {\bibfnamefont {L.}~\bibnamefont {Ben Kamri~Ahmed}},\ }\bibfield  {title} {\bibinfo {title} {Unveiling the thermoelectric potential of perthioborate {{MBS}}{$_3$}({{M}} = {{Rb}};{{Cs}}) compounds: Insights from {{DFT}} calculations},\ }\href {https://doi.org/10.1140/epjb/s10051-025-00896-4} {\bibfield  {journal} {\bibinfo  {journal} {The European Physical Journal B}\ }\textbf {\bibinfo {volume} {98}},\ \bibinfo {pages} {50} (\bibinfo {year} {2025})}\BibitemShut {NoStop}%
\bibitem [{\citenamefont {Gurunani}\ and\ \citenamefont {Gupta}(2025)}]{Gurunani_2025}%
  \BibitemOpen
  \bibfield  {author} {\bibinfo {author} {\bibfnamefont {B.}~\bibnamefont {Gurunani}}\ and\ \bibinfo {author} {\bibfnamefont {D.~C.}\ \bibnamefont {Gupta}},\ }\bibfield  {title} {\bibinfo {title} {A {{Comprehensive Analysis}} of {{CoHfX}} ({{X}} = {{As}}, {{Bi}}) {{Half Heusler Alloys}}: {{Implications}} for {{High-Performance Thermoelectric}} and {{Photovoltaic Devices}}},\ }\href {https://doi.org/10.1007/s10904-024-03279-x} {\bibfield  {journal} {\bibinfo  {journal} {Journal of Inorganic and Organometallic Polymers and Materials}\ }\textbf {\bibinfo {volume} {35}},\ \bibinfo {pages} {2327} (\bibinfo {year} {2025})}\BibitemShut {NoStop}%
\bibitem [{\citenamefont {Hongjie}\ \emph {et~al.}(2025{\natexlab{a}})\citenamefont {Hongjie}, \citenamefont {Nazir}, \citenamefont {Panse}, \citenamefont {Khera}, \citenamefont {Talha}, \citenamefont {Manzoor}, \citenamefont {Ghodhbani},\ and\ \citenamefont {Sharma}}]{Hongjie_2025}%
  \BibitemOpen
  \bibfield  {author} {\bibinfo {author} {\bibfnamefont {L.}~\bibnamefont {Hongjie}}, \bibinfo {author} {\bibfnamefont {A.}~\bibnamefont {Nazir}}, \bibinfo {author} {\bibfnamefont {V.~R.}\ \bibnamefont {Panse}}, \bibinfo {author} {\bibfnamefont {E.~A.}\ \bibnamefont {Khera}}, \bibinfo {author} {\bibfnamefont {H.~M.}\ \bibnamefont {Talha}}, \bibinfo {author} {\bibfnamefont {M.}~\bibnamefont {Manzoor}}, \bibinfo {author} {\bibfnamefont {R.}~\bibnamefont {Ghodhbani}},\ and\ \bibinfo {author} {\bibfnamefont {R.}~\bibnamefont {Sharma}},\ }\bibfield  {title} {\bibinfo {title} {{{DFT Insights}} on the {{Optoelectronic}} and {{Thermoelectric Characteristics}} of {{Lead-Free Na2InGaX6}}({{X}} = {{Cl}}, {{I}}) {{Double Perovskites}} for {{Environmentally Friendly Energy Applications}}},\ }\bibfield  {journal} {\bibinfo  {journal} {Journal of Inorganic and Organometallic Polymers and Materials}\ }\href {https://doi.org/10.1007/s10904-025-03711-w} {10.1007/s10904-025-03711-w} (\bibinfo {year}
  {2025}{\natexlab{a}})\BibitemShut {NoStop}%
\bibitem [{\citenamefont {Hongjie}\ \emph {et~al.}(2025{\natexlab{b}})\citenamefont {Hongjie}, \citenamefont {Dixit}, \citenamefont {Kumari}, \citenamefont {Abraham}, \citenamefont {Mishra}, \citenamefont {Sharma}, \citenamefont {Kumari},\ and\ \citenamefont {Kallel}}]{Hongjie_2025a}%
  \BibitemOpen
  \bibfield  {author} {\bibinfo {author} {\bibfnamefont {L.}~\bibnamefont {Hongjie}}, \bibinfo {author} {\bibfnamefont {A.}~\bibnamefont {Dixit}}, \bibinfo {author} {\bibfnamefont {A.}~\bibnamefont {Kumari}}, \bibinfo {author} {\bibfnamefont {J.~A.}\ \bibnamefont {Abraham}}, \bibinfo {author} {\bibfnamefont {A.~K.}\ \bibnamefont {Mishra}}, \bibinfo {author} {\bibfnamefont {R.}~\bibnamefont {Sharma}}, \bibinfo {author} {\bibfnamefont {P.}~\bibnamefont {Kumari}},\ and\ \bibinfo {author} {\bibfnamefont {M.}~\bibnamefont {Kallel}},\ }\bibfield  {title} {\bibinfo {title} {Optoelectronic and {{Thermoelectric Attributes}} in {{Lead Free Hybrid Double Perovskite Semiconductors Rb2SbAuX6}} ({{X}} = {{Br}}, {{Cl}}) via {{DFT Study}}},\ }\bibfield  {journal} {\bibinfo  {journal} {Journal of Inorganic and Organometallic Polymers and Materials}\ }\href {https://doi.org/10.1007/s10904-025-03814-4} {10.1007/s10904-025-03814-4} (\bibinfo {year} {2025}{\natexlab{b}})\BibitemShut {NoStop}%
\bibitem [{\citenamefont {Iram}\ \emph {et~al.}(2025{\natexlab{a}})\citenamefont {Iram}, \citenamefont {Sharma}, \citenamefont {Gouadria}, \citenamefont {Kumar},\ and\ \citenamefont {Asiri}}]{Iram_2025}%
  \BibitemOpen
  \bibfield  {author} {\bibinfo {author} {\bibfnamefont {N.}~\bibnamefont {Iram}}, \bibinfo {author} {\bibfnamefont {R.}~\bibnamefont {Sharma}}, \bibinfo {author} {\bibfnamefont {S.}~\bibnamefont {Gouadria}}, \bibinfo {author} {\bibfnamefont {A.}~\bibnamefont {Kumar}},\ and\ \bibinfo {author} {\bibfnamefont {M.}~\bibnamefont {Asiri}},\ }\bibfield  {title} {\bibinfo {title} {First-principles study of lead-free flouro double perovskites {{A2LiAsF6}} ({{A}} = {{Na}}, {{K}}) for optoelectronic and renewable energy applications},\ }\href {https://doi.org/10.1016/j.mseb.2025.118218} {\bibfield  {journal} {\bibinfo  {journal} {Materials Science and Engineering: B}\ }\textbf {\bibinfo {volume} {317}},\ \bibinfo {pages} {118218} (\bibinfo {year} {2025}{\natexlab{a}})}\BibitemShut {NoStop}%
\bibitem [{\citenamefont {Iram}\ \emph {et~al.}(2025{\natexlab{b}})\citenamefont {Iram}, \citenamefont {Sharma}, \citenamefont {Ahmad}, \citenamefont {Khalid},\ and\ \citenamefont {Alanazi}}]{Iram_2025a}%
  \BibitemOpen
  \bibfield  {author} {\bibinfo {author} {\bibfnamefont {N.}~\bibnamefont {Iram}}, \bibinfo {author} {\bibfnamefont {R.}~\bibnamefont {Sharma}}, \bibinfo {author} {\bibfnamefont {J.}~\bibnamefont {Ahmad}}, \bibinfo {author} {\bibfnamefont {S.}~\bibnamefont {Khalid}},\ and\ \bibinfo {author} {\bibfnamefont {M.~M.}\ \bibnamefont {Alanazi}},\ }\bibfield  {title} {\bibinfo {title} {Theoretical investigation on the multifunctional attributes of {{RhHfX}} ({{X}}={{P}}, {{As}}) half {{Heusler}} semiconductor for advanced technological applications},\ }\href {https://doi.org/10.1016/j.jpcs.2025.112829} {\bibfield  {journal} {\bibinfo  {journal} {Journal of Physics and Chemistry of Solids}\ }\textbf {\bibinfo {volume} {206}},\ \bibinfo {pages} {112829} (\bibinfo {year} {2025}{\natexlab{b}})}\BibitemShut {NoStop}%
\bibitem [{\citenamefont {Jolayemi}\ \emph {et~al.}(2025)\citenamefont {Jolayemi}, \citenamefont {Mule}, \citenamefont {Uto},\ and\ \citenamefont {Olawole}}]{Jolayemi_2025}%
  \BibitemOpen
  \bibfield  {author} {\bibinfo {author} {\bibfnamefont {O.~R.}\ \bibnamefont {Jolayemi}}, \bibinfo {author} {\bibfnamefont {G.~M.}\ \bibnamefont {Mule}}, \bibinfo {author} {\bibfnamefont {O.~T.}\ \bibnamefont {Uto}},\ and\ \bibinfo {author} {\bibfnamefont {O.~C.}\ \bibnamefont {Olawole}},\ }\bibfield  {title} {\bibinfo {title} {Highly efficient {{XCoSi}} ({{X}}={{V}}, {{Nb}}, {{Ta}}) compounds for thermoelectricity: A density functional theory approach},\ }\href {https://doi.org/10.1007/s10825-024-02273-3} {\bibfield  {journal} {\bibinfo  {journal} {Journal of Computational Electronics}\ }\textbf {\bibinfo {volume} {24}},\ \bibinfo {pages} {35} (\bibinfo {year} {2025})}\BibitemShut {NoStop}%
\bibitem [{\citenamefont {Karthikraja}\ \emph {et~al.}(2025)\citenamefont {Karthikraja}, \citenamefont {Nulakani}, \citenamefont {Devi}, \citenamefont {Murugan}, \citenamefont {Ramanujam}, \citenamefont {Vaidyanathan},\ and\ \citenamefont {Subramanian}}]{Karthikraja_2025}%
  \BibitemOpen
  \bibfield  {author} {\bibinfo {author} {\bibfnamefont {E.}~\bibnamefont {Karthikraja}}, \bibinfo {author} {\bibfnamefont {N.~V.~R.}\ \bibnamefont {Nulakani}}, \bibinfo {author} {\bibfnamefont {P.}~\bibnamefont {Devi}}, \bibinfo {author} {\bibfnamefont {P.}~\bibnamefont {Murugan}}, \bibinfo {author} {\bibfnamefont {K.}~\bibnamefont {Ramanujam}}, \bibinfo {author} {\bibfnamefont {V.~G.}\ \bibnamefont {Vaidyanathan}},\ and\ \bibinfo {author} {\bibfnamefont {V.}~\bibnamefont {Subramanian}},\ }\bibfield  {title} {\bibinfo {title} {First-principles insights into biphenylene-based graphynes: Promising novel two-dimensional carbon allotropes for thermoelectric applications},\ }\href {https://doi.org/10.1007/s12039-025-02361-2} {\bibfield  {journal} {\bibinfo  {journal} {Journal of Chemical Sciences}\ }\textbf {\bibinfo {volume} {137}},\ \bibinfo {pages} {29} (\bibinfo {year} {2025})}\BibitemShut {NoStop}%
\bibitem [{\citenamefont {Ketfi}\ \emph {et~al.}(2025)\citenamefont {Ketfi}, \citenamefont {Berri}, \citenamefont {Maouche},\ and\ \citenamefont {Bouarissa}}]{Ketfi_2025}%
  \BibitemOpen
  \bibfield  {author} {\bibinfo {author} {\bibfnamefont {M.}~\bibnamefont {Ketfi}}, \bibinfo {author} {\bibfnamefont {S.}~\bibnamefont {Berri}}, \bibinfo {author} {\bibfnamefont {D.}~\bibnamefont {Maouche}},\ and\ \bibinfo {author} {\bibfnamefont {N.}~\bibnamefont {Bouarissa}},\ }\bibfield  {title} {\bibinfo {title} {Comprehensive analysis of the structural, electronic, half-metalic, and thermoelectric properties of the quaternary {{Heusler}} compounds {{CoMnPtAl}} and {{CoMnIrGe}}},\ }\href {https://doi.org/10.1016/j.cocom.2025.e01021} {\bibfield  {journal} {\bibinfo  {journal} {Computational Condensed Matter}\ }\textbf {\bibinfo {volume} {43}},\ \bibinfo {pages} {e01021} (\bibinfo {year} {2025})}\BibitemShut {NoStop}%
\bibitem [{\citenamefont {Khan}\ \emph {et~al.}(2025)\citenamefont {Khan}, \citenamefont {Gul}, \citenamefont {Ullah}, \citenamefont {Ifseisi}, \citenamefont {Aziz},\ and\ \citenamefont {Abbas}}]{Khan_2025}%
  \BibitemOpen
  \bibfield  {author} {\bibinfo {author} {\bibfnamefont {M.~S.}\ \bibnamefont {Khan}}, \bibinfo {author} {\bibfnamefont {B.}~\bibnamefont {Gul}}, \bibinfo {author} {\bibfnamefont {Z.}~\bibnamefont {Ullah}}, \bibinfo {author} {\bibfnamefont {A.~A.}\ \bibnamefont {Ifseisi}}, \bibinfo {author} {\bibfnamefont {S.~M.}\ \bibnamefont {Aziz}},\ and\ \bibinfo {author} {\bibfnamefont {F.}~\bibnamefont {Abbas}},\ }\bibfield  {title} {\bibinfo {title} {Electronic {{Structure}}, {{Optical}} and {{Thermoelectric Properties}} of {{Novel Rb2XAgF6}} ({{X}} = {{Ga}} and {{In}}) halide-based {{Perovskites}}},\ }\href {https://doi.org/10.1007/s10904-024-03579-2} {\bibfield  {journal} {\bibinfo  {journal} {Journal of Inorganic and Organometallic Polymers and Materials}\ }\textbf {\bibinfo {volume} {35}},\ \bibinfo {pages} {5120} (\bibinfo {year} {2025})}\BibitemShut {NoStop}%
\bibitem [{\citenamefont {Khatar}\ \emph {et~al.}(2025{\natexlab{a}})\citenamefont {Khatar}, \citenamefont {Houari}, \citenamefont {Lantri}, \citenamefont {Bentata}, \citenamefont {Bouadjemi}, \citenamefont {Aziz}, \citenamefont {Boucherdoud},\ and\ \citenamefont {Boudjelal}}]{Khatar_2025}%
  \BibitemOpen
  \bibfield  {author} {\bibinfo {author} {\bibfnamefont {A.}~\bibnamefont {Khatar}}, \bibinfo {author} {\bibfnamefont {M.}~\bibnamefont {Houari}}, \bibinfo {author} {\bibfnamefont {T.}~\bibnamefont {Lantri}}, \bibinfo {author} {\bibfnamefont {S.}~\bibnamefont {Bentata}}, \bibinfo {author} {\bibfnamefont {B.}~\bibnamefont {Bouadjemi}}, \bibinfo {author} {\bibfnamefont {Z.}~\bibnamefont {Aziz}}, \bibinfo {author} {\bibfnamefont {A.}~\bibnamefont {Boucherdoud}},\ and\ \bibinfo {author} {\bibfnamefont {M.}~\bibnamefont {Boudjelal}},\ }\bibfield  {title} {\bibinfo {title} {Advanced materials for next-generation devices: Insights into the structural, optical, and thermoelectric properties of {{Hf2Pd2AlBi}} and {{Hf2Pd2AlSb}} alloys},\ }\bibfield  {journal} {\bibinfo  {journal} {Indian Journal of Physics}\ }\href {https://doi.org/10.1007/s12648-025-03643-8} {10.1007/s12648-025-03643-8} (\bibinfo {year} {2025}{\natexlab{a}})\BibitemShut {NoStop}%
\bibitem [{\citenamefont {Kumari}\ \emph {et~al.}(2025{\natexlab{a}})\citenamefont {Kumari}, \citenamefont {Abraham}, \citenamefont {Mishra}, \citenamefont {Sharma},\ and\ \citenamefont {Alshehri}}]{Kumari_2025}%
  \BibitemOpen
  \bibfield  {author} {\bibinfo {author} {\bibfnamefont {A.}~\bibnamefont {Kumari}}, \bibinfo {author} {\bibfnamefont {J.~A.}\ \bibnamefont {Abraham}}, \bibinfo {author} {\bibfnamefont {A.~K.}\ \bibnamefont {Mishra}}, \bibinfo {author} {\bibfnamefont {R.}~\bibnamefont {Sharma}},\ and\ \bibinfo {author} {\bibfnamefont {A.}~\bibnamefont {Alshehri}},\ }\bibfield  {title} {\bibinfo {title} {Ab-{{Initio}} based investigation of the optoelectronic, transport, mechanical properties of vacancy-ordered double perovskites {{A2PtI6}} ({{A}} = {{Na}}, {{Li}}): {{An}} emerging class of solar cell and thermoelectric materials},\ }\href {https://doi.org/10.1016/j.optlastec.2025.112609} {\bibfield  {journal} {\bibinfo  {journal} {Optics \& Laser Technology}\ }\textbf {\bibinfo {volume} {185}},\ \bibinfo {pages} {112609} (\bibinfo {year} {2025}{\natexlab{a}})}\BibitemShut {NoStop}%
\bibitem [{\citenamefont {Kumari}\ \emph {et~al.}(2025{\natexlab{b}})\citenamefont {Kumari}, \citenamefont {Ali}, \citenamefont {Abraham}, \citenamefont {Mishra}, \citenamefont {Kallel}, \citenamefont {Shewakh}, \citenamefont {Formanova},\ and\ \citenamefont {Sharma}}]{Kumari_2025a}%
  \BibitemOpen
  \bibfield  {author} {\bibinfo {author} {\bibfnamefont {A.}~\bibnamefont {Kumari}}, \bibinfo {author} {\bibfnamefont {A.~B.~M.}\ \bibnamefont {Ali}}, \bibinfo {author} {\bibfnamefont {J.~A.}\ \bibnamefont {Abraham}}, \bibinfo {author} {\bibfnamefont {A.~K.}\ \bibnamefont {Mishra}}, \bibinfo {author} {\bibfnamefont {M.}~\bibnamefont {Kallel}}, \bibinfo {author} {\bibfnamefont {W.~M.}\ \bibnamefont {Shewakh}}, \bibinfo {author} {\bibfnamefont {S.}~\bibnamefont {Formanova}},\ and\ \bibinfo {author} {\bibfnamefont {R.}~\bibnamefont {Sharma}},\ }\bibfield  {title} {\bibinfo {title} {A {{Comprehensive DFT Analysis}} of {{Novel Vacancy-Ordered Double Perovskites Na2SnX6}} ({{X}} = {{Br}}, {{I}}) for the {{Opto-electronic}} and {{Thermoelectric Properties Applications}}},\ }\bibfield  {journal} {\bibinfo  {journal} {Journal of Inorganic and Organometallic Polymers and Materials}\ }\href {https://doi.org/10.1007/s10904-025-03843-z} {10.1007/s10904-025-03843-z} (\bibinfo {year} {2025}{\natexlab{b}})\BibitemShut
  {NoStop}%
\bibitem [{\citenamefont {Luo}\ \emph {et~al.}(2025)\citenamefont {Luo}, \citenamefont {Chen}, \citenamefont {Wu}, \citenamefont {Wu}, \citenamefont {Qin},\ and\ \citenamefont {Ding}}]{Luo_2025}%
  \BibitemOpen
  \bibfield  {author} {\bibinfo {author} {\bibfnamefont {X.}~\bibnamefont {Luo}}, \bibinfo {author} {\bibfnamefont {P.}~\bibnamefont {Chen}}, \bibinfo {author} {\bibfnamefont {H.}~\bibnamefont {Wu}}, \bibinfo {author} {\bibfnamefont {X.}~\bibnamefont {Wu}}, \bibinfo {author} {\bibfnamefont {D.}~\bibnamefont {Qin}},\ and\ \bibinfo {author} {\bibfnamefont {G.}~\bibnamefont {Ding}},\ }\bibfield  {title} {\bibinfo {title} {Thermoelectric transport in {{Weyl}} semiconductor tellurium: {{The}} role of {{Weyl}} fermions},\ }\href {https://doi.org/10.1016/j.ssc.2025.115922} {\bibfield  {journal} {\bibinfo  {journal} {Solid State Communications}\ }\textbf {\bibinfo {volume} {401}},\ \bibinfo {pages} {115922} (\bibinfo {year} {2025})}\BibitemShut {NoStop}%
\bibitem [{\citenamefont {Manzoor}\ \emph {et~al.}(2025{\natexlab{a}})\citenamefont {Manzoor}, \citenamefont {Jamil}, \citenamefont {Sehgal}, \citenamefont {Sharma},\ and\ \citenamefont {Ibrahim}}]{Manzoor_2025}%
  \BibitemOpen
  \bibfield  {author} {\bibinfo {author} {\bibfnamefont {M.}~\bibnamefont {Manzoor}}, \bibinfo {author} {\bibfnamefont {M.}~\bibnamefont {Jamil}}, \bibinfo {author} {\bibfnamefont {L.}~\bibnamefont {Sehgal}}, \bibinfo {author} {\bibfnamefont {R.}~\bibnamefont {Sharma}},\ and\ \bibinfo {author} {\bibfnamefont {A.~A.}\ \bibnamefont {Ibrahim}},\ }\bibfield  {title} {\bibinfo {title} {Density functional theory based modelling on the physical properties of {{CsInAgAsX6}} ({{X}} = {{Cl}}, {{Br}}) double perovskite for green energy applications},\ }\href {https://doi.org/10.1016/j.mtcomm.2025.111586} {\bibfield  {journal} {\bibinfo  {journal} {Materials Today Communications}\ }\textbf {\bibinfo {volume} {43}},\ \bibinfo {pages} {111586} (\bibinfo {year} {2025}{\natexlab{a}})}\BibitemShut {NoStop}%
\bibitem [{\citenamefont {Manzoor}\ \emph {et~al.}(2025{\natexlab{b}})\citenamefont {Manzoor}, \citenamefont {Sharma}, \citenamefont {Kumari}, \citenamefont {{Al-Asbahi}}, \citenamefont {Kumar},\ and\ \citenamefont {Ullah}}]{Manzoor_2025a}%
  \BibitemOpen
  \bibfield  {author} {\bibinfo {author} {\bibfnamefont {M.}~\bibnamefont {Manzoor}}, \bibinfo {author} {\bibfnamefont {R.}~\bibnamefont {Sharma}}, \bibinfo {author} {\bibfnamefont {P.}~\bibnamefont {Kumari}}, \bibinfo {author} {\bibfnamefont {B.~A.}\ \bibnamefont {{Al-Asbahi}}}, \bibinfo {author} {\bibfnamefont {Y.~A.}\ \bibnamefont {Kumar}},\ and\ \bibinfo {author} {\bibfnamefont {H.}~\bibnamefont {Ullah}},\ }\bibfield  {title} {\bibinfo {title} {A {{Comprehensive DFT Insight}} on {{Thermoelectric Response}} of {{Halide Perovskites TlGeZ3}} ({{Z}} = {{Cl}}, {{Br}} and {{I}}) for {{Renewable Energy Devices}}},\ }\href {https://doi.org/10.1007/s10904-024-03587-2} {\bibfield  {journal} {\bibinfo  {journal} {Journal of Inorganic and Organometallic Polymers and Materials}\ }\textbf {\bibinfo {volume} {35}},\ \bibinfo {pages} {5287} (\bibinfo {year} {2025}{\natexlab{b}})}\BibitemShut {NoStop}%
\bibitem [{\citenamefont {Manzoor}\ \emph {et~al.}(2025{\natexlab{c}})\citenamefont {Manzoor}, \citenamefont {Abraham}, \citenamefont {Aslam}, \citenamefont {{Al-Asbahi}},\ and\ \citenamefont {Sharma}}]{Manzoor_2025b}%
  \BibitemOpen
  \bibfield  {author} {\bibinfo {author} {\bibfnamefont {M.}~\bibnamefont {Manzoor}}, \bibinfo {author} {\bibfnamefont {J.~A.}\ \bibnamefont {Abraham}}, \bibinfo {author} {\bibfnamefont {M.}~\bibnamefont {Aslam}}, \bibinfo {author} {\bibfnamefont {B.~A.}\ \bibnamefont {{Al-Asbahi}}},\ and\ \bibinfo {author} {\bibfnamefont {R.}~\bibnamefont {Sharma}},\ }\bibfield  {title} {\bibinfo {title} {A comprehensive {{DFT}} study on the physical attributes of {{Li}} based {{Half-Heusler}} compounds {{LiYXZ}} ({{X}} = {{Pt}}, {{Pd}}; {{Z}} = {{Si}}, {{Ge}})},\ }\href {https://doi.org/10.1016/j.ssc.2025.115875} {\bibfield  {journal} {\bibinfo  {journal} {Solid State Communications}\ }\textbf {\bibinfo {volume} {400}},\ \bibinfo {pages} {115875} (\bibinfo {year} {2025}{\natexlab{c}})}\BibitemShut {NoStop}%
\bibitem [{\citenamefont {Nazir}\ \emph {et~al.}(2025{\natexlab{a}})\citenamefont {Nazir}, \citenamefont {Saxena}, \citenamefont {Khera}, \citenamefont {Manzoor}, \citenamefont {Sharma}, \citenamefont {Ibrahim}, \citenamefont {{El-Meligy}},\ and\ \citenamefont {Abdullaeva}}]{Nazir_2025}%
  \BibitemOpen
  \bibfield  {author} {\bibinfo {author} {\bibfnamefont {A.}~\bibnamefont {Nazir}}, \bibinfo {author} {\bibfnamefont {A.}~\bibnamefont {Saxena}}, \bibinfo {author} {\bibfnamefont {E.~A.}\ \bibnamefont {Khera}}, \bibinfo {author} {\bibfnamefont {M.}~\bibnamefont {Manzoor}}, \bibinfo {author} {\bibfnamefont {R.}~\bibnamefont {Sharma}}, \bibinfo {author} {\bibfnamefont {A.~A.}\ \bibnamefont {Ibrahim}}, \bibinfo {author} {\bibfnamefont {M.~A.}\ \bibnamefont {{El-Meligy}}},\ and\ \bibinfo {author} {\bibfnamefont {B.}~\bibnamefont {Abdullaeva}},\ }\bibfield  {title} {\bibinfo {title} {Lead-{{Free Double Halide Perovskite Compounds}}: {{Unveiling}} the {{Structural}}, {{Optoelectronic}}, and {{Transport Properties}} of {{A2TlRhF6}} ({{A}} = {{K}}, {{Rb}}) for {{Robust}} and {{Sustainable Green Energy Applications}}},\ }\bibfield  {journal} {\bibinfo  {journal} {Journal of Inorganic and Organometallic Polymers and Materials}\ }\href {https://doi.org/10.1007/s10904-025-03666-y} {10.1007/s10904-025-03666-y} (\bibinfo
  {year} {2025}{\natexlab{a}})\BibitemShut {NoStop}%
\bibitem [{\citenamefont {Nazir}\ \emph {et~al.}(2025{\natexlab{b}})\citenamefont {Nazir}, \citenamefont {Khera}, \citenamefont {Manzoor}, \citenamefont {Sahid}, \citenamefont {Sharma}, \citenamefont {Khan}, \citenamefont {Kamolova},\ and\ \citenamefont {Saidani}}]{Nazir_2025a}%
  \BibitemOpen
  \bibfield  {author} {\bibinfo {author} {\bibfnamefont {A.}~\bibnamefont {Nazir}}, \bibinfo {author} {\bibfnamefont {E.~A.}\ \bibnamefont {Khera}}, \bibinfo {author} {\bibfnamefont {M.}~\bibnamefont {Manzoor}}, \bibinfo {author} {\bibfnamefont {S.}~\bibnamefont {Sahid}}, \bibinfo {author} {\bibfnamefont {R.}~\bibnamefont {Sharma}}, \bibinfo {author} {\bibfnamefont {R.}~\bibnamefont {Khan}}, \bibinfo {author} {\bibfnamefont {N.}~\bibnamefont {Kamolova}},\ and\ \bibinfo {author} {\bibfnamefont {T.}~\bibnamefont {Saidani}},\ }\bibfield  {title} {\bibinfo {title} {Role of {{Anion Variation}} in {{Physical Properties}} of {{Novel Vacancy Ordered Ga2PtX6}} ({{X}} = {{Br}} and {{I}}) for {{Solar Cell}} and {{Thermoelectric Applications}}},\ }\href {https://doi.org/10.1007/s10904-025-03625-7} {\bibfield  {journal} {\bibinfo  {journal} {Journal of Inorganic and Organometallic Polymers and Materials}\ }\textbf {\bibinfo {volume} {35}},\ \bibinfo {pages} {5832} (\bibinfo {year} {2025}{\natexlab{b}})}\BibitemShut
  {NoStop}%
\bibitem [{\citenamefont {Nazir}\ \emph {et~al.}(2025{\natexlab{c}})\citenamefont {Nazir}, \citenamefont {Khera}, \citenamefont {Manzoor}, \citenamefont {Sharma}, \citenamefont {Benabdallah},\ and\ \citenamefont {Ghodhbani}}]{Nazir_2025b}%
  \BibitemOpen
  \bibfield  {author} {\bibinfo {author} {\bibfnamefont {A.}~\bibnamefont {Nazir}}, \bibinfo {author} {\bibfnamefont {E.~A.}\ \bibnamefont {Khera}}, \bibinfo {author} {\bibfnamefont {M.}~\bibnamefont {Manzoor}}, \bibinfo {author} {\bibfnamefont {R.}~\bibnamefont {Sharma}}, \bibinfo {author} {\bibfnamefont {F.}~\bibnamefont {Benabdallah}},\ and\ \bibinfo {author} {\bibfnamefont {R.}~\bibnamefont {Ghodhbani}},\ }\bibfield  {title} {\bibinfo {title} {A {{DFT Insight}} on the physical, optoelectronic and thermoelectric characteristics of half-{{Heusler NaZn}}({{N}}/{{P}}) compounds for power generation applications},\ }\href {https://doi.org/10.1016/j.ssc.2025.115896} {\bibfield  {journal} {\bibinfo  {journal} {Solid State Communications}\ }\textbf {\bibinfo {volume} {400}},\ \bibinfo {pages} {115896} (\bibinfo {year} {2025}{\natexlab{c}})}\BibitemShut {NoStop}%
\bibitem [{\citenamefont {Nazir}\ \emph {et~al.}(2025{\natexlab{d}})\citenamefont {Nazir}, \citenamefont {Dixit}, \citenamefont {Ali}, \citenamefont {Khera}, \citenamefont {Manzoor}, \citenamefont {Sharma}, \citenamefont {Sardor}, \citenamefont {Hayitov}, \citenamefont {Al~Otaibi},\ and\ \citenamefont {Althubeiti}}]{Nazir_2025c}%
  \BibitemOpen
  \bibfield  {author} {\bibinfo {author} {\bibfnamefont {A.}~\bibnamefont {Nazir}}, \bibinfo {author} {\bibfnamefont {A.}~\bibnamefont {Dixit}}, \bibinfo {author} {\bibfnamefont {A.~B.~M.}\ \bibnamefont {Ali}}, \bibinfo {author} {\bibfnamefont {E.~A.}\ \bibnamefont {Khera}}, \bibinfo {author} {\bibfnamefont {M.}~\bibnamefont {Manzoor}}, \bibinfo {author} {\bibfnamefont {R.}~\bibnamefont {Sharma}}, \bibinfo {author} {\bibfnamefont {S.}~\bibnamefont {Sardor}}, \bibinfo {author} {\bibfnamefont {A.}~\bibnamefont {Hayitov}}, \bibinfo {author} {\bibfnamefont {S.}~\bibnamefont {Al~Otaibi}},\ and\ \bibinfo {author} {\bibfnamefont {K.}~\bibnamefont {Althubeiti}},\ }\bibfield  {title} {\bibinfo {title} {Exploring the {{Multifunctional Properties}} of {{Novel Full-Heusler AcCuZ2}} ({{Z}} = {{Se}}, {{Te}}) for {{High-Efficiency Photovoltaic}} and {{Renewable Energy Technologies}}},\ }\bibfield  {journal} {\bibinfo  {journal} {Journal of Electronic Materials}\ }\href {https://doi.org/10.1007/s11664-025-12207-9}
  {10.1007/s11664-025-12207-9} (\bibinfo {year} {2025}{\natexlab{d}})\BibitemShut {NoStop}%
\bibitem [{\citenamefont {Nazir}\ \emph {et~al.}(2025{\natexlab{e}})\citenamefont {Nazir}, \citenamefont {Ghodhbani}, \citenamefont {Khera}, \citenamefont {Manzoor}, \citenamefont {Sardor}, \citenamefont {Hayitov}, \citenamefont {Sharma},\ and\ \citenamefont {Boukhris}}]{Nazir_2025d}%
  \BibitemOpen
  \bibfield  {author} {\bibinfo {author} {\bibfnamefont {A.}~\bibnamefont {Nazir}}, \bibinfo {author} {\bibfnamefont {R.}~\bibnamefont {Ghodhbani}}, \bibinfo {author} {\bibfnamefont {E.~A.}\ \bibnamefont {Khera}}, \bibinfo {author} {\bibfnamefont {M.}~\bibnamefont {Manzoor}}, \bibinfo {author} {\bibfnamefont {S.}~\bibnamefont {Sardor}}, \bibinfo {author} {\bibfnamefont {A.}~\bibnamefont {Hayitov}}, \bibinfo {author} {\bibfnamefont {R.}~\bibnamefont {Sharma}},\ and\ \bibinfo {author} {\bibfnamefont {I.}~\bibnamefont {Boukhris}},\ }\bibfield  {title} {\bibinfo {title} {Probing the physical attributes of calcium-based perovskites for {{ACaI3}} ({{A}} = {{Li}} and {{Na}}) perovskites for next-generation renewable energy devices via {{TB-mBJ}} approach},\ }\bibfield  {journal} {\bibinfo  {journal} {Journal of Electroceramics}\ }\href {https://doi.org/10.1007/s10832-025-00415-y} {10.1007/s10832-025-00415-y} (\bibinfo {year} {2025}{\natexlab{e}})\BibitemShut {NoStop}%
\bibitem [{\citenamefont {Owais}\ \emph {et~al.}(2025)\citenamefont {Owais}, \citenamefont {Luo}, \citenamefont {Rehman}, \citenamefont {Mushtaq},\ and\ \citenamefont {Alkahtani}}]{Owais_2025}%
  \BibitemOpen
  \bibfield  {author} {\bibinfo {author} {\bibfnamefont {M.}~\bibnamefont {Owais}}, \bibinfo {author} {\bibfnamefont {X.}~\bibnamefont {Luo}}, \bibinfo {author} {\bibfnamefont {M.}~\bibnamefont {Rehman}}, \bibinfo {author} {\bibfnamefont {R.~T.}\ \bibnamefont {Mushtaq}},\ and\ \bibinfo {author} {\bibfnamefont {M.}~\bibnamefont {Alkahtani}},\ }\bibfield  {title} {\bibinfo {title} {Investigating {{Fe}} and {{Cr}} doping effects on thermoelectric efficiency in {{Mg3Sb2}} through first-principles calculations for sustainable energy solutions},\ }\href {https://doi.org/10.1038/s41598-025-92809-9} {\bibfield  {journal} {\bibinfo  {journal} {Scientific Reports}\ }\textbf {\bibinfo {volume} {15}},\ \bibinfo {pages} {9419} (\bibinfo {year} {2025})}\BibitemShut {NoStop}%
\bibitem [{\citenamefont {Rached}\ \emph {et~al.}(2025)\citenamefont {Rached}, \citenamefont {Rached}, \citenamefont {Caid}, \citenamefont {Amrani}, \citenamefont {Rached}, \citenamefont {Mansour}, \citenamefont {Mahmoud}, \citenamefont {{Al-Qaisi}}, \citenamefont {Alyami},\ and\ \citenamefont {Belkacem}}]{Rached_2025}%
  \BibitemOpen
  \bibfield  {author} {\bibinfo {author} {\bibfnamefont {H.}~\bibnamefont {Rached}}, \bibinfo {author} {\bibfnamefont {D.}~\bibnamefont {Rached}}, \bibinfo {author} {\bibfnamefont {M.}~\bibnamefont {Caid}}, \bibinfo {author} {\bibfnamefont {L.}~\bibnamefont {Amrani}}, \bibinfo {author} {\bibfnamefont {Y.}~\bibnamefont {Rached}}, \bibinfo {author} {\bibfnamefont {H.}~\bibnamefont {Mansour}}, \bibinfo {author} {\bibfnamefont {N.~T.}\ \bibnamefont {Mahmoud}}, \bibinfo {author} {\bibfnamefont {S.}~\bibnamefont {{Al-Qaisi}}}, \bibinfo {author} {\bibfnamefont {M.}~\bibnamefont {Alyami}},\ and\ \bibinfo {author} {\bibfnamefont {A.~A.~A.}\ \bibnamefont {Belkacem}},\ }\bibfield  {title} {\bibinfo {title} {Future {{Insights}} into {{Double Perovskites Ba2AlTMO6}} ({{TM}} = {{W}}, {{Re}}, and {{Os}}) for {{Sustainable}} and {{Clean Energy Production}}: {{A DFT Investigation Using GGA}}, {{TB-mBJ}}, and {{HSE06 Methods}}},\ }\href {https://doi.org/10.1007/s10904-024-03584-5} {\bibfield  {journal} {\bibinfo  {journal}
  {Journal of Inorganic and Organometallic Polymers and Materials}\ }\textbf {\bibinfo {volume} {35}},\ \bibinfo {pages} {5239} (\bibinfo {year} {2025})}\BibitemShut {NoStop}%
\bibitem [{\citenamefont {Ramzan}\ \emph {et~al.}(2025{\natexlab{a}})\citenamefont {Ramzan}, \citenamefont {Sofi}, \citenamefont {{Ishfaq-ul-Islam}}, \citenamefont {Khan},\ and\ \citenamefont {Khan}}]{Ramzan_2025a}%
  \BibitemOpen
  \bibfield  {author} {\bibinfo {author} {\bibfnamefont {A.}~\bibnamefont {Ramzan}}, \bibinfo {author} {\bibfnamefont {M.~Y.}\ \bibnamefont {Sofi}}, \bibinfo {author} {\bibfnamefont {M.}~\bibnamefont {{Ishfaq-ul-Islam}}}, \bibinfo {author} {\bibfnamefont {M.~S.}\ \bibnamefont {Khan}},\ and\ \bibinfo {author} {\bibfnamefont {M.~A.}\ \bibnamefont {Khan}},\ }\bibfield  {title} {\bibinfo {title} {Half-metallic ferromagnetism and thermoelectric-efficient behavior in chalcogenide spinels {{MgNi}}{\textsubscript{2}} {{X}}{\textsubscript{4}} ({{X}} = {{S}}, {{Se}}): A first-principles approach},\ }\href {https://doi.org/10.1039/D5RA03555D} {\bibfield  {journal} {\bibinfo  {journal} {RSC Advances}\ }\textbf {\bibinfo {volume} {15}},\ \bibinfo {pages} {24002} (\bibinfo {year} {2025}{\natexlab{a}})}\BibitemShut {NoStop}%
\bibitem [{\citenamefont {Righi}\ \emph {et~al.}(2025)\citenamefont {Righi}, \citenamefont {Bendahma}, \citenamefont {Labdelli}, \citenamefont {Mana}, \citenamefont {Bessaha}, \citenamefont {Bessaha}, \citenamefont {Khenata}, \citenamefont {Singh}, \citenamefont {Eithiraj}, \citenamefont {Ul~Haq},\ and\ \citenamefont {{Bin-Omran}}}]{Righi_2025}%
  \BibitemOpen
  \bibfield  {author} {\bibinfo {author} {\bibfnamefont {A.}~\bibnamefont {Righi}}, \bibinfo {author} {\bibfnamefont {F.}~\bibnamefont {Bendahma}}, \bibinfo {author} {\bibfnamefont {A.}~\bibnamefont {Labdelli}}, \bibinfo {author} {\bibfnamefont {M.}~\bibnamefont {Mana}}, \bibinfo {author} {\bibfnamefont {F.}~\bibnamefont {Bessaha}}, \bibinfo {author} {\bibfnamefont {G.}~\bibnamefont {Bessaha}}, \bibinfo {author} {\bibfnamefont {R.}~\bibnamefont {Khenata}}, \bibinfo {author} {\bibfnamefont {D.}~\bibnamefont {Singh}}, \bibinfo {author} {\bibfnamefont {R.}~\bibnamefont {Eithiraj}}, \bibinfo {author} {\bibfnamefont {B.}~\bibnamefont {Ul~Haq}},\ and\ \bibinfo {author} {\bibfnamefont {S.}~\bibnamefont {{Bin-Omran}}},\ }\bibfield  {title} {\bibinfo {title} {Structural, optoelectronic, thermodynamic, and thermoelectric properties of {{LiScNiZ}} ({{Z}} = {{Si}}, {{Ge}}, {{Sn}}) quaternary {{Heusler}} compounds via {{DFT}} approach},\ }\href {https://doi.org/10.1016/j.cocom.2025.e01092} {\bibfield  {journal} {\bibinfo
  {journal} {Computational Condensed Matter}\ }\textbf {\bibinfo {volume} {44}},\ \bibinfo {pages} {e01092} (\bibinfo {year} {2025})}\BibitemShut {NoStop}%
\bibitem [{\citenamefont {Sofi}\ \emph {et~al.}(2025)\citenamefont {Sofi}, \citenamefont {Khan},\ and\ \citenamefont {Khan}}]{Sofi_2025}%
  \BibitemOpen
  \bibfield  {author} {\bibinfo {author} {\bibfnamefont {M.~Y.}\ \bibnamefont {Sofi}}, \bibinfo {author} {\bibfnamefont {M.~S.}\ \bibnamefont {Khan}},\ and\ \bibinfo {author} {\bibfnamefont {M.~A.}\ \bibnamefont {Khan}},\ }\bibfield  {title} {\bibinfo {title} {Semiconducting ferromagnetism and thermoelectric performance of {{Rb}}{\textsubscript{2}} {{GeMI}}{\textsubscript{6}} ({{M}} = {{V}}, {{Ni}}, {{Mn}}): A computational perspective},\ }\href {https://doi.org/10.1039/D4MA01091D} {\bibfield  {journal} {\bibinfo  {journal} {Materials Advances}\ }\textbf {\bibinfo {volume} {6}},\ \bibinfo {pages} {2071} (\bibinfo {year} {2025})}\BibitemShut {NoStop}%
\bibitem [{\citenamefont {Srivastava}\ \emph {et~al.}(2025)\citenamefont {Srivastava}, \citenamefont {Rani}, \citenamefont {Rani}, \citenamefont {Toual}, \citenamefont {Dubey}, \citenamefont {Pandit}, \citenamefont {Verma}, \citenamefont {Nehra},\ and\ \citenamefont {Kamlesh}}]{Srivastava_2025}%
  \BibitemOpen
  \bibfield  {author} {\bibinfo {author} {\bibfnamefont {S.}~\bibnamefont {Srivastava}}, \bibinfo {author} {\bibfnamefont {U.}~\bibnamefont {Rani}}, \bibinfo {author} {\bibfnamefont {M.}~\bibnamefont {Rani}}, \bibinfo {author} {\bibfnamefont {Y.}~\bibnamefont {Toual}}, \bibinfo {author} {\bibfnamefont {A.}~\bibnamefont {Dubey}}, \bibinfo {author} {\bibfnamefont {N.}~\bibnamefont {Pandit}}, \bibinfo {author} {\bibfnamefont {A.~S.}\ \bibnamefont {Verma}}, \bibinfo {author} {\bibfnamefont {S.}~\bibnamefont {Nehra}},\ and\ \bibinfo {author} {\bibfnamefont {P.~K.}\ \bibnamefont {Kamlesh}},\ }\bibfield  {title} {\bibinfo {title} {Computational investigation of {{CaZnGe}} and {{CaZnSn}} half-heusler compounds: {{Potential}} candidates for thermoelectric devices},\ }\href {https://doi.org/10.1016/j.nxmate.2025.100724} {\bibfield  {journal} {\bibinfo  {journal} {Next Materials}\ }\textbf {\bibinfo {volume} {8}},\ \bibinfo {pages} {100724} (\bibinfo {year} {2025})}\BibitemShut {NoStop}%
\bibitem [{\citenamefont {Tufail}\ \emph {et~al.}(2025)\citenamefont {Tufail}, \citenamefont {Ahmed}, \citenamefont {Haleem}, \citenamefont {Amin},\ and\ \citenamefont {Shafiq}}]{Tufail_2025}%
  \BibitemOpen
  \bibfield  {author} {\bibinfo {author} {\bibfnamefont {D.}~\bibnamefont {Tufail}}, \bibinfo {author} {\bibfnamefont {U.}~\bibnamefont {Ahmed}}, \bibinfo {author} {\bibfnamefont {M.}~\bibnamefont {Haleem}}, \bibinfo {author} {\bibfnamefont {B.}~\bibnamefont {Amin}},\ and\ \bibinfo {author} {\bibfnamefont {M.}~\bibnamefont {Shafiq}},\ }\bibfield  {title} {\bibinfo {title} {{{DFT}} study of alkaline earth metals {{NaXH}}{\textsubscript{3}} ({{X}} = {{Be}}, {{Mg}}, {{Ca}}, {{Sr}}) for hydrogen storage capacity},\ }\href {https://doi.org/10.1039/D4RA05327C} {\bibfield  {journal} {\bibinfo  {journal} {RSC Advances}\ }\textbf {\bibinfo {volume} {15}},\ \bibinfo {pages} {337} (\bibinfo {year} {2025})}\BibitemShut {NoStop}%
\bibitem [{\citenamefont {Ullah}\ \emph {et~al.}(2025)\citenamefont {Ullah}, \citenamefont {Khan}, \citenamefont {Khan}, \citenamefont {Otaibi}, \citenamefont {Althubeiti},\ and\ \citenamefont {Abdullaev}}]{Ullah_2025}%
  \BibitemOpen
  \bibfield  {author} {\bibinfo {author} {\bibfnamefont {Z.}~\bibnamefont {Ullah}}, \bibinfo {author} {\bibfnamefont {R.}~\bibnamefont {Khan}}, \bibinfo {author} {\bibfnamefont {M.~A.}\ \bibnamefont {Khan}}, \bibinfo {author} {\bibfnamefont {S.~A.}\ \bibnamefont {Otaibi}}, \bibinfo {author} {\bibfnamefont {K.}~\bibnamefont {Althubeiti}},\ and\ \bibinfo {author} {\bibfnamefont {S.}~\bibnamefont {Abdullaev}},\ }\bibfield  {title} {\bibinfo {title} {High-temperature thermoelectric performance of spinel {{MgGa2O4}} through a first-principles and {{Boltzmann}} transport study},\ }\href {https://doi.org/10.1016/j.commatsci.2025.114163} {\bibfield  {journal} {\bibinfo  {journal} {Computational Materials Science}\ }\textbf {\bibinfo {volume} {259}},\ \bibinfo {pages} {114163} (\bibinfo {year} {2025})}\BibitemShut {NoStop}%
\bibitem [{\citenamefont {Yasin}\ \emph {et~al.}(2025{\natexlab{a}})\citenamefont {Yasin}, \citenamefont {Ullah}, \citenamefont {Abualnaja},\ and\ \citenamefont {Murtaza}}]{Yasin_2025a}%
  \BibitemOpen
  \bibfield  {author} {\bibinfo {author} {\bibfnamefont {S.}~\bibnamefont {Yasin}}, \bibinfo {author} {\bibfnamefont {H.}~\bibnamefont {Ullah}}, \bibinfo {author} {\bibfnamefont {K.~M.}\ \bibnamefont {Abualnaja}},\ and\ \bibinfo {author} {\bibfnamefont {G.}~\bibnamefont {Murtaza}},\ }\bibfield  {title} {\bibinfo {title} {Structural {{Stability}}, {{Half-Metallic Ferromagnetism}}, {{Magneto-Optical}}, and {{Thermoelectric Properties}} of {{Europium-Based Ternary Zintl Compounds EuZn2C2}}({{C}} = {{P}}, {{As}}): {{A Promising Alternative}} for {{Spintronics}} and {{Thermoelectric Applications}}},\ }\bibfield  {journal} {\bibinfo  {journal} {Journal of Inorganic and Organometallic Polymers and Materials}\ }\href {https://doi.org/10.1007/s10904-025-03692-w} {10.1007/s10904-025-03692-w} (\bibinfo {year} {2025}{\natexlab{a}})\BibitemShut {NoStop}%
\bibitem [{\citenamefont {Yutomo}\ \emph {et~al.}(2025)\citenamefont {Yutomo}, \citenamefont {Noor},\ and\ \citenamefont {Winata}}]{Yutomo_2025}%
  \BibitemOpen
  \bibfield  {author} {\bibinfo {author} {\bibfnamefont {E.~B.}\ \bibnamefont {Yutomo}}, \bibinfo {author} {\bibfnamefont {F.~A.}\ \bibnamefont {Noor}},\ and\ \bibinfo {author} {\bibfnamefont {T.}~\bibnamefont {Winata}},\ }\bibfield  {title} {\bibinfo {title} {Uniaxial strain effects on the electronic and thermoelectric properties of {{SnSe}} monolayer: {{A}} density functional theory study},\ }\href {https://doi.org/10.1016/j.comptc.2025.115184} {\bibfield  {journal} {\bibinfo  {journal} {Computational and Theoretical Chemistry}\ }\textbf {\bibinfo {volume} {1248}},\ \bibinfo {pages} {115184} (\bibinfo {year} {2025})}\BibitemShut {NoStop}%
\bibitem [{\citenamefont {Zayed}\ \emph {et~al.}(2025)\citenamefont {Zayed}, \citenamefont {Mustafa}, \citenamefont {{Al-Daraghmeh}}, \citenamefont {Younas}, \citenamefont {Algethami}, \citenamefont {Bouzgarrou}, \citenamefont {Khan}, \citenamefont {{Al-Buriahi}},\ and\ \citenamefont {Mahmood}}]{Zayed_2025}%
  \BibitemOpen
  \bibfield  {author} {\bibinfo {author} {\bibfnamefont {O.}~\bibnamefont {Zayed}}, \bibinfo {author} {\bibfnamefont {G.~M.}\ \bibnamefont {Mustafa}}, \bibinfo {author} {\bibfnamefont {T.~M.}\ \bibnamefont {{Al-Daraghmeh}}}, \bibinfo {author} {\bibfnamefont {B.}~\bibnamefont {Younas}}, \bibinfo {author} {\bibfnamefont {N.}~\bibnamefont {Algethami}}, \bibinfo {author} {\bibfnamefont {S.}~\bibnamefont {Bouzgarrou}}, \bibinfo {author} {\bibfnamefont {M.~T.}\ \bibnamefont {Khan}}, \bibinfo {author} {\bibfnamefont {M.}~\bibnamefont {{Al-Buriahi}}},\ and\ \bibinfo {author} {\bibfnamefont {Q.}~\bibnamefont {Mahmood}},\ }\bibfield  {title} {\bibinfo {title} {Study of electronics, optoelectronic and thermoelectric aspects of novel {{Zintl-phase}} alloys {{CaCd2X2}} ({{X}} = {{P}}, {{As}}, {{Sb}}) for solar cells and renewable energy},\ }\href {https://doi.org/10.1016/j.ssc.2025.116020} {\bibfield  {journal} {\bibinfo  {journal} {Solid State Communications}\ }\textbf {\bibinfo {volume} {403}},\ \bibinfo {pages}
  {116020} (\bibinfo {year} {2025})}\BibitemShut {NoStop}%
\bibitem [{\citenamefont {Abbas}\ \emph {et~al.}(2025)\citenamefont {Abbas}, \citenamefont {Xu}, \citenamefont {Riaz}, \citenamefont {Ishfaq}, \citenamefont {Nasarullah}, \citenamefont {Nazar}, \citenamefont {Alqorashi},\ and\ \citenamefont {Faizan}}]{Abbas_2025}%
  \BibitemOpen
  \bibfield  {author} {\bibinfo {author} {\bibfnamefont {J.}~\bibnamefont {Abbas}}, \bibinfo {author} {\bibfnamefont {Y.}~\bibnamefont {Xu}}, \bibinfo {author} {\bibfnamefont {K.}~\bibnamefont {Riaz}}, \bibinfo {author} {\bibfnamefont {M.}~\bibnamefont {Ishfaq}}, \bibinfo {author} {\bibfnamefont {{\relax Mr}.}~\bibnamefont {Nasarullah}}, \bibinfo {author} {\bibfnamefont {M.}~\bibnamefont {Nazar}}, \bibinfo {author} {\bibfnamefont {A.~K.}\ \bibnamefont {Alqorashi}},\ and\ \bibinfo {author} {\bibfnamefont {M.}~\bibnamefont {Faizan}},\ }\href {https://doi.org/10.2139/ssrn.5211836} {\bibinfo {title} {Comprehensive {{Study}} of the {{Energetics}}, {{Thermal}}, {{Electronic}}, {{Optical}}, {{Mechanical}}, {{Magnetic}}, and {{Thermoelectric Characteristics}} of {{Halide Double Perovskites X2nacrcl6}} ({{X}} = {{K}}, {{Rb}}) {{Using First-Principles Approach}}}} (\bibinfo {year} {2025})\BibitemShut {NoStop}%
\bibitem [{\citenamefont {{Al-Anazy}}\ \emph {et~al.}(2025{\natexlab{a}})\citenamefont {{Al-Anazy}}, \citenamefont {Ali}, \citenamefont {Ayyaz}, \citenamefont {Zayed}, \citenamefont {{Al-Daraghmeh}}, \citenamefont {{El-Rayyes}}, \citenamefont {Anbarasan},\ and\ \citenamefont {Mahmood}}]{Al-Anazy_2025}%
  \BibitemOpen
  \bibfield  {author} {\bibinfo {author} {\bibfnamefont {M.~M.}\ \bibnamefont {{Al-Anazy}}}, \bibinfo {author} {\bibfnamefont {H.~I.}\ \bibnamefont {Ali}}, \bibinfo {author} {\bibfnamefont {A.}~\bibnamefont {Ayyaz}}, \bibinfo {author} {\bibfnamefont {O.}~\bibnamefont {Zayed}}, \bibinfo {author} {\bibfnamefont {T.~M.}\ \bibnamefont {{Al-Daraghmeh}}}, \bibinfo {author} {\bibfnamefont {A.}~\bibnamefont {{El-Rayyes}}}, \bibinfo {author} {\bibfnamefont {R.}~\bibnamefont {Anbarasan}},\ and\ \bibinfo {author} {\bibfnamefont {Q.}~\bibnamefont {Mahmood}},\ }\bibfield  {title} {\bibinfo {title} {{{DFT Simulation}} of {{Eco-friendly Halide Double Perovskites A2AuInCl6}} ({{A}} = {{K}} and {{Rb}}) for {{Optoelectronic}} and {{Thermoelectric Devices}}},\ }\bibfield  {journal} {\bibinfo  {journal} {Journal of Inorganic and Organometallic Polymers and Materials}\ }\href {https://doi.org/10.1007/s10904-025-03724-5} {10.1007/s10904-025-03724-5} (\bibinfo {year} {2025}{\natexlab{a}})\BibitemShut {NoStop}%
\bibitem [{\citenamefont {{Al-Anazy}}\ \emph {et~al.}(2025{\natexlab{b}})\citenamefont {{Al-Anazy}}, \citenamefont {Zaman}, \citenamefont {Ayyaz}, \citenamefont {Almashnowi}, \citenamefont {Boukhris}, \citenamefont {Mahmood}, \citenamefont {Bouzgarrou},\ and\ \citenamefont {{Al-Buriahi}}}]{Al-Anazy_2025a}%
  \BibitemOpen
  \bibfield  {author} {\bibinfo {author} {\bibfnamefont {M.~M.}\ \bibnamefont {{Al-Anazy}}}, \bibinfo {author} {\bibfnamefont {M.}~\bibnamefont {Zaman}}, \bibinfo {author} {\bibfnamefont {A.}~\bibnamefont {Ayyaz}}, \bibinfo {author} {\bibfnamefont {M.~Y.}\ \bibnamefont {Almashnowi}}, \bibinfo {author} {\bibfnamefont {I.}~\bibnamefont {Boukhris}}, \bibinfo {author} {\bibfnamefont {Q.}~\bibnamefont {Mahmood}}, \bibinfo {author} {\bibfnamefont {S.}~\bibnamefont {Bouzgarrou}},\ and\ \bibinfo {author} {\bibfnamefont {M.~S.}\ \bibnamefont {{Al-Buriahi}}},\ }\bibfield  {title} {\bibinfo {title} {First {{Principles Investigation}} of {{Optoelectronic}} and {{Thermoelectric Functionality}} of {{Double Perovskites K2AuAlX6}} ({{X}} = {{Cl}} or {{Br}}) for {{Energy Conversion Applications}}},\ }\bibfield  {journal} {\bibinfo  {journal} {Journal of Inorganic and Organometallic Polymers and Materials}\ }\href {https://doi.org/10.1007/s10904-025-03859-5} {10.1007/s10904-025-03859-5} (\bibinfo {year}
  {2025}{\natexlab{b}})\BibitemShut {NoStop}%
\bibitem [{\citenamefont {{Al-Hmoud}}\ \emph {et~al.}(2025)\citenamefont {{Al-Hmoud}}, \citenamefont {Gul}, \citenamefont {Khan}, \citenamefont {Benabdellah}, \citenamefont {Ullah}, \citenamefont {Binzowaimil},\ and\ \citenamefont {Aziz}}]{Al-Hmoud_2025}%
  \BibitemOpen
  \bibfield  {author} {\bibinfo {author} {\bibfnamefont {M.}~\bibnamefont {{Al-Hmoud}}}, \bibinfo {author} {\bibfnamefont {B.}~\bibnamefont {Gul}}, \bibinfo {author} {\bibfnamefont {M.~S.}\ \bibnamefont {Khan}}, \bibinfo {author} {\bibfnamefont {G.}~\bibnamefont {Benabdellah}}, \bibinfo {author} {\bibfnamefont {Z.}~\bibnamefont {Ullah}}, \bibinfo {author} {\bibfnamefont {A.~M.}\ \bibnamefont {Binzowaimil}},\ and\ \bibinfo {author} {\bibfnamefont {S.~M.}\ \bibnamefont {Aziz}},\ }\bibfield  {title} {\bibinfo {title} {Exploring the impact of {{Sc}} and {{Y}} substitution on the physical properties of novel quaternary chalcogenides: {{A}} first-principles insight},\ }\href {https://doi.org/10.1016/j.cplett.2025.142311} {\bibfield  {journal} {\bibinfo  {journal} {Chemical Physics Letters}\ }\textbf {\bibinfo {volume} {877}},\ \bibinfo {pages} {142311} (\bibinfo {year} {2025})}\BibitemShut {NoStop}%
\bibitem [{\citenamefont {Albalawi}(2025{\natexlab{a}})}]{Albalawi_2025}%
  \BibitemOpen
  \bibfield  {author} {\bibinfo {author} {\bibfnamefont {H.}~\bibnamefont {Albalawi}},\ }\bibfield  {title} {\bibinfo {title} {Study of physical aspects of lead-free double perovskites {{Cs2AuXZ6}} ({{X}} = {{Al}}/{{Ga}}; {{Z}} = {{Cl}}/{{Br}}) for solar cells and thermoelectric applications},\ }\href {https://doi.org/10.1007/s11082-025-08230-z} {\bibfield  {journal} {\bibinfo  {journal} {Optical and Quantum Electronics}\ }\textbf {\bibinfo {volume} {57}},\ \bibinfo {pages} {313} (\bibinfo {year} {2025}{\natexlab{a}})}\BibitemShut {NoStop}%
\bibitem [{\citenamefont {Ali}\ \emph {et~al.}(2025{\natexlab{b}})\citenamefont {Ali}, \citenamefont {Ahmed}, \citenamefont {Saeed}, \citenamefont {Khan}, \citenamefont {Ali}, \citenamefont {Elansary}, \citenamefont {Mahmoud}, \citenamefont {Moussa}, \citenamefont {Bacha},\ and\ \citenamefont {Saeed}}]{Ali_2025}%
  \BibitemOpen
  \bibfield  {author} {\bibinfo {author} {\bibfnamefont {Z.}~\bibnamefont {Ali}}, \bibinfo {author} {\bibfnamefont {M.}~\bibnamefont {Ahmed}}, \bibinfo {author} {\bibfnamefont {M.~U.}\ \bibnamefont {Saeed}}, \bibinfo {author} {\bibfnamefont {A.~A.}\ \bibnamefont {Khan}}, \bibinfo {author} {\bibfnamefont {S.}~\bibnamefont {Ali}}, \bibinfo {author} {\bibfnamefont {H.~O.}\ \bibnamefont {Elansary}}, \bibinfo {author} {\bibfnamefont {E.~A.}\ \bibnamefont {Mahmoud}}, \bibinfo {author} {\bibfnamefont {I.~M.}\ \bibnamefont {Moussa}}, \bibinfo {author} {\bibfnamefont {A.}~\bibnamefont {Bacha}},\ and\ \bibinfo {author} {\bibfnamefont {Y.}~\bibnamefont {Saeed}},\ }\bibfield  {title} {\bibinfo {title} {Investigation of structural, magnetic, electronic, and optical properties for transition metal doped {{Cs2Sb2Cl6}} using {{DFT}} calculation},\ }\href {https://doi.org/10.1016/j.comptc.2025.115206} {\bibfield  {journal} {\bibinfo  {journal} {Computational and Theoretical Chemistry}\ }\textbf {\bibinfo {volume} {1248}},\
  \bibinfo {pages} {115206} (\bibinfo {year} {2025}{\natexlab{b}})}\BibitemShut {NoStop}%
\bibitem [{\citenamefont {Almeshal}\ \emph {et~al.}(2025)\citenamefont {Almeshal}, \citenamefont {{Saad H.-E.}},\ and\ \citenamefont {Alsobhi}}]{Almeshal_2025}%
  \BibitemOpen
  \bibfield  {author} {\bibinfo {author} {\bibfnamefont {A.}~\bibnamefont {Almeshal}}, \bibinfo {author} {\bibfnamefont {M.~M.}\ \bibnamefont {{Saad H.-E.}}},\ and\ \bibinfo {author} {\bibfnamefont {B.}~\bibnamefont {Alsobhi}},\ }\bibfield  {title} {\bibinfo {title} {A comprehensive study of physical properties of {{TMP2}} ({{TM}} = {{Fe}}, {{Ru}}, {{Os}}): {{As}} semiconductors for optoelectronic and thermoelectronic applications},\ }\href {https://doi.org/10.1016/j.rinp.2025.108211} {\bibfield  {journal} {\bibinfo  {journal} {Results in Physics}\ }\textbf {\bibinfo {volume} {72}},\ \bibinfo {pages} {108211} (\bibinfo {year} {2025})}\BibitemShut {NoStop}%
\bibitem [{\citenamefont {Ayyaz}\ \emph {et~al.}(2025{\natexlab{a}})\citenamefont {Ayyaz}, \citenamefont {{El-Rayyes}}, \citenamefont {Ali}, \citenamefont {Kebaili}, \citenamefont {Shakir}, \citenamefont {Syed},\ and\ \citenamefont {Mahmood}}]{Ayyaz_2025}%
  \BibitemOpen
  \bibfield  {author} {\bibinfo {author} {\bibfnamefont {A.}~\bibnamefont {Ayyaz}}, \bibinfo {author} {\bibfnamefont {A.}~\bibnamefont {{El-Rayyes}}}, \bibinfo {author} {\bibfnamefont {H.~I.}\ \bibnamefont {Ali}}, \bibinfo {author} {\bibfnamefont {I.}~\bibnamefont {Kebaili}}, \bibinfo {author} {\bibfnamefont {M.~B.}\ \bibnamefont {Shakir}}, \bibinfo {author} {\bibfnamefont {I.~S.}\ \bibnamefont {Syed}},\ and\ \bibinfo {author} {\bibfnamefont {Q.}~\bibnamefont {Mahmood}},\ }\bibfield  {title} {\bibinfo {title} {Study of electronic, optical, and thermoelectric aspects of double perovskites {{Rb2TlAsX6}} ({{X}} = {{Cl}}, {{Br}}) for green energy applications: A {{DFT}} approach},\ }\href {https://doi.org/10.1007/s43207-025-00484-3} {\bibfield  {journal} {\bibinfo  {journal} {Journal of the Korean Ceramic Society}\ }\textbf {\bibinfo {volume} {62}},\ \bibinfo {pages} {643} (\bibinfo {year} {2025}{\natexlab{a}})}\BibitemShut {NoStop}%
\bibitem [{\citenamefont {Bellahcene}\ \emph {et~al.}(2025)\citenamefont {Bellahcene}, \citenamefont {Bencherif}, \citenamefont {Chiker}, \citenamefont {Missoum},\ and\ \citenamefont {Bensaid}}]{Bellahcene_2025}%
  \BibitemOpen
  \bibfield  {author} {\bibinfo {author} {\bibfnamefont {F.~Z.~Z.}\ \bibnamefont {Bellahcene}}, \bibinfo {author} {\bibfnamefont {K.}~\bibnamefont {Bencherif}}, \bibinfo {author} {\bibfnamefont {F.}~\bibnamefont {Chiker}}, \bibinfo {author} {\bibfnamefont {D.-E.}\ \bibnamefont {Missoum}},\ and\ \bibinfo {author} {\bibfnamefont {D.}~\bibnamefont {Bensaid}},\ }\bibfield  {title} {\bibinfo {title} {{{DFT}} study of the novel double perovskite {{Sr2PrRuO6}}: Structural, electronic, optical, magnetic, and thermoelectric properties},\ }\href {https://doi.org/10.1140/epjb/s10051-025-00887-5} {\bibfield  {journal} {\bibinfo  {journal} {The European Physical Journal B}\ }\textbf {\bibinfo {volume} {98}},\ \bibinfo {pages} {45} (\bibinfo {year} {2025})}\BibitemShut {NoStop}%
\bibitem [{\citenamefont {Benahmedi}\ \emph {et~al.}(2025{\natexlab{b}})\citenamefont {Benahmedi}, \citenamefont {Besbes},\ and\ \citenamefont {Djelti}}]{Benahmedi_2025}%
  \BibitemOpen
  \bibfield  {author} {\bibinfo {author} {\bibfnamefont {L.}~\bibnamefont {Benahmedi}}, \bibinfo {author} {\bibfnamefont {A.}~\bibnamefont {Besbes}},\ and\ \bibinfo {author} {\bibfnamefont {R.}~\bibnamefont {Djelti}},\ }\bibfield  {title} {\bibinfo {title} {First-principles investigation of physical, mechanical, thermodynamics and transport properties of tetragonal double perovskite {{Sr2MnSbO6}}: {{A DFT}}+{{U}}+{{SOC}} study},\ }\href {https://doi.org/10.1016/j.matchemphys.2025.130520} {\bibfield  {journal} {\bibinfo  {journal} {Materials Chemistry and Physics}\ }\textbf {\bibinfo {volume} {334}},\ \bibinfo {pages} {130520} (\bibinfo {year} {2025}{\natexlab{b}})}\BibitemShut {NoStop}%
\bibitem [{\citenamefont {Dahane}\ and\ \citenamefont {{Ez-Zahraouy}}(2025)}]{Dahane_2025}%
  \BibitemOpen
  \bibfield  {author} {\bibinfo {author} {\bibfnamefont {L.}~\bibnamefont {Dahane}}\ and\ \bibinfo {author} {\bibfnamefont {H.}~\bibnamefont {{Ez-Zahraouy}}},\ }\bibfield  {title} {\bibinfo {title} {First-principles study of the structural, elastic, optoelectronic, thermoelectric, and photocatalytic properties of lead-free halide perovskite {{Cs2SnGeX6}} for hydrogen production and photovoltaic applications},\ }\href {https://doi.org/10.1016/j.ijhydene.2025.06.105} {\bibfield  {journal} {\bibinfo  {journal} {International Journal of Hydrogen Energy}\ }\textbf {\bibinfo {volume} {145}},\ \bibinfo {pages} {589} (\bibinfo {year} {2025})}\BibitemShut {NoStop}%
\bibitem [{\citenamefont {Das}\ and\ \citenamefont {Kalita}(2025)}]{Das_2025}%
  \BibitemOpen
  \bibfield  {author} {\bibinfo {author} {\bibfnamefont {G.}~\bibnamefont {Das}}\ and\ \bibinfo {author} {\bibfnamefont {B.}~\bibnamefont {Kalita}},\ }\bibfield  {title} {\bibinfo {title} {Lowering of lattice thermal conductivity through strain application on {{LiCaB}} half-heusler alloys in presence of aliovalent doping},\ }\href {https://doi.org/10.1016/j.jpcs.2025.112809} {\bibfield  {journal} {\bibinfo  {journal} {Journal of Physics and Chemistry of Solids}\ }\textbf {\bibinfo {volume} {205}},\ \bibinfo {pages} {112809} (\bibinfo {year} {2025})}\BibitemShut {NoStop}%
\bibitem [{\citenamefont {Farah}\ \emph {et~al.}(2025)\citenamefont {Farah}, \citenamefont {Radja}, \citenamefont {Ibrahim}, \citenamefont {Khadra}, \citenamefont {Mohammed},\ and\ \citenamefont {{Al-Douri}}}]{Farah_2025}%
  \BibitemOpen
  \bibfield  {author} {\bibinfo {author} {\bibfnamefont {B.~L.}\ \bibnamefont {Farah}}, \bibinfo {author} {\bibfnamefont {K.}~\bibnamefont {Radja}}, \bibinfo {author} {\bibfnamefont {A.}~\bibnamefont {Ibrahim}}, \bibinfo {author} {\bibfnamefont {K.}~\bibnamefont {Khadra}}, \bibinfo {author} {\bibfnamefont {A.}~\bibnamefont {Mohammed}},\ and\ \bibinfo {author} {\bibfnamefont {Y.}~\bibnamefont {{Al-Douri}}},\ }\bibfield  {title} {\bibinfo {title} {{{RbNbX}}{$_3$} ({{X}} = cl, {{Br}}, {{I}}) lead-free halide perovskites: {{A DFT}} study of structural, elastic, electronic, and thermoelectric properties for energy applications},\ }\href {https://doi.org/10.1016/j.chemphys.2025.112843} {\bibfield  {journal} {\bibinfo  {journal} {Chemical Physics}\ }\textbf {\bibinfo {volume} {598}},\ \bibinfo {pages} {112843} (\bibinfo {year} {2025})}\BibitemShut {NoStop}%
\bibitem [{\citenamefont {Lemnawar}\ \emph {et~al.}(2025)\citenamefont {Lemnawar}, \citenamefont {El~Bakkali}, \citenamefont {Bouri}, \citenamefont {Labrim}, \citenamefont {Amraoui}, \citenamefont {Kibbou}, \citenamefont {Louzazni},\ and\ \citenamefont {Nouneh}}]{Lemnawar_2025}%
  \BibitemOpen
  \bibfield  {author} {\bibinfo {author} {\bibfnamefont {A.}~\bibnamefont {Lemnawar}}, \bibinfo {author} {\bibfnamefont {I.}~\bibnamefont {El~Bakkali}}, \bibinfo {author} {\bibfnamefont {N.}~\bibnamefont {Bouri}}, \bibinfo {author} {\bibfnamefont {H.}~\bibnamefont {Labrim}}, \bibinfo {author} {\bibfnamefont {S.}~\bibnamefont {Amraoui}}, \bibinfo {author} {\bibfnamefont {M.}~\bibnamefont {Kibbou}}, \bibinfo {author} {\bibfnamefont {M.}~\bibnamefont {Louzazni}},\ and\ \bibinfo {author} {\bibfnamefont {K.}~\bibnamefont {Nouneh}},\ }\bibfield  {title} {\bibinfo {title} {First principles study of the physical properties of {{Y2FeCrO6}} double perovskite: {{Optoelectronic}} and thermoelectric applications},\ }\href {https://doi.org/10.1016/j.rineng.2025.104530} {\bibfield  {journal} {\bibinfo  {journal} {Results in Engineering}\ }\textbf {\bibinfo {volume} {25}},\ \bibinfo {pages} {104530} (\bibinfo {year} {2025})}\BibitemShut {NoStop}%
\bibitem [{\citenamefont {Mechehoud}\ \emph {et~al.}(2025{\natexlab{a}})\citenamefont {Mechehoud}, \citenamefont {Zitouni}, \citenamefont {Cherif}, \citenamefont {Bouadjemi}, \citenamefont {Houari}, \citenamefont {Haid}, \citenamefont {Matougui}, \citenamefont {Lantri}, \citenamefont {Bentata}, \citenamefont {Aziz},\ and\ \citenamefont {Bouhafs}}]{Mechehoud_2025}%
  \BibitemOpen
  \bibfield  {author} {\bibinfo {author} {\bibfnamefont {N.}~\bibnamefont {Mechehoud}}, \bibinfo {author} {\bibfnamefont {A.}~\bibnamefont {Zitouni}}, \bibinfo {author} {\bibfnamefont {M.~H.}\ \bibnamefont {Cherif}}, \bibinfo {author} {\bibfnamefont {B.}~\bibnamefont {Bouadjemi}}, \bibinfo {author} {\bibfnamefont {M.}~\bibnamefont {Houari}}, \bibinfo {author} {\bibfnamefont {S.}~\bibnamefont {Haid}}, \bibinfo {author} {\bibfnamefont {M.}~\bibnamefont {Matougui}}, \bibinfo {author} {\bibfnamefont {T.}~\bibnamefont {Lantri}}, \bibinfo {author} {\bibfnamefont {S.}~\bibnamefont {Bentata}}, \bibinfo {author} {\bibfnamefont {Z.}~\bibnamefont {Aziz}},\ and\ \bibinfo {author} {\bibfnamefont {B.}~\bibnamefont {Bouhafs}},\ }\bibfield  {title} {\bibinfo {title} {Multifunctional halide double perovskites: {{Cs2AgMoCl6}} and {{K2AgMoCl6}} for renewable energy and spintronic technologies},\ }\href {https://doi.org/10.1016/j.cocom.2025.e01087} {\bibfield  {journal} {\bibinfo  {journal} {Computational Condensed Matter}\
  }\textbf {\bibinfo {volume} {44}},\ \bibinfo {pages} {e01087} (\bibinfo {year} {2025}{\natexlab{a}})}\BibitemShut {NoStop}%
\bibitem [{\citenamefont {Mechehoud}\ \emph {et~al.}(2025{\natexlab{b}})\citenamefont {Mechehoud}, \citenamefont {Zitouni}, \citenamefont {Bouadjemi}, \citenamefont {Houari}, \citenamefont {Haid}, \citenamefont {Matougui}, \citenamefont {Hamdi~Cherif}, \citenamefont {Lantri}, \citenamefont {Bentata},\ and\ \citenamefont {Bouhafs}}]{Mechehoud_2025a}%
  \BibitemOpen
  \bibfield  {author} {\bibinfo {author} {\bibfnamefont {N.}~\bibnamefont {Mechehoud}}, \bibinfo {author} {\bibfnamefont {A.}~\bibnamefont {Zitouni}}, \bibinfo {author} {\bibfnamefont {B.}~\bibnamefont {Bouadjemi}}, \bibinfo {author} {\bibfnamefont {M.}~\bibnamefont {Houari}}, \bibinfo {author} {\bibfnamefont {S.}~\bibnamefont {Haid}}, \bibinfo {author} {\bibfnamefont {M.}~\bibnamefont {Matougui}}, \bibinfo {author} {\bibfnamefont {M.}~\bibnamefont {Hamdi~Cherif}}, \bibinfo {author} {\bibfnamefont {T.}~\bibnamefont {Lantri}}, \bibinfo {author} {\bibfnamefont {S.}~\bibnamefont {Bentata}},\ and\ \bibinfo {author} {\bibfnamefont {B.}~\bibnamefont {Bouhafs}},\ }\bibfield  {title} {\bibinfo {title} {A comprehensive {{DFT}}+{{U}} study of optoelectronic and thermoelectric properties of a new half semiconductor ({{HSC}}) chloro-elpasolite {{Rb2AgMoCl6}} for green energy},\ }\href {https://doi.org/10.1016/j.physb.2025.417603} {\bibfield  {journal} {\bibinfo  {journal} {Physica B: Condensed Matter}\ }\textbf {\bibinfo
  {volume} {715}},\ \bibinfo {pages} {417603} (\bibinfo {year} {2025}{\natexlab{b}})}\BibitemShut {NoStop}%
\bibitem [{\citenamefont {Mohamed}\ \emph {et~al.}(2025)\citenamefont {Mohamed}, \citenamefont {Gul}, \citenamefont {Khan}, \citenamefont {Aziz},\ and\ \citenamefont {Ahmad}}]{Mohamed_2025}%
  \BibitemOpen
  \bibfield  {author} {\bibinfo {author} {\bibfnamefont {A.~S.}\ \bibnamefont {Mohamed}}, \bibinfo {author} {\bibfnamefont {B.}~\bibnamefont {Gul}}, \bibinfo {author} {\bibfnamefont {M.~S.}\ \bibnamefont {Khan}}, \bibinfo {author} {\bibfnamefont {S.~M.}\ \bibnamefont {Aziz}},\ and\ \bibinfo {author} {\bibfnamefont {H.}~\bibnamefont {Ahmad}},\ }\bibfield  {title} {\bibinfo {title} {Unveiling the optoelectronic and thermoelectric properties of {{Cs2HgM}} ({{M}} = {{Se}}, {{Te}}) semiconductors for sustainable optoelectronic technologies},\ }\href {https://doi.org/10.1016/j.inoche.2025.114837} {\bibfield  {journal} {\bibinfo  {journal} {Inorganic Chemistry Communications}\ }\textbf {\bibinfo {volume} {179}},\ \bibinfo {pages} {114837} (\bibinfo {year} {2025})}\BibitemShut {NoStop}%
\bibitem [{\citenamefont {Ramzan}\ \emph {et~al.}(2025{\natexlab{b}})\citenamefont {Ramzan}, \citenamefont {Sofi}, \citenamefont {Khan}, \citenamefont {Ali},\ and\ \citenamefont {Khan}}]{Ramzan_2025}%
  \BibitemOpen
  \bibfield  {author} {\bibinfo {author} {\bibfnamefont {A.}~\bibnamefont {Ramzan}}, \bibinfo {author} {\bibfnamefont {M.~Y.}\ \bibnamefont {Sofi}}, \bibinfo {author} {\bibfnamefont {M.~S.}\ \bibnamefont {Khan}}, \bibinfo {author} {\bibfnamefont {J.}~\bibnamefont {Ali}},\ and\ \bibinfo {author} {\bibfnamefont {M.~A.}\ \bibnamefont {Khan}},\ }\bibfield  {title} {\bibinfo {title} {First-principles investigation of {{Mg-based MgAl2X4}} ({{X}} = {{S}}, {{Se}}) spinels for optoelectronic and energy harvesting applications},\ }\href {https://doi.org/10.1140/epjb/s10051-025-00937-y} {\bibfield  {journal} {\bibinfo  {journal} {The European Physical Journal B}\ }\textbf {\bibinfo {volume} {98}},\ \bibinfo {pages} {102} (\bibinfo {year} {2025}{\natexlab{b}})}\BibitemShut {NoStop}%
\bibitem [{\citenamefont {{Saad H.-E.}}\ and\ \citenamefont {Alsobhi}(2025{\natexlab{a}})}]{SaadH.-E._2025}%
  \BibitemOpen
  \bibfield  {author} {\bibinfo {author} {\bibfnamefont {M.~M.}\ \bibnamefont {{Saad H.-E.}}}\ and\ \bibinfo {author} {\bibfnamefont {B.}~\bibnamefont {Alsobhi}},\ }\bibfield  {title} {\bibinfo {title} {Theoretical study of vacancy-ordered rubidium double perovskites ({{Rb2VBr6}}, {{Rb2CrBr6}}, and {{Rb2MnBr6}}) for spintronics, optoelectronics and thermoelectronics applications},\ }\href {https://doi.org/10.1016/j.physleta.2025.130276} {\bibfield  {journal} {\bibinfo  {journal} {Physics Letters A}\ }\textbf {\bibinfo {volume} {535}},\ \bibinfo {pages} {130276} (\bibinfo {year} {2025}{\natexlab{a}})}\BibitemShut {NoStop}%
\bibitem [{\citenamefont {{Saad H.-E.}}\ and\ \citenamefont {Alsobhi}(2025{\natexlab{b}})}]{SaadH.-E._2025a}%
  \BibitemOpen
  \bibfield  {author} {\bibinfo {author} {\bibfnamefont {M.~M.}\ \bibnamefont {{Saad H.-E.}}}\ and\ \bibinfo {author} {\bibfnamefont {B.}~\bibnamefont {Alsobhi}},\ }\bibfield  {title} {\bibinfo {title} {{{DFT}} insights on the chloride double perovskites {{X2AuBiCl6}} ({{X}} = {{K}}, {{Rb}}, and {{Cs}}) with semiconductor nature for {{PV}} and optoelectronic applications},\ }\href {https://doi.org/10.1016/j.cocom.2025.e01040} {\bibfield  {journal} {\bibinfo  {journal} {Computational Condensed Matter}\ }\textbf {\bibinfo {volume} {43}},\ \bibinfo {pages} {e01040} (\bibinfo {year} {2025}{\natexlab{b}})}\BibitemShut {NoStop}%
\bibitem [{\citenamefont {Shah}\ \emph {et~al.}(2025)\citenamefont {Shah}, \citenamefont {Afzal}, \citenamefont {Khan}, \citenamefont {Zafar}, \citenamefont {Sifuna}, \citenamefont {Nassani}, \citenamefont {Asif},\ and\ \citenamefont {Abbas}}]{Shah_2025}%
  \BibitemOpen
  \bibfield  {author} {\bibinfo {author} {\bibfnamefont {S.~Z.~A.}\ \bibnamefont {Shah}}, \bibinfo {author} {\bibfnamefont {A.}~\bibnamefont {Afzal}}, \bibinfo {author} {\bibfnamefont {F.}~\bibnamefont {Khan}}, \bibinfo {author} {\bibfnamefont {S.}~\bibnamefont {Zafar}}, \bibinfo {author} {\bibfnamefont {J.}~\bibnamefont {Sifuna}}, \bibinfo {author} {\bibfnamefont {A.~A.}\ \bibnamefont {Nassani}}, \bibinfo {author} {\bibfnamefont {S.~U.}\ \bibnamefont {Asif}},\ and\ \bibinfo {author} {\bibfnamefont {Z.}~\bibnamefont {Abbas}},\ }\bibfield  {title} {\bibinfo {title} {Small band gap {{Pb-free}} double perovskites {{X2NaSbBr6}} ({{X}}={{Na}}, {{Li}}): {{A}} study of the stabilities, opto-electronic and thermoelectric aspects from the first-principles approach},\ }\href {https://doi.org/10.1016/j.cocom.2024.e01001} {\bibfield  {journal} {\bibinfo  {journal} {Computational Condensed Matter}\ }\textbf {\bibinfo {volume} {42}},\ \bibinfo {pages} {e01001} (\bibinfo {year} {2025})}\BibitemShut {NoStop}%
\bibitem [{\citenamefont {Yasin}\ \emph {et~al.}(2025{\natexlab{b}})\citenamefont {Yasin}, \citenamefont {Ali}, \citenamefont {Muddassir}, \citenamefont {Iqbal}, \citenamefont {Bano},\ and\ \citenamefont {Shakil}}]{Yasin_2025}%
  \BibitemOpen
  \bibfield  {author} {\bibinfo {author} {\bibfnamefont {S.}~\bibnamefont {Yasin}}, \bibinfo {author} {\bibfnamefont {A.}~\bibnamefont {Ali}}, \bibinfo {author} {\bibfnamefont {M.}~\bibnamefont {Muddassir}}, \bibinfo {author} {\bibfnamefont {N.}~\bibnamefont {Iqbal}}, \bibinfo {author} {\bibfnamefont {N.}~\bibnamefont {Bano}},\ and\ \bibinfo {author} {\bibfnamefont {M.}~\bibnamefont {Shakil}},\ }\bibfield  {title} {\bibinfo {title} {First-principles calculations to investigate electronic, optical, mechanical and thermoelectric properties of lead-free halide double perovskites {{Na2InBiX6}} ({{X}} = cl, {{Br}} and {{I}}) for optoelectronic and thermoelectric applications},\ }\href {https://doi.org/10.1016/j.comptc.2025.115107} {\bibfield  {journal} {\bibinfo  {journal} {Computational and Theoretical Chemistry}\ }\textbf {\bibinfo {volume} {1245}},\ \bibinfo {pages} {115107} (\bibinfo {year} {2025}{\natexlab{b}})}\BibitemShut {NoStop}%
\bibitem [{\citenamefont {Ahmed}\ \emph {et~al.}(2025)\citenamefont {Ahmed}, \citenamefont {Ullah}, \citenamefont {Rehman}, \citenamefont {Murtza}, \citenamefont {Ayyaz}, \citenamefont {Irfan}, \citenamefont {Kebaili},\ and\ \citenamefont {{El-Rayyes}}}]{Ahmed_2025}%
  \BibitemOpen
  \bibfield  {author} {\bibinfo {author} {\bibfnamefont {M.~A.}\ \bibnamefont {Ahmed}}, \bibinfo {author} {\bibfnamefont {S.}~\bibnamefont {Ullah}}, \bibinfo {author} {\bibfnamefont {K.~U.}\ \bibnamefont {Rehman}}, \bibinfo {author} {\bibfnamefont {G.}~\bibnamefont {Murtza}}, \bibinfo {author} {\bibfnamefont {A.}~\bibnamefont {Ayyaz}}, \bibinfo {author} {\bibfnamefont {M.}~\bibnamefont {Irfan}}, \bibinfo {author} {\bibfnamefont {I.}~\bibnamefont {Kebaili}},\ and\ \bibinfo {author} {\bibfnamefont {A.}~\bibnamefont {{El-Rayyes}}},\ }\bibfield  {title} {\bibinfo {title} {A {{First-Principles Study}} of the {{Structural}}, {{Optoelectronic}} and {{Thermoelectric Properties}} of {{Ba2XIO6}} ({{X}} = {{Li}}, {{K}} and {{Rb}}) for {{Energy Harvesting}}},\ }\href {https://doi.org/10.1007/s13538-025-01847-1} {\bibfield  {journal} {\bibinfo  {journal} {Brazilian Journal of Physics}\ }\textbf {\bibinfo {volume} {55}},\ \bibinfo {pages} {221} (\bibinfo {year} {2025})}\BibitemShut {NoStop}%
\bibitem [{\citenamefont {Albalawi}(2025{\natexlab{b}})}]{Albalawi_2025a}%
  \BibitemOpen
  \bibfield  {author} {\bibinfo {author} {\bibfnamefont {H.}~\bibnamefont {Albalawi}},\ }\bibfield  {title} {\bibinfo {title} {Study of {{Electronic}}, {{Optical}}, {{Mechanical}}, and {{Thermoelectric Aspects}} of {{Double Perovskite Oxides Ba2CaMO6}} ({{M}} = {{S}}/{{Se}}) for {{Solar Cells}} and {{Transport Applications}}},\ }\bibfield  {journal} {\bibinfo  {journal} {Journal of Inorganic and Organometallic Polymers and Materials}\ }\href {https://doi.org/10.1007/s10904-025-03800-w} {10.1007/s10904-025-03800-w} (\bibinfo {year} {2025}{\natexlab{b}})\BibitemShut {NoStop}%
\bibitem [{\citenamefont {Algethami}\ \emph {et~al.}(2025)\citenamefont {Algethami}, \citenamefont {Salem}, \citenamefont {Khudair}, \citenamefont {Ditta}, \citenamefont {{Nasarullah}}, \citenamefont {Ghani}, \citenamefont {Faizan}, \citenamefont {Nazar}, \citenamefont {Yousif}, \citenamefont {Kudhair},\ and\ \citenamefont {Muninir}}]{Algethami_2025}%
  \BibitemOpen
  \bibfield  {author} {\bibinfo {author} {\bibfnamefont {N.}~\bibnamefont {Algethami}}, \bibinfo {author} {\bibfnamefont {K.~H.}\ \bibnamefont {Salem}}, \bibinfo {author} {\bibfnamefont {Z.~F.}\ \bibnamefont {Khudair}}, \bibinfo {author} {\bibfnamefont {A.~A.}\ \bibnamefont {Ditta}}, \bibinfo {author} {\bibnamefont {{Nasarullah}}}, \bibinfo {author} {\bibfnamefont {M.~U.}\ \bibnamefont {Ghani}}, \bibinfo {author} {\bibfnamefont {M.}~\bibnamefont {Faizan}}, \bibinfo {author} {\bibfnamefont {M.}~\bibnamefont {Nazar}}, \bibinfo {author} {\bibfnamefont {Z.~S.}\ \bibnamefont {Yousif}}, \bibinfo {author} {\bibfnamefont {S.~A.}\ \bibnamefont {Kudhair}},\ and\ \bibinfo {author} {\bibfnamefont {M.~A.}\ \bibnamefont {Muninir}},\ }\bibfield  {title} {\bibinfo {title} {Theoretical analysis of rare earth-based halide double perovskites {{Cs2REAgCl6}} ({{RE}}={{La}}, {{Lu}}): {{Unveiling}} physical properties},\ }\href {https://doi.org/10.1016/j.jre.2025.06.018} {\bibfield  {journal} {\bibinfo  {journal} {Journal of Rare
  Earths}\ ,\ \bibinfo {pages} {S1002072125002303}} (\bibinfo {year} {2025})}\BibitemShut {NoStop}%
\bibitem [{\citenamefont {Alsaiari}\ \emph {et~al.}(2025)\citenamefont {Alsaiari}, \citenamefont {Ahmed}, \citenamefont {Hanf}, \citenamefont {Rekha}, \citenamefont {Kundlas}, \citenamefont {Ouladsmane}, \citenamefont {Muhammad},\ and\ \citenamefont {Rehman}}]{Alsaiari_2025}%
  \BibitemOpen
  \bibfield  {author} {\bibinfo {author} {\bibfnamefont {N.~S.}\ \bibnamefont {Alsaiari}}, \bibinfo {author} {\bibfnamefont {I.}~\bibnamefont {Ahmed}}, \bibinfo {author} {\bibfnamefont {S.}~\bibnamefont {Hanf}}, \bibinfo {author} {\bibfnamefont {M.~M.}\ \bibnamefont {Rekha}}, \bibinfo {author} {\bibfnamefont {M.}~\bibnamefont {Kundlas}}, \bibinfo {author} {\bibfnamefont {M.}~\bibnamefont {Ouladsmane}}, \bibinfo {author} {\bibfnamefont {K.}~\bibnamefont {Muhammad}},\ and\ \bibinfo {author} {\bibfnamefont {J.}~\bibnamefont {Rehman}},\ }\bibfield  {title} {\bibinfo {title} {Unveiling the {{Potential}} of {{X2RbAsI6}} ({{X}} = {{K}}, {{Cs}}) {{Halide Double Perovskites}} for {{Advanced Applications Through First-Principles Modeling}}},\ }\bibfield  {journal} {\bibinfo  {journal} {Journal of Inorganic and Organometallic Polymers and Materials}\ }\href {https://doi.org/10.1007/s10904-025-03746-z} {10.1007/s10904-025-03746-z} (\bibinfo {year} {2025})\BibitemShut {NoStop}%
\bibitem [{\citenamefont {AlShaikh~Mohammad}\ \emph {et~al.}(2025)\citenamefont {AlShaikh~Mohammad}, \citenamefont {Badaoui}, \citenamefont {Manzoor}, \citenamefont {Larguech}, \citenamefont {Asad},\ and\ \citenamefont {Menni}}]{AlShaikhMohammad_2025}%
  \BibitemOpen
  \bibfield  {author} {\bibinfo {author} {\bibfnamefont {N.~F.}\ \bibnamefont {AlShaikh~Mohammad}}, \bibinfo {author} {\bibfnamefont {A.}~\bibnamefont {Badaoui}}, \bibinfo {author} {\bibfnamefont {M.}~\bibnamefont {Manzoor}}, \bibinfo {author} {\bibfnamefont {S.}~\bibnamefont {Larguech}}, \bibinfo {author} {\bibfnamefont {J.}~\bibnamefont {Asad}},\ and\ \bibinfo {author} {\bibfnamefont {Y.}~\bibnamefont {Menni}},\ }\bibfield  {title} {\bibinfo {title} {First principles investigation of semiconducting and thermoelectric behavior in {{TiIrBi}} half-{{Heusler}} compound},\ }\href {https://doi.org/10.1063/5.0279869} {\bibfield  {journal} {\bibinfo  {journal} {AIP Advances}\ }\textbf {\bibinfo {volume} {15}},\ \bibinfo {pages} {075015} (\bibinfo {year} {2025})}\BibitemShut {NoStop}%
\bibitem [{\citenamefont {Ashik}\ \emph {et~al.}(2025)\citenamefont {Ashik}, \citenamefont {Biswas}, \citenamefont {Fahim},\ and\ \citenamefont {Amin}}]{Ashik_2025}%
  \BibitemOpen
  \bibfield  {author} {\bibinfo {author} {\bibfnamefont {A.}~\bibnamefont {Ashik}}, \bibinfo {author} {\bibfnamefont {P.}~\bibnamefont {Biswas}}, \bibinfo {author} {\bibfnamefont {M.~H.}\ \bibnamefont {Fahim}},\ and\ \bibinfo {author} {\bibfnamefont {M.~R.}\ \bibnamefont {Amin}},\ }\bibfield  {title} {\bibinfo {title} {A first-principles investigation of lead-free novel direct band gap double perovskite oxides {{X}} 2 {{AlBiO}} 6 ({{X}} = {{Mg}}, {{Ca}}, {{Ba}}) for implementation in optoelectronic and thermoelectric technologies},\ }\href {https://doi.org/10.1016/j.matchemphys.2025.130911} {\bibfield  {journal} {\bibinfo  {journal} {Materials Chemistry and Physics}\ }\textbf {\bibinfo {volume} {342}},\ \bibinfo {pages} {130911} (\bibinfo {year} {2025})}\BibitemShut {NoStop}%
\bibitem [{\citenamefont {Ayyaz}\ \emph {et~al.}(2025{\natexlab{b}})\citenamefont {Ayyaz}, \citenamefont {Alkhaldi}, \citenamefont {Dar}, \citenamefont {Ali}, \citenamefont {Mahmood}, \citenamefont {{Al-Buriahi}}, \citenamefont {Boukhris},\ and\ \citenamefont {{Al-Anazy}}}]{Ayyaz_2025a}%
  \BibitemOpen
  \bibfield  {author} {\bibinfo {author} {\bibfnamefont {A.}~\bibnamefont {Ayyaz}}, \bibinfo {author} {\bibfnamefont {N.~D.}\ \bibnamefont {Alkhaldi}}, \bibinfo {author} {\bibfnamefont {S.~A.}\ \bibnamefont {Dar}}, \bibinfo {author} {\bibfnamefont {S.~K.}\ \bibnamefont {Ali}}, \bibinfo {author} {\bibfnamefont {Q.}~\bibnamefont {Mahmood}}, \bibinfo {author} {\bibfnamefont {M.~S.}\ \bibnamefont {{Al-Buriahi}}}, \bibinfo {author} {\bibfnamefont {I.}~\bibnamefont {Boukhris}},\ and\ \bibinfo {author} {\bibfnamefont {M.~M.}\ \bibnamefont {{Al-Anazy}}},\ }\bibfield  {title} {\bibinfo {title} {Unveiling theoretical findings on optoelectronic and transport characteristics of {{Na2ScAu}}({{Cl}}/{{Br}}/{{I}})6 for energy conversion applications},\ }\href {https://doi.org/10.1007/s11082-025-08282-1} {\bibfield  {journal} {\bibinfo  {journal} {Optical and Quantum Electronics}\ }\textbf {\bibinfo {volume} {57}},\ \bibinfo {pages} {362} (\bibinfo {year} {2025}{\natexlab{b}})}\BibitemShut {NoStop}%
\bibitem [{\citenamefont {Ayyaz}\ \emph {et~al.}(2025{\natexlab{c}})\citenamefont {Ayyaz}, \citenamefont {Zaman}, \citenamefont {Alkhaldi}, \citenamefont {Ali}, \citenamefont {Boukhris}, \citenamefont {Bouzgarrou}, \citenamefont {{Al-Anazy}},\ and\ \citenamefont {Mahmood}}]{Ayyaz_2025b}%
  \BibitemOpen
  \bibfield  {author} {\bibinfo {author} {\bibfnamefont {A.}~\bibnamefont {Ayyaz}}, \bibinfo {author} {\bibfnamefont {M.}~\bibnamefont {Zaman}}, \bibinfo {author} {\bibfnamefont {H.~D.}\ \bibnamefont {Alkhaldi}}, \bibinfo {author} {\bibfnamefont {H.~I.}\ \bibnamefont {Ali}}, \bibinfo {author} {\bibfnamefont {I.}~\bibnamefont {Boukhris}}, \bibinfo {author} {\bibfnamefont {S.}~\bibnamefont {Bouzgarrou}}, \bibinfo {author} {\bibfnamefont {M.~M.}\ \bibnamefont {{Al-Anazy}}},\ and\ \bibinfo {author} {\bibfnamefont {Q.}~\bibnamefont {Mahmood}},\ }\bibfield  {title} {\bibinfo {title} {Computational screening of appealing perspectives of indium-based halide double perovskites {{In}}{\textsubscript{2}} {{AgSbX}}{\textsubscript{6}} ({{X}} = {{Cl}}, {{Br}}, and {{I}}) for energy harvesting technologies},\ }\href {https://doi.org/10.1039/D5RA00242G} {\bibfield  {journal} {\bibinfo  {journal} {RSC Advances}\ }\textbf {\bibinfo {volume} {15}},\ \bibinfo {pages} {11128} (\bibinfo {year} {2025}{\natexlab{c}})}\BibitemShut
  {NoStop}%
\bibitem [{\citenamefont {Ayyaz}\ \emph {et~al.}(2025{\natexlab{d}})\citenamefont {Ayyaz}, \citenamefont {Mahmood}, \citenamefont {Dar}, \citenamefont {Touqir}, \citenamefont {Amin}, \citenamefont {Boukhris},\ and\ \citenamefont {Bouzgarrou}}]{Ayyaz_2025c}%
  \BibitemOpen
  \bibfield  {author} {\bibinfo {author} {\bibfnamefont {A.}~\bibnamefont {Ayyaz}}, \bibinfo {author} {\bibfnamefont {Q.}~\bibnamefont {Mahmood}}, \bibinfo {author} {\bibfnamefont {S.~A.}\ \bibnamefont {Dar}}, \bibinfo {author} {\bibfnamefont {M.}~\bibnamefont {Touqir}}, \bibinfo {author} {\bibfnamefont {L.~G.}\ \bibnamefont {Amin}}, \bibinfo {author} {\bibfnamefont {I.}~\bibnamefont {Boukhris}},\ and\ \bibinfo {author} {\bibfnamefont {S.}~\bibnamefont {Bouzgarrou}},\ }\bibfield  {title} {\bibinfo {title} {{{DFT-based}} comparative study of mechanical, electro-optic, and transport response of halide double perovskites {{Na2MAlZ6}} ({{M}} = {{Ag}}, {{Cu}}; {{Z}} = {{Br}}, {{I}}) for green energy applications},\ }\href {https://doi.org/10.1016/j.mseb.2025.118250} {\bibfield  {journal} {\bibinfo  {journal} {Materials Science and Engineering: B}\ }\textbf {\bibinfo {volume} {317}},\ \bibinfo {pages} {118250} (\bibinfo {year} {2025}{\natexlab{d}})}\BibitemShut {NoStop}%
\bibitem [{\citenamefont {Baaalla}\ \emph {et~al.}(2025)\citenamefont {Baaalla}, \citenamefont {Rehman}, \citenamefont {Absike}, \citenamefont {Waqas~Iqbal}, \citenamefont {Sarwar}, \citenamefont {Ismayilova}, \citenamefont {Alrobei},\ and\ \citenamefont {Mohammad}}]{Baaalla_2025}%
  \BibitemOpen
  \bibfield  {author} {\bibinfo {author} {\bibfnamefont {N.}~\bibnamefont {Baaalla}}, \bibinfo {author} {\bibfnamefont {I.~U.}\ \bibnamefont {Rehman}}, \bibinfo {author} {\bibfnamefont {H.}~\bibnamefont {Absike}}, \bibinfo {author} {\bibfnamefont {M.}~\bibnamefont {Waqas~Iqbal}}, \bibinfo {author} {\bibfnamefont {S.}~\bibnamefont {Sarwar}}, \bibinfo {author} {\bibfnamefont {N.}~\bibnamefont {Ismayilova}}, \bibinfo {author} {\bibfnamefont {H.}~\bibnamefont {Alrobei}},\ and\ \bibinfo {author} {\bibfnamefont {A.}~\bibnamefont {Mohammad}},\ }\bibfield  {title} {\bibinfo {title} {Computational study of {{Rb2LiWX6}} ({{X}} = {{Cl}}, {{Br}}): {{Electronic}}, mechanical, optical, and thermoelectric properties for energy applications},\ }\href {https://doi.org/10.1016/j.micrna.2025.208285} {\bibfield  {journal} {\bibinfo  {journal} {Micro and Nanostructures}\ }\textbf {\bibinfo {volume} {207}},\ \bibinfo {pages} {208285} (\bibinfo {year} {2025})}\BibitemShut {NoStop}%
\bibitem [{\citenamefont {Benahmedi}\ \emph {et~al.}(2025{\natexlab{c}})\citenamefont {Benahmedi}, \citenamefont {Besbes},\ and\ \citenamefont {Djelti}}]{Benahmedi_2025b}%
  \BibitemOpen
  \bibfield  {author} {\bibinfo {author} {\bibfnamefont {L.}~\bibnamefont {Benahmedi}}, \bibinfo {author} {\bibfnamefont {A.}~\bibnamefont {Besbes}},\ and\ \bibinfo {author} {\bibfnamefont {R.}~\bibnamefont {Djelti}},\ }\bibfield  {title} {\bibinfo {title} {First-principles study of electro-structural, mechanical, optical, and thermal properties of hexagonal chalcogenide perovskites {{CsTaX3}} ({{X}} = {{S}}, {{Se}})},\ }\href {https://doi.org/10.1016/j.physb.2025.417452} {\bibfield  {journal} {\bibinfo  {journal} {Physica B: Condensed Matter}\ }\textbf {\bibinfo {volume} {714}},\ \bibinfo {pages} {417452} (\bibinfo {year} {2025}{\natexlab{c}})}\BibitemShut {NoStop}%
\bibitem [{\citenamefont {Bouferrache}\ \emph {et~al.}(2025)\citenamefont {Bouferrache}, \citenamefont {Ghebouli}, \citenamefont {Ghebouli}, \citenamefont {Fatmi},\ and\ \citenamefont {Ahmed}}]{Bouferrache_2025}%
  \BibitemOpen
  \bibfield  {author} {\bibinfo {author} {\bibfnamefont {K.}~\bibnamefont {Bouferrache}}, \bibinfo {author} {\bibfnamefont {M.~A.}\ \bibnamefont {Ghebouli}}, \bibinfo {author} {\bibfnamefont {B.}~\bibnamefont {Ghebouli}}, \bibinfo {author} {\bibfnamefont {M.}~\bibnamefont {Fatmi}},\ and\ \bibinfo {author} {\bibfnamefont {S.~I.}\ \bibnamefont {Ahmed}},\ }\bibfield  {title} {\bibinfo {title} {Organic--inorganic hexahalometalate-crystal semiconductor {{K}}{\textsubscript{2}} ({{Sn}}, {{Se}}, {{Te}}){{Br}}{\textsubscript{6}} hybrid double perovskites for solar energy applications},\ }\href {https://doi.org/10.1039/D5RA00862J} {\bibfield  {journal} {\bibinfo  {journal} {RSC Advances}\ }\textbf {\bibinfo {volume} {15}},\ \bibinfo {pages} {11923} (\bibinfo {year} {2025})}\BibitemShut {NoStop}%
\bibitem [{\citenamefont {Bourahla}\ \emph {et~al.}(2025)\citenamefont {Bourahla}, \citenamefont {Chiker}, \citenamefont {Khachai}, \citenamefont {Khenata}, \citenamefont {Bouhemadou}, \citenamefont {Singh}, \citenamefont {{Bin-Omran}}, \citenamefont {Eithiraj}, \citenamefont {Jappor},\ and\ \citenamefont {Khan}}]{Bourahla_2025}%
  \BibitemOpen
  \bibfield  {author} {\bibinfo {author} {\bibfnamefont {C.}~\bibnamefont {Bourahla}}, \bibinfo {author} {\bibfnamefont {F.}~\bibnamefont {Chiker}}, \bibinfo {author} {\bibfnamefont {H.}~\bibnamefont {Khachai}}, \bibinfo {author} {\bibfnamefont {R.}~\bibnamefont {Khenata}}, \bibinfo {author} {\bibfnamefont {A.}~\bibnamefont {Bouhemadou}}, \bibinfo {author} {\bibfnamefont {D.}~\bibnamefont {Singh}}, \bibinfo {author} {\bibfnamefont {S.}~\bibnamefont {{Bin-Omran}}}, \bibinfo {author} {\bibfnamefont {R.}~\bibnamefont {Eithiraj}}, \bibinfo {author} {\bibfnamefont {H.~R.}\ \bibnamefont {Jappor}},\ and\ \bibinfo {author} {\bibfnamefont {S.~A.}\ \bibnamefont {Khan}},\ }\bibfield  {title} {\bibinfo {title} {Unveiling the structural, optical coating and thermoelectric characteristics of kesterite-quaternary chalcogenides {{Ag2InGaX4}} ({{X}} = {{S}}, {{Se}}, {{Te}}) via {{DFT}} study},\ }\href {https://doi.org/10.1016/j.jpcs.2025.112970} {\bibfield  {journal} {\bibinfo  {journal} {Journal of Physics and Chemistry of
  Solids}\ }\textbf {\bibinfo {volume} {207}},\ \bibinfo {pages} {112970} (\bibinfo {year} {2025})}\BibitemShut {NoStop}%
\bibitem [{\citenamefont {Chelh}\ \emph {et~al.}(2025)\citenamefont {Chelh}, \citenamefont {Dahbi},\ and\ \citenamefont {{Ez-Zahraouy}}}]{Chelh_2025}%
  \BibitemOpen
  \bibfield  {author} {\bibinfo {author} {\bibfnamefont {A.}~\bibnamefont {Chelh}}, \bibinfo {author} {\bibfnamefont {S.}~\bibnamefont {Dahbi}},\ and\ \bibinfo {author} {\bibfnamefont {H.}~\bibnamefont {{Ez-Zahraouy}}},\ }\bibfield  {title} {\bibinfo {title} {Ab-initio study of the structural, electronic, optical, and thermoelectric properties of chalcogenide-doped {{Sr2UZnO6}}},\ }\href {https://doi.org/10.1016/j.ssc.2025.116024} {\bibfield  {journal} {\bibinfo  {journal} {Solid State Communications}\ }\textbf {\bibinfo {volume} {404}},\ \bibinfo {pages} {116024} (\bibinfo {year} {2025})}\BibitemShut {NoStop}%
\bibitem [{\citenamefont {El~Akkel}\ and\ \citenamefont {{Ez-Zahraouy}}(2025)}]{ElAkkel_2025}%
  \BibitemOpen
  \bibfield  {author} {\bibinfo {author} {\bibfnamefont {M.}~\bibnamefont {El~Akkel}}\ and\ \bibinfo {author} {\bibfnamefont {H.}~\bibnamefont {{Ez-Zahraouy}}},\ }\bibfield  {title} {\bibinfo {title} {Multifunctional double perovskites {{Cs2BI6}} ({{B}} = {{Ti}}, {{Ge}}, {{Se}}, {{Sn}}, and {{Te}}) for solar energy harvesting: Photovoltaic, photocatalytic, and thermoelectric pathways},\ }\href {https://doi.org/10.1016/j.energy.2025.137928} {\bibfield  {journal} {\bibinfo  {journal} {Energy}\ }\textbf {\bibinfo {volume} {335}},\ \bibinfo {pages} {137928} (\bibinfo {year} {2025})}\BibitemShut {NoStop}%
\bibitem [{\citenamefont {Farooq}\ \emph {et~al.}(2025)\citenamefont {Farooq}, \citenamefont {Moussa}, \citenamefont {Mumtaz},\ and\ \citenamefont {Nazir}}]{Farooq_2025}%
  \BibitemOpen
  \bibfield  {author} {\bibinfo {author} {\bibfnamefont {A.}~\bibnamefont {Farooq}}, \bibinfo {author} {\bibfnamefont {I.~M.}\ \bibnamefont {Moussa}}, \bibinfo {author} {\bibfnamefont {S.}~\bibnamefont {Mumtaz}},\ and\ \bibinfo {author} {\bibfnamefont {S.}~\bibnamefont {Nazir}},\ }\bibfield  {title} {\bibinfo {title} {Stable large half-metallic energy-gap, high thermoelectric performance, and enhanced {{T C}} in {{LaMn}} 3 {{Ru}} 2 {{M}} 2 {{O12}} ({{M}} = {{Mn}} and {{Fe}}): {{Strain}} modulations},\ }\href {https://doi.org/10.1016/j.mtcomm.2025.112306} {\bibfield  {journal} {\bibinfo  {journal} {Materials Today Communications}\ }\textbf {\bibinfo {volume} {45}},\ \bibinfo {pages} {112306} (\bibinfo {year} {2025})}\BibitemShut {NoStop}%
\bibitem [{\citenamefont {Hamza}\ \emph {et~al.}(2025)\citenamefont {Hamza}, \citenamefont {Ishfaq}, \citenamefont {Aldaghfag}, \citenamefont {Yaseen}, \citenamefont {Murtaza},\ and\ \citenamefont {{Nasarullah}}}]{Hamza_2025}%
  \BibitemOpen
  \bibfield  {author} {\bibinfo {author} {\bibfnamefont {M.}~\bibnamefont {Hamza}}, \bibinfo {author} {\bibfnamefont {M.}~\bibnamefont {Ishfaq}}, \bibinfo {author} {\bibfnamefont {S.~A.}\ \bibnamefont {Aldaghfag}}, \bibinfo {author} {\bibfnamefont {M.}~\bibnamefont {Yaseen}}, \bibinfo {author} {\bibfnamefont {S.}~\bibnamefont {Murtaza}},\ and\ \bibinfo {author} {\bibnamefont {{Nasarullah}}},\ }\bibfield  {title} {\bibinfo {title} {First-principles insights into {{X2MgS4}} ({{X}}= {{Pm}}, {{Pr}}) spinels for energy harvesting and spintronic applications},\ }\href {https://doi.org/10.1016/j.physb.2025.417435} {\bibfield  {journal} {\bibinfo  {journal} {Physica B: Condensed Matter}\ }\textbf {\bibinfo {volume} {714}},\ \bibinfo {pages} {417435} (\bibinfo {year} {2025})}\BibitemShut {NoStop}%
\bibitem [{\citenamefont {Israr}\ \emph {et~al.}(2025)\citenamefont {Israr}, \citenamefont {Rehman}, \citenamefont {Jehangir}, \citenamefont {{EL-Gawaad}},\ and\ \citenamefont {Farooq}}]{Israr_2025}%
  \BibitemOpen
  \bibfield  {author} {\bibinfo {author} {\bibfnamefont {N.}~\bibnamefont {Israr}}, \bibinfo {author} {\bibfnamefont {W.~U.}\ \bibnamefont {Rehman}}, \bibinfo {author} {\bibfnamefont {M.~A.}\ \bibnamefont {Jehangir}}, \bibinfo {author} {\bibfnamefont {N.~S.~A.}\ \bibnamefont {{EL-Gawaad}}},\ and\ \bibinfo {author} {\bibfnamefont {U.}~\bibnamefont {Farooq}},\ }\bibfield  {title} {\bibinfo {title} {First-{{Principles Investigation}} of {{Narrow Bandgap Halide Double Perovskites A2AgSbI6}} ({{A}} = {{K}}, {{Rb}})},\ }\bibfield  {journal} {\bibinfo  {journal} {Journal of Inorganic and Organometallic Polymers and Materials}\ }\href {https://doi.org/10.1007/s10904-025-03695-7} {10.1007/s10904-025-03695-7} (\bibinfo {year} {2025})\BibitemShut {NoStop}%
\bibitem [{\citenamefont {Khatar}\ \emph {et~al.}(2025{\natexlab{b}})\citenamefont {Khatar}, \citenamefont {Houari}, \citenamefont {Lantri}, \citenamefont {Bentata}, \citenamefont {Bouadjemi}, \citenamefont {Aziz}, \citenamefont {Boucherdoud},\ and\ \citenamefont {Boudjelal}}]{Khatar_2025a}%
  \BibitemOpen
  \bibfield  {author} {\bibinfo {author} {\bibfnamefont {A.}~\bibnamefont {Khatar}}, \bibinfo {author} {\bibfnamefont {M.}~\bibnamefont {Houari}}, \bibinfo {author} {\bibfnamefont {T.}~\bibnamefont {Lantri}}, \bibinfo {author} {\bibfnamefont {S.}~\bibnamefont {Bentata}}, \bibinfo {author} {\bibfnamefont {B.}~\bibnamefont {Bouadjemi}}, \bibinfo {author} {\bibfnamefont {Z.}~\bibnamefont {Aziz}}, \bibinfo {author} {\bibfnamefont {A.}~\bibnamefont {Boucherdoud}},\ and\ \bibinfo {author} {\bibfnamefont {M.}~\bibnamefont {Boudjelal}},\ }\bibfield  {title} {\bibinfo {title} {Advanced materials for next-generation devices: Insights into the structural, optical, and thermoelectric properties of {{Hf2Pd2AlBi}} and {{Hf2Pd2AlSb}} alloys},\ }\bibfield  {journal} {\bibinfo  {journal} {Indian Journal of Physics}\ }\href {https://doi.org/10.1007/s12648-025-03643-8} {10.1007/s12648-025-03643-8} (\bibinfo {year} {2025}{\natexlab{b}})\BibitemShut {NoStop}%
\bibitem [{\citenamefont {Mbilo}\ \emph {et~al.}(2025{\natexlab{a}})\citenamefont {Mbilo}, \citenamefont {Musembi}, \citenamefont {Kachira}, \citenamefont {Nyamunga}, \citenamefont {Musanyi}, \citenamefont {Wafula},\ and\ \citenamefont {Yusuf}}]{Mbilo_2025}%
  \BibitemOpen
  \bibfield  {author} {\bibinfo {author} {\bibfnamefont {M.}~\bibnamefont {Mbilo}}, \bibinfo {author} {\bibfnamefont {R.}~\bibnamefont {Musembi}}, \bibinfo {author} {\bibfnamefont {J.~P.}\ \bibnamefont {Kachira}}, \bibinfo {author} {\bibfnamefont {M.}~\bibnamefont {Nyamunga}}, \bibinfo {author} {\bibfnamefont {I.}~\bibnamefont {Musanyi}}, \bibinfo {author} {\bibfnamefont {S.}~\bibnamefont {Wafula}},\ and\ \bibinfo {author} {\bibfnamefont {M.}~\bibnamefont {Yusuf}},\ }\bibfield  {title} {\bibinfo {title} {First {{Principles Study}} of the {{Properties}} of {{Cs2GaAgF6 Double Halide Perovskite Compound}} for {{Optoelectronic}} and {{Thermoelectric Applications}}},\ }\bibfield  {journal} {\bibinfo  {journal} {Journal of Inorganic and Organometallic Polymers and Materials}\ }\href {https://doi.org/10.1007/s10904-025-03838-w} {10.1007/s10904-025-03838-w} (\bibinfo {year} {2025}{\natexlab{a}})\BibitemShut {NoStop}%
\bibitem [{\citenamefont {Mbilo}\ \emph {et~al.}(2025{\natexlab{b}})\citenamefont {Mbilo}, \citenamefont {Musembi}, \citenamefont {Kachira}, \citenamefont {Onsate}, \citenamefont {Keheze},\ and\ \citenamefont {Mapasha}}]{Mbilo_2025a}%
  \BibitemOpen
  \bibfield  {author} {\bibinfo {author} {\bibfnamefont {M.}~\bibnamefont {Mbilo}}, \bibinfo {author} {\bibfnamefont {R.}~\bibnamefont {Musembi}}, \bibinfo {author} {\bibfnamefont {J.~P.}\ \bibnamefont {Kachira}}, \bibinfo {author} {\bibfnamefont {W.~N.}\ \bibnamefont {Onsate}}, \bibinfo {author} {\bibfnamefont {F.~M.}\ \bibnamefont {Keheze}},\ and\ \bibinfo {author} {\bibfnamefont {R.~E.}\ \bibnamefont {Mapasha}},\ }\bibfield  {title} {\bibinfo {title} {Insights into the optoelectronic and thermoelectric properties of lead-free {{Rb2NaIrF6}} double perovskite compound: {{A}} first-principles study},\ }\href {https://doi.org/10.1016/j.rinp.2025.108349} {\bibfield  {journal} {\bibinfo  {journal} {Results in Physics}\ }\textbf {\bibinfo {volume} {75}},\ \bibinfo {pages} {108349} (\bibinfo {year} {2025}{\natexlab{b}})}\BibitemShut {NoStop}%
\bibitem [{\citenamefont {Musanyi}\ \emph {et~al.}(2025)\citenamefont {Musanyi}, \citenamefont {Mbilo}, \citenamefont {Musembi}, \citenamefont {Kachira}, \citenamefont {Nyongesa}, \citenamefont {Nyamunga},\ and\ \citenamefont {Wafula}}]{Musanyi_2025}%
  \BibitemOpen
  \bibfield  {author} {\bibinfo {author} {\bibfnamefont {I.}~\bibnamefont {Musanyi}}, \bibinfo {author} {\bibfnamefont {M.}~\bibnamefont {Mbilo}}, \bibinfo {author} {\bibfnamefont {R.}~\bibnamefont {Musembi}}, \bibinfo {author} {\bibfnamefont {J.}~\bibnamefont {Kachira}}, \bibinfo {author} {\bibfnamefont {F.}~\bibnamefont {Nyongesa}}, \bibinfo {author} {\bibfnamefont {M.}~\bibnamefont {Nyamunga}},\ and\ \bibinfo {author} {\bibfnamefont {S.}~\bibnamefont {Wafula}},\ }\bibfield  {title} {\bibinfo {title} {Optoelectronic and thermoelectric properties of the {{K2SbAu}} zintl phase ternary compound using first principles methods},\ }\href {https://doi.org/10.1016/j.cocom.2025.e01073} {\bibfield  {journal} {\bibinfo  {journal} {Computational Condensed Matter}\ }\textbf {\bibinfo {volume} {44}},\ \bibinfo {pages} {e01073} (\bibinfo {year} {2025})}\BibitemShut {NoStop}%
\bibitem [{\citenamefont {Mustafa}\ \emph {et~al.}(2025)\citenamefont {Mustafa}, \citenamefont {Younas}, \citenamefont {Yasir}, \citenamefont {Noor}, \citenamefont {Mumtaz}, \citenamefont {Moussa},\ and\ \citenamefont {Elansary}}]{Mustafa_2025}%
  \BibitemOpen
  \bibfield  {author} {\bibinfo {author} {\bibfnamefont {G.~M.}\ \bibnamefont {Mustafa}}, \bibinfo {author} {\bibfnamefont {B.}~\bibnamefont {Younas}}, \bibinfo {author} {\bibfnamefont {M.~A.}\ \bibnamefont {Yasir}}, \bibinfo {author} {\bibfnamefont {N.}~\bibnamefont {Noor}}, \bibinfo {author} {\bibfnamefont {S.}~\bibnamefont {Mumtaz}}, \bibinfo {author} {\bibfnamefont {I.~M.}\ \bibnamefont {Moussa}},\ and\ \bibinfo {author} {\bibfnamefont {H.~O.}\ \bibnamefont {Elansary}},\ }\bibfield  {title} {\bibinfo {title} {Theoretical analysis of structural, optoelectronic and transport features of {{Tl2Re}}({{Cl}}/{{Br}})6 double perovskites for energy harvesting applications},\ }\href {https://doi.org/10.1016/j.cplett.2025.142169} {\bibfield  {journal} {\bibinfo  {journal} {Chemical Physics Letters}\ }\textbf {\bibinfo {volume} {874--875}},\ \bibinfo {pages} {142169} (\bibinfo {year} {2025})}\BibitemShut {NoStop}%
\bibitem [{\citenamefont {Nasarullah}\ \emph {et~al.}(2025)\citenamefont {Nasarullah}, \citenamefont {Kanjariya}, \citenamefont {Manjunath}, \citenamefont {Kumar}, \citenamefont {Choudhury}, \citenamefont {Alqorashi}, \citenamefont {{Al-Hasnaawei}}, \citenamefont {Shah}, \citenamefont {Shankhyan}, \citenamefont {Shabir},\ and\ \citenamefont {Faizan}}]{Nasarullah_2025}%
  \BibitemOpen
  \bibfield  {author} {\bibinfo {author} {\bibfnamefont {{\relax Mr}.}~\bibnamefont {Nasarullah}}, \bibinfo {author} {\bibfnamefont {P.}~\bibnamefont {Kanjariya}}, \bibinfo {author} {\bibfnamefont {H.~R.}\ \bibnamefont {Manjunath}}, \bibinfo {author} {\bibfnamefont {A.}~\bibnamefont {Kumar}}, \bibinfo {author} {\bibfnamefont {S.}~\bibnamefont {Choudhury}}, \bibinfo {author} {\bibfnamefont {A.~K.}\ \bibnamefont {Alqorashi}}, \bibinfo {author} {\bibfnamefont {S.}~\bibnamefont {{Al-Hasnaawei}}}, \bibinfo {author} {\bibfnamefont {S.~K.}\ \bibnamefont {Shah}}, \bibinfo {author} {\bibfnamefont {A.}~\bibnamefont {Shankhyan}}, \bibinfo {author} {\bibfnamefont {A.}~\bibnamefont {Shabir}},\ and\ \bibinfo {author} {\bibfnamefont {M.}~\bibnamefont {Faizan}},\ }\href {https://doi.org/10.2139/ssrn.5227378} {\bibinfo {title} {Probing the {{Physical Properties}} of {{Double Perovskites Sr2ce}}({{Ni}},{{Sn}}){{O6 Through First-Principles Investigations}} for {{Renewable Energy Applications}}}} (\bibinfo {year}
  {2025})\BibitemShut {NoStop}%
\bibitem [{\citenamefont {{Nasarullah}}\ \emph {et~al.}(2025)\citenamefont {{Nasarullah}}, \citenamefont {Nazar}, \citenamefont {Sajid}, \citenamefont {Naqvi}, \citenamefont {{Al-Hazmi}},\ and\ \citenamefont {Alawaideh}}]{Nasarullah_2025a}%
  \BibitemOpen
  \bibfield  {author} {\bibinfo {author} {\bibnamefont {{Nasarullah}}}, \bibinfo {author} {\bibfnamefont {M.}~\bibnamefont {Nazar}}, \bibinfo {author} {\bibfnamefont {M.}~\bibnamefont {Sajid}}, \bibinfo {author} {\bibfnamefont {S.~M. K.~A.}\ \bibnamefont {Naqvi}}, \bibinfo {author} {\bibfnamefont {G.~A. A.~M.}\ \bibnamefont {{Al-Hazmi}}},\ and\ \bibinfo {author} {\bibfnamefont {Y.~M.}\ \bibnamefont {Alawaideh}},\ }\bibfield  {title} {\bibinfo {title} {First-principles study of {{A-site}} cation effects in halide double perovskites {{A2NaMoI6}} ({{A}} = {{Cs}}, {{Rb}}) for photovoltaic applications},\ }\bibfield  {journal} {\bibinfo  {journal} {Journal of Sol-Gel Science and Technology}\ }\href {https://doi.org/10.1007/s10971-025-06713-9} {10.1007/s10971-025-06713-9} (\bibinfo {year} {2025})\BibitemShut {NoStop}%
\bibitem [{\citenamefont {Nazir}\ \emph {et~al.}(2025{\natexlab{f}})\citenamefont {Nazir}, \citenamefont {Noor}, \citenamefont {Majeed}, \citenamefont {Mumtaz},\ and\ \citenamefont {Elhindi}}]{Nazir_2025e}%
  \BibitemOpen
  \bibfield  {author} {\bibinfo {author} {\bibfnamefont {S.}~\bibnamefont {Nazir}}, \bibinfo {author} {\bibfnamefont {N.~A.}\ \bibnamefont {Noor}}, \bibinfo {author} {\bibfnamefont {F.}~\bibnamefont {Majeed}}, \bibinfo {author} {\bibfnamefont {S.}~\bibnamefont {Mumtaz}},\ and\ \bibinfo {author} {\bibfnamefont {K.~M.}\ \bibnamefont {Elhindi}},\ }\bibfield  {title} {\bibinfo {title} {First-{{Principles Calculations}} of {{Halide Double Perovskite K2InBiX6}} ({{X}} = {{Cl}}, {{Br}}, {{I}}) for {{Solar Cell Applications}}},\ }\href {https://doi.org/10.1007/s10904-025-03618-6} {\bibfield  {journal} {\bibinfo  {journal} {Journal of Inorganic and Organometallic Polymers and Materials}\ }\textbf {\bibinfo {volume} {35}},\ \bibinfo {pages} {5721} (\bibinfo {year} {2025}{\natexlab{f}})}\BibitemShut {NoStop}%
\bibitem [{\citenamefont {Noor}\ \emph {et~al.}(2025{\natexlab{a}})\citenamefont {Noor}, \citenamefont {{Abu-Farsakh}},\ and\ \citenamefont {Bououdina}}]{Noor_2025}%
  \BibitemOpen
  \bibfield  {author} {\bibinfo {author} {\bibfnamefont {N.}~\bibnamefont {Noor}}, \bibinfo {author} {\bibfnamefont {H.}~\bibnamefont {{Abu-Farsakh}}},\ and\ \bibinfo {author} {\bibfnamefont {M.}~\bibnamefont {Bououdina}},\ }\bibfield  {title} {\bibinfo {title} {Unveiling the optoelectronic and thermoelectric characteristics of {{Rb-based}} double perovskites for solar cell applications},\ }\href {https://doi.org/10.1016/j.inoche.2025.115077} {\bibfield  {journal} {\bibinfo  {journal} {Inorganic Chemistry Communications}\ }\textbf {\bibinfo {volume} {180}},\ \bibinfo {pages} {115077} (\bibinfo {year} {2025}{\natexlab{a}})}\BibitemShut {NoStop}%
\bibitem [{\citenamefont {Noor}\ \emph {et~al.}(2025{\natexlab{b}})\citenamefont {Noor}, \citenamefont {Boota}, \citenamefont {Maqsood}, \citenamefont {Khan}, \citenamefont {Niaz}, \citenamefont {Mumtaz},\ and\ \citenamefont {Elansary}}]{Noor_2025a}%
  \BibitemOpen
  \bibfield  {author} {\bibinfo {author} {\bibfnamefont {N.~A.}\ \bibnamefont {Noor}}, \bibinfo {author} {\bibfnamefont {M.}~\bibnamefont {Boota}}, \bibinfo {author} {\bibfnamefont {S.}~\bibnamefont {Maqsood}}, \bibinfo {author} {\bibfnamefont {M.~A.}\ \bibnamefont {Khan}}, \bibinfo {author} {\bibfnamefont {S.}~\bibnamefont {Niaz}}, \bibinfo {author} {\bibfnamefont {S.}~\bibnamefont {Mumtaz}},\ and\ \bibinfo {author} {\bibfnamefont {H.~O.}\ \bibnamefont {Elansary}},\ }\bibfield  {title} {\bibinfo {title} {{{DFT Calculations}} of {{Structural}}, {{Opto-Electronic}}, and {{Transport Properties}} of {{HgLu2}}({{S}}/{{Se}})4 {{Spinel Compounds}}},\ }\bibfield  {journal} {\bibinfo  {journal} {Journal of Inorganic and Organometallic Polymers and Materials}\ }\href {https://doi.org/10.1007/s10904-025-03827-z} {10.1007/s10904-025-03827-z} (\bibinfo {year} {2025}{\natexlab{b}})\BibitemShut {NoStop}%
\bibitem [{\citenamefont {Saeed}\ \emph {et~al.}(2025)\citenamefont {Saeed}, \citenamefont {Khan}, \citenamefont {Pervaiz}, \citenamefont {Ali}, \citenamefont {Mumtaz}, \citenamefont {Ali}, \citenamefont {Elansary}, \citenamefont {Mahmoud}, \citenamefont {Moussa},\ and\ \citenamefont {Saeed}}]{Saeed_2025}%
  \BibitemOpen
  \bibfield  {author} {\bibinfo {author} {\bibfnamefont {M.~U.}\ \bibnamefont {Saeed}}, \bibinfo {author} {\bibfnamefont {A.}~\bibnamefont {Khan}}, \bibinfo {author} {\bibfnamefont {S.}~\bibnamefont {Pervaiz}}, \bibinfo {author} {\bibfnamefont {S.}~\bibnamefont {Ali}}, \bibinfo {author} {\bibfnamefont {S.}~\bibnamefont {Mumtaz}}, \bibinfo {author} {\bibfnamefont {H.}~\bibnamefont {Ali}}, \bibinfo {author} {\bibfnamefont {H.~O.}\ \bibnamefont {Elansary}}, \bibinfo {author} {\bibfnamefont {E.~A.}\ \bibnamefont {Mahmoud}}, \bibinfo {author} {\bibfnamefont {I.~M.}\ \bibnamefont {Moussa}},\ and\ \bibinfo {author} {\bibfnamefont {Y.}~\bibnamefont {Saeed}},\ }\bibfield  {title} {\bibinfo {title} {First-principles investigation of structural, electronic, optical, elastic, and thermal properties of double perovskites {{Cs}} 2 {{AgMX}} 6 ({{M}} = {{Al}}, {{In}}, {{Ga}}; {{X}} = {{Br}}, {{Cl}}) for thermoelectric and water splitting applications},\ }\href {https://doi.org/10.1016/j.jpcs.2025.112968} {\bibfield
  {journal} {\bibinfo  {journal} {Journal of Physics and Chemistry of Solids}\ }\textbf {\bibinfo {volume} {207}},\ \bibinfo {pages} {112968} (\bibinfo {year} {2025})}\BibitemShut {NoStop}%
\bibitem [{\citenamefont {Safdar}\ \emph {et~al.}(2025)\citenamefont {Safdar}, \citenamefont {Mumtaz}, \citenamefont {Moussa},\ and\ \citenamefont {Nazir}}]{Safdar_2025}%
  \BibitemOpen
  \bibfield  {author} {\bibinfo {author} {\bibfnamefont {H.}~\bibnamefont {Safdar}}, \bibinfo {author} {\bibfnamefont {S.}~\bibnamefont {Mumtaz}}, \bibinfo {author} {\bibfnamefont {I.~M.}\ \bibnamefont {Moussa}},\ and\ \bibinfo {author} {\bibfnamefont {S.}~\bibnamefont {Nazir}},\ }\bibfield  {title} {\bibinfo {title} {Above room temperature {{T C}} , 100\% spin-polarization, and a high thermoelectric response in the ordered {{CaCu}} 3 {{Mn}} 2 {{Ir}} 2 {{O12}}},\ }\href {https://doi.org/10.1016/j.physb.2025.417149} {\bibfield  {journal} {\bibinfo  {journal} {Physica B: Condensed Matter}\ }\textbf {\bibinfo {volume} {707}},\ \bibinfo {pages} {417149} (\bibinfo {year} {2025})}\BibitemShut {NoStop}%
\bibitem [{\citenamefont {Yaseen}\ \emph {et~al.}(2025)\citenamefont {Yaseen}, \citenamefont {Ain}, \citenamefont {Murtaza}, \citenamefont {Munir}, \citenamefont {Ahmed}, \citenamefont {Qaid},\ and\ \citenamefont {Aldwayyan}}]{Yaseen_2025}%
  \BibitemOpen
  \bibfield  {author} {\bibinfo {author} {\bibfnamefont {H.}~\bibnamefont {Yaseen}}, \bibinfo {author} {\bibfnamefont {Q.}~\bibnamefont {Ain}}, \bibinfo {author} {\bibfnamefont {H.}~\bibnamefont {Murtaza}}, \bibinfo {author} {\bibfnamefont {J.}~\bibnamefont {Munir}}, \bibinfo {author} {\bibfnamefont {A.~A.~A.}\ \bibnamefont {Ahmed}}, \bibinfo {author} {\bibfnamefont {S.~M.~H.}\ \bibnamefont {Qaid}},\ and\ \bibinfo {author} {\bibfnamefont {A.~S.}\ \bibnamefont {Aldwayyan}},\ }\bibfield  {title} {\bibinfo {title} {The {{Influence}} of {{Multiple Potentials}} on the {{Bandgap Improvement}} and the {{Physical Properties}} of {{Cs2InSbF6 Fluoroperovskite}}: A {{DFT Study}}},\ }\bibfield  {journal} {\bibinfo  {journal} {Journal of Inorganic and Organometallic Polymers and Materials}\ }\href {https://doi.org/10.1007/s10904-025-03903-4} {10.1007/s10904-025-03903-4} (\bibinfo {year} {2025})\BibitemShut {NoStop}%
\bibitem [{\citenamefont {Yasir}\ \emph {et~al.}(2025{\natexlab{a}})\citenamefont {Yasir}, \citenamefont {Shahzadi}, \citenamefont {Saeed}, \citenamefont {Gassoumi}, \citenamefont {Noor}, \citenamefont {Mumtaz},\ and\ \citenamefont {Laref}}]{Yasir_2025}%
  \BibitemOpen
  \bibfield  {author} {\bibinfo {author} {\bibfnamefont {M.~A.}\ \bibnamefont {Yasir}}, \bibinfo {author} {\bibfnamefont {S.}~\bibnamefont {Shahzadi}}, \bibinfo {author} {\bibfnamefont {U.}~\bibnamefont {Saeed}}, \bibinfo {author} {\bibfnamefont {A.}~\bibnamefont {Gassoumi}}, \bibinfo {author} {\bibfnamefont {N.~A.}\ \bibnamefont {Noor}}, \bibinfo {author} {\bibfnamefont {S.}~\bibnamefont {Mumtaz}},\ and\ \bibinfo {author} {\bibfnamefont {A.}~\bibnamefont {Laref}},\ }\bibfield  {title} {\bibinfo {title} {A {{First-Principles Insight}} to {{Explore}} the {{Optoelectronic}}, {{Thermodynamic}}, and {{Transport Properties}} of {{Ferromagnetic Double Perovskites Li}}{$_{2}$}{{CuOsX}}{$_6$} ({{X}} = {{Cl}}, {{Br}}) for {{Energy Conversion Applications}}},\ }\bibfield  {journal} {\bibinfo  {journal} {Journal of Inorganic and Organometallic Polymers and Materials}\ }\href {https://doi.org/10.1007/s10904-025-03937-8} {10.1007/s10904-025-03937-8} (\bibinfo {year} {2025}{\natexlab{a}})\BibitemShut {NoStop}%
\bibitem [{\citenamefont {Yasir}\ \emph {et~al.}(2025{\natexlab{b}})\citenamefont {Yasir}, \citenamefont {Saeed}, \citenamefont {Noor}, \citenamefont {Mumtaz},\ and\ \citenamefont {Elhindi}}]{Yasir_2025a}%
  \BibitemOpen
  \bibfield  {author} {\bibinfo {author} {\bibfnamefont {M.~A.}\ \bibnamefont {Yasir}}, \bibinfo {author} {\bibfnamefont {U.}~\bibnamefont {Saeed}}, \bibinfo {author} {\bibfnamefont {N.}~\bibnamefont {Noor}}, \bibinfo {author} {\bibfnamefont {S.}~\bibnamefont {Mumtaz}},\ and\ \bibinfo {author} {\bibfnamefont {K.~M.}\ \bibnamefont {Elhindi}},\ }\bibfield  {title} {\bibinfo {title} {Computational strain engineering and {{Hubbard}}'s effect investigation on {{Li2CuWX6}} ({{X}} = {{Cl}}, {{Br}}) double perovskites: {{A}} promising {{Route}} to spintronic {{Innovation}}},\ }\href {https://doi.org/10.1016/j.jpcs.2025.112733} {\bibfield  {journal} {\bibinfo  {journal} {Journal of Physics and Chemistry of Solids}\ }\textbf {\bibinfo {volume} {203}},\ \bibinfo {pages} {112733} (\bibinfo {year} {2025}{\natexlab{b}})}\BibitemShut {NoStop}%
\bibitem [{\citenamefont {Limelette}(2021)}]{limeletteThermopowerFigureMerit2021}%
  \BibitemOpen
  \bibfield  {author} {\bibinfo {author} {\bibfnamefont {P.}~\bibnamefont {Limelette}},\ }\bibfield  {title} {\bibinfo {title} {Thermopower, figure of merit and {{Fermi}} integrals},\ }\href {https://doi.org/10.1038/s41598-021-03760-4} {\bibfield  {journal} {\bibinfo  {journal} {Sci Rep}\ }\textbf {\bibinfo {volume} {11}},\ \bibinfo {pages} {24323} (\bibinfo {year} {2021})}\BibitemShut {NoStop}%
\end{thebibliography}%

\end{document}